\documentclass[preprint,12pt,3p]{elsarticle}
\usepackage{graphicx,pslatex,float,color,array,amssymb,amsmath,euscript,multirow}
\usepackage{subcaption}
\usepackage{lipsum}
\usepackage{siunitx}
\usepackage{adjustbox}

\usepackage{gensymb}
\usepackage{siunitx}

\usepackage[english]{babel}
\usepackage[utf8]{inputenc}
\title{Fatigue life prediction at mesoscopic scale of samples containing casting defects: A novel energy based non-local model}
\author[1,2,cor]{Arjun Kalkur Matpadi Raghavendra}
\author[1]{Vincent Maurel}
\author[2]{Lionel Marcin}
\author[1]{Henry Proudhon}

\address[1]{MINES Paris, PSL University, MAT - Centre des Matériaux, CNRS UMR 7633, BP 87 91003 Evry, France}
\address[2]{Safran Aircraft Engines Villaroche, Rond-Point René Ravaud, 77550 Moissy-Cramayel, France}
\cortext[cor]{Corresponding author: arjun-kalkur.matpadi\_raghavendra@minesparis.psl.eu}

\begin{document}
	\begin{frontmatter}

		\begin{abstract}
			Fatigue failure driven by stress gradients associated to casting defects was studied in two cast nickel-based superalloys. The experimental campaign revealed complex damage phenomena linked to spongeous shrinkages, characterized by their intricate arrangement of defects in the material medium, forming defect clusters. Multiple cracks were observed to initiate from defect volumes, coalescing with neighboring void surfaces along crystallographic planes. Defects were characterized using X-ray computed tomography, and image-based finite element (FE) models were constructed as digital representations of each experimental sample explicitly containing all real casting defects. Numerical simulations of these FE models under the same conditions as the experiments revealed that tortuous defects contain small ligaments where very high local stresses develop. These ligaments initiate early cracks, but due to the limited stressed volumes, these cracks do not drive the life of the material. A thorough comparison of simulations with experiments led to the development of an original method to define stressed volumes and address small ligaments. Finally, a novel energy-based non-local model was proposed, using two parameters to predict the fatigue lives of samples containing casting defects at the mesoscopic scale. The model was validated against samples with varying porosity levels, sizes, and geometries, accurately predicting fatigue lives within a factor of 3 compared to experimental results. This new approach generalizes the application of non-local methods to real casting defects by considering their shape and stressed volumes to estimate fatigue properties.
		\end{abstract}
		
		%	\begin{graphicalabstract}
		%		\includegraphics{grabs}
		%	\end{graphicalabstract}
		
	
		
	\end{frontmatter}

\section{Introduction}
Cast nickel-based superalloys possess high mechanical strength and creep resistance at temperature over \SI{650}{\celsius} making it an excellent material to be used in aircraft engines. Predominantly employed in the production of turbine blades and disks, these alloys, despite their favorable casting characteristics, are susceptible to the formation of cavity defects such as shrinkages and pores, which can significantly impact the mechanical properties. Pores formed by trapped gases are usually smaller in size and spherical in shape whilst shrinkages are large tortuous cavities formed due to the contraction of molten metal during freezing \cite{buffiereExperimentalStudyPorosity2001,NADOT2022106531, rotellaInfluenceNaturalDefects2017,KUNZ201247,wangFatigueBehaviorA356T62001,elkhoukhi2019,bercelliProbabilisticApproachHigh2021}. The characteristics of induced cavities like size, morphology etc., depends on the fabrication process; for instance, additively manufactured materials exhibit cavities arising from unmelted powders, insufficient fusion of powders, and entrapped gases \cite{tammas-williamsXCTAnalysisInfluence2015,murrEFFECTBUILDPARAMETERS}. All of these cavity defects can degrade material's fatigue performance drastically by promoting initiation and propagation of crack driven by stress concentration. Magnitude of this degradation depends on each defect's features. In general, the size and position of defects relative to the free surface are considered as a primary factor influencing the fatigue life. Empirical relationships, as established by El-Haddad, Murakami, and other researchers, indicate an inverse relationship between fatigue limit and defect size \cite{murakamiEffectsDefectsInclusions1994,le_investigation_2018}. On the other hand, surface defects are found to be more deleterious than internal defects, as they are subjected to environmental and additional external factors \cite{yamashitaDefectAnalysisFatigue2018, rotellaInfluenceNaturalDefects2017,NADOT2022106531}. This understanding highlights the critical role of defect characteristics in determining the fatigue performance of nickel-based superalloys in high-temperature applications. Utilising this inter-dependency between defect features and fatigue properties, a few researchers already have proposed parametric fatigue models where the properties are estimated from the prior knowledge of defect characteristics \cite{yang2022,tijani2013,tijani2013detection,szmytka2020}. \\

Sphericity is one of the defect characteristics that estimates how spherical a defect is: a perfect sphere has a value of one and vice versa. Several prior sphericity investigations shows that the shrinkages are tortuous and larger in size with a very low sphericity \cite{matpadi2023generation,buffiereExperimentalStudyPorosity2001,raghavendra2022role}. Unlike dendritic shrinkage that appear as a distinct well defined cavity, spongeous shrinkages appear in the form of clusters \cite{matpadi2023generation,hangaiClusteredShrinkagePores2010}. In a previous work dealing also with casted Ni-based superalloys, spatial point pattern analysis of spongeous shrinkages were used to capture the spatial distribution of defects in material medium \cite{matpadi2023generation}. Two important conclusions were drawn from this analysis: a) there are two different nucleation mechanisms associated to two different types of defects (shrinkages and pores). One nucleating from absence of molten metal while the other due to trapped gases, b) large defects are localised in a specific region of the studied samples, resulting in defect clusters where the smaller defects which are pores are attracted towards larger defects to form defect clusters. \\

This clustered arrangement increases the magnitude of stress concentration as compared to the case where a defect is isolated and is not influenced by its surroundings. Vincent et al \cite{Nadot_ellipsoid} demonstrated that fatigue estimations can be made taking into account the gradients by replacing pores with equivalent ellipsoids whilst other researchers replaced the pores with spheres \cite{bercelliProbabilisticApproachHigh2021, afroz2023}. While these methods prove efficient when defects are far off from each other, they are impractical for defects clustered within a confined volume. Furthermore, there are other methods proposed in the literature that incorporates a porosity field obtained from X-ray radiography techniques and subsequently modifying the local elastic modulus to predict the fatigue strengths of the materials \cite{bleicher2017,hardin2009}. However, the cracks initiating from cavities occur due to a very localised plasticity and therefore information about the local morphology is crucial for a better fatigue life estimation.\\
Despite numerous numerical models utilizing empirical relationships and linear elastic fracture mechanics (LEFM) to estimate fatigue life based on defect size and distance from the free surface, no existing model takes into account the real morphologies of defects. The same argument was raised by Serrano Munoz et al. \cite{serrano-munoz_casting_2018} who showed that the tortuous morphology of shrinkage can pose regions with very high stress concentration factors (SCF) compared to those of a sphere and demonstrated that a natural defect is much more hazardous than an artificially induced spherical defect experimentally. However, these existing models fail drastically when applied on real tortuous defects like inter-dendritic shrinkages and it is even worse in the case of spongeous shrinkages which gives rise to defect clusters. 
Nevertheless, there has been some attempts made in the past to make fatigue estimations via non-local approaches by including real morphologies of defects into FE model via X-ray computed tomography (XCT) which provided impressive insight into defect morphology and the level of criticality induced by defects \cite{le_investigation_2018,raghavendra2022role,pedranz_new_2023,maireQuantitativeXrayTomography2014}. Dezecot et al \cite{dezecot3DCharacterizationModeling2017,dezecotSitu3DCharacterization2016} conducted an in-situ low cycle fatigue (LCF) tomography analysis on a cast porous aluminum alloy, demonstrating that, in the presence of pores, crack development is influenced not only by microstructure but also by the mechanical fields surrounding the pores. To evaluate damage accumulation in porous sample around defects non-local approaches are generally employed to estimate an equivalent field value at the mesoscopic scale. Notably, there exists three common non-local methods namely, the point method, the line method and the volume method which were compiled together in the theory of critical distances by Taylor \cite{taylorGeometricalEffectsFatigue1999,taylor_theory_2008}. For the case of 3D defects, the volumetric homogenization non-local method is generally used where a sphere is centered on a hotspot and the local field values within this sphere are averaged. Many numerical models have been developed using this approach to predict stress at fatigue endurance limit (considered at $10^6$ cycles generally) and has been demonstrated to give impressive results for notched specimens \cite{le_investigation_2018,krzyzak_fatigue_2014,pedranz_new_2023,berto2011multiaxial,lazzarin_neubers_2005,moghtaderi_review_2023,liao2020}. Nevertheless, applying non-local approach to treat real defects produces errors and uncertainties due to lack of definition of certain principal aspects which are usually predefined when treating a notch: A notch contains one hotspot and the stressed volume is around this hotspot. On the contrary, notions of stressed volume and hotspots in the case of real defects is not straightforward. There can be one or more hotspots for each defect in the material medium depending on the tortuosity of the defect, particularly in the case of clustered defects. \\
This paper addresses these challenges and proposes a generalized energy-based (EB) non-local model applicable to real tortuous defects or extreme cases such as defect clusters. The model takes into account the morphology of defects, leveraging X-ray tomography to access defect morphology and construct image-based FE models containing spongeous shrinkages. Full-field numerical simulations of these digital replicas are then compared to experimental results for model calibration and validation.

\section{Experimental methods and results}
\subsection{Material, Microstructure and HCF testing}
\begin{figure}[H]
	\centering
	\begin{subfigure}{0.35\textwidth}
		\includegraphics[width=\textwidth]{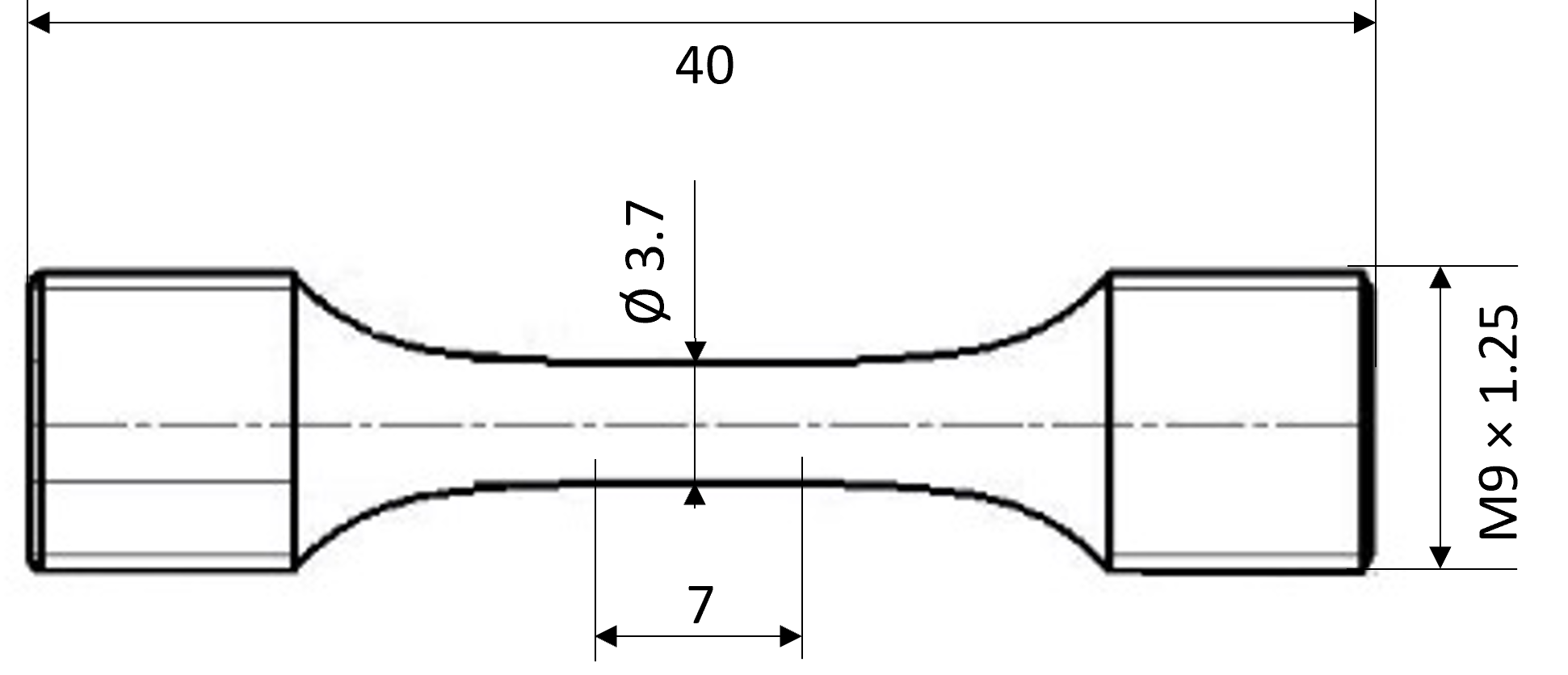}
		\caption{}\label{fig:Dia_IN100}
	\end{subfigure}
	\begin{subfigure}{0.35\textwidth}
		\includegraphics[width=\textwidth]{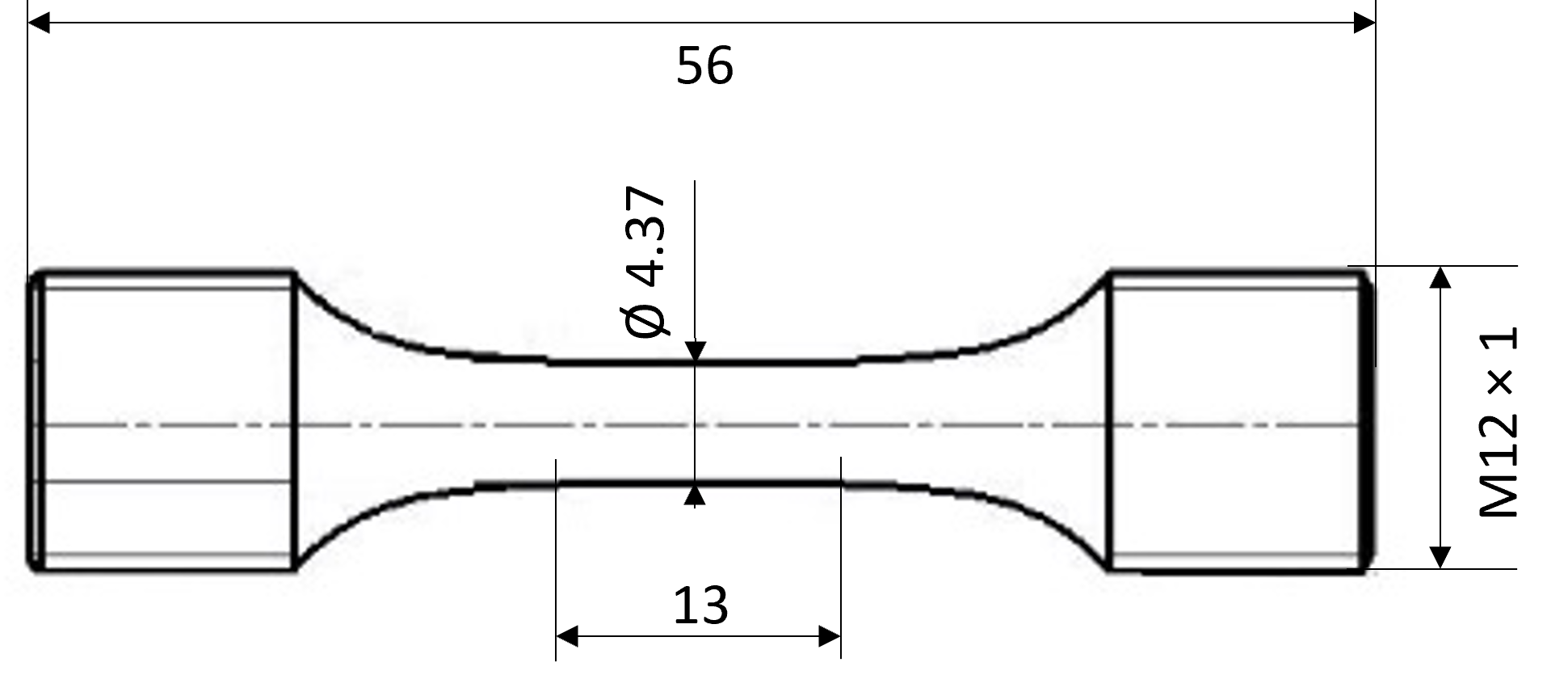}
		\caption{}\label{fig:Dia_R125}
	\end{subfigure}
		\begin{subfigure}{0.35\textwidth}
			\includegraphics[width=\textwidth]{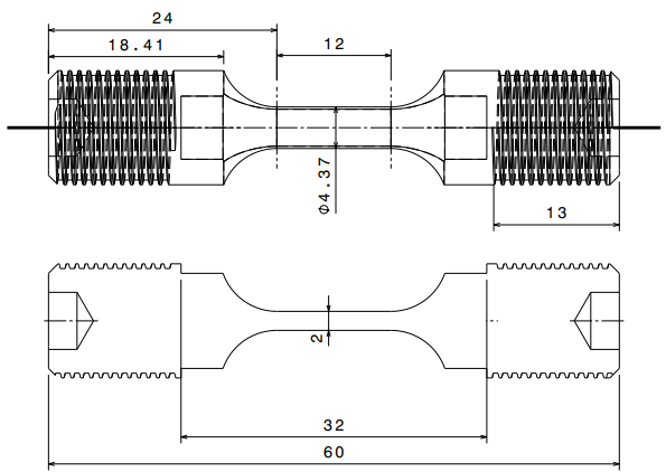}
			\caption{}\label{fig:Dia_Rf125}
		\end{subfigure}
		 \caption{Sketch diagram of the fatigue samples geometry for axysymmetrical a) IN100 samples with gauge section diameter of 3.7 mm, b) R125 samples with gauge section diameter of 4.37 mm and c) R125 samples with flat faces in the gauge section.}
		 \label{fig:Dia}
		\end{figure}

\begin{table}[t]
    \centering
    \caption{Chemical composition of IN100 and R125 materials used in the present work (wt.\%)}
    \label{table:Composition}
    \begin{tabular}{c ccc ccc ccc ccc c} 
    %\begin{tabular}{c c c c c c c c} 
        \hline
        \textbf{} & \textbf{Al} & \textbf{B} & \textbf{C} & \textbf{Co} & \textbf{Cr} & \textbf{Mo} & \textbf{Ti} & \textbf{V} & \textbf{Zr} & \textbf{Ta} & \textbf{Hf} & \textbf{W} & \textbf{Ni} \\ [0.5ex] 
        %\hline
        %\textbf{Material: IN100} \\
        \hline
        \textbf{Inconel 100} & 4.90 & 0.02 & 0.07 & 18.20 & 12.10 & 3.22 & 4.20 & 0.70 & 0.07 & - & - & - & Bal.\\
        \textbf{René 125}    & 5.0 & 0.02 & 0.1 & 10.0 & 9.0 & 2.0 & 2.5 & 0.70 & 0.05 & 4.0 & 1.5 & 7 & Bal.\\
        \hline
    \end{tabular}
\end{table}

In the following analysis, two different as-cast Ni-based superalloys Inconel 100 (IN100) and René 125 (R125) are used to demonstrate the application of the proposed non-local approach to predict fatigue lives. These materials are commonly used in aeronautical engines as turbine discs and blades operating at intermediate temperatures of 300 - \SI{750}{\celsius} \cite{adair_identification_2013}. Both these materials are precipitation strengthened by $\gamma '$ precipitates (primary, secondary and tertiary) which are an ordered $\mathrm{L1_{2}}$ type phase coherently precipitated in solid solution of $\gamma$ matrix (Face centered cubic (FCC) structure) \cite{wang2016dissolution,li2004crystallographic}. Material properties therefore depends strongly on the $\gamma '$ characteristics which evolves primarily with respect to temperature \cite{JOUIAD2016}. IN100 and R125 contain high amount of refractory elements such as Molybdenum, Chromium and Cobalt to prevent corrosion. However, R125 consists of noble metal elements to increase strength and creep resistance. Carbides appear as minor phases in both these materials. Due to the presence of high amounts of hafnium, R125 materials are prone to the formation of hafnium rich oxides ($\mathrm{HfO_2}$) \cite{kantzos2000high}.  The chemical composition of IN100 and R125 are shown in Table \ref{table:Composition}. \\

Four IN100 samples (labelled as I) and eight R125 samples (labelled as R) with cylindrical cross sections were machined from ingot bars containing spongeous shrinkages of ASTM grades between 5 and 8 (as per ASTM E2660 standard). Additionally, five cylindrcal R125 samples and two R125 samples (labelled as Rf) of different geometry (flat samples of nearly rectangular cross sections) with lesser porosity levels or containing no shrinkage were machined to validate the proposed non-local model. IN100 samples were 40 mm long with a gauge section diameter of 3.7 mm (see Fig. \ref{fig:Dia_IN100}) whilst R125 samples were 54 mm long with gauge section diameter of 4.37 mm (see Fig. \ref{fig:Dia_R125}). The flat R125 samples on the other hand had a flat geometry at the gauge section with a thickness of 2 mm, see Fig. \ref{fig:Dia_Rf125}. \\
All samples were tested using a Schenck hydraulic machine which has a maximum load capacity of 50 kN equipped with a 4-zone lamp furnace. A K-type thermocouple was welded on the samples to control the temperature at gauge section during the test. IN100 samples were tested at \SI{750}{\celsius} whilst R125 samples were tested at \SI{700}{\celsius} which are the optimum operating temperature for these materials in an aircraft engine. The endurance limit of both materials respectively at \SI{750}{\celsius} and \SI{700}{\celsius} is taken from the literature. Endurance limit of IN100 at \SI{750}{\celsius} is found to be 228 MPa \cite{raghavendra2022role} and that of R125 at \SI{700}{\celsius} is 240 MPa \cite{kantzos2000high} respectively at $2 \times 10^6$ cycles for IN100 and at $2 \times 10^7$ cycles for R125. The properties of both materials at the respective temperatures are reported in table~\ref{table:Params} \cite{vincentIdentificationElastoviscoplasticityDamage2009,wojcikIdentificationChabocheLemaitre2021,gongDeterminationMaterialProperties2010}. After heating the sample to the imposed temperature, uniaxial sinusoidal cyclic loads (force-controlled tests) were applied at a load ratio R = 0 until failure. A frequency of 80 Hz was used for all tests except for the validation samples (R9-R13 and Rf1-Rf2) for which a frequency of 20 Hz was used. The stress amplitude for each test is reported later in table~\ref{table:Exp_results}. \\

\begin{table}
    \centering
    \caption{Material properties of IN100 at \SI{750}{\celsius} and René 125 at \SI{700}{\celsius}.}
    \label{table:Params}
% …

\begin{adjustbox}{width=\textwidth,keepaspectratio}    \begin{tabular}{c ccc ccc ccc ccc c} 
    %\begin{tabular}{c c c c c c c c} 
        \hline
        \textbf{} & \textbf{Young's Modulus} & \textbf{Poisson's ratio} & \textbf{Yield's stress} & \textbf{Norton flow coefficient} & \textbf{Norton flow exponent} & \textbf{Endurance limit of healthy sample}\\ 
         &(GPa)&&  $R_0$ (MPa)& K (MPa)&n&$\sigma _{D,healthy}$ (MPa)\\[0.5ex] 
        \hline
        \textbf{Inconel 100} & 162 & 0.3 & 235 & 640 & 21 & 228\\
        \textbf{René 125}    & 172 & 0.3 & 313 & 494 & 34 & 240 \\
        \hline
    \end{tabular}
    \end{adjustbox}
\end{table}

\subsection{X-ray Computed Tomography (XCT) and image-based FE model}\label{XCT_intro}
\begin{figure}[H]
    \centering
    
    \begin{subfigure}{0.48\textwidth}
    	\includegraphics[width=\textwidth]{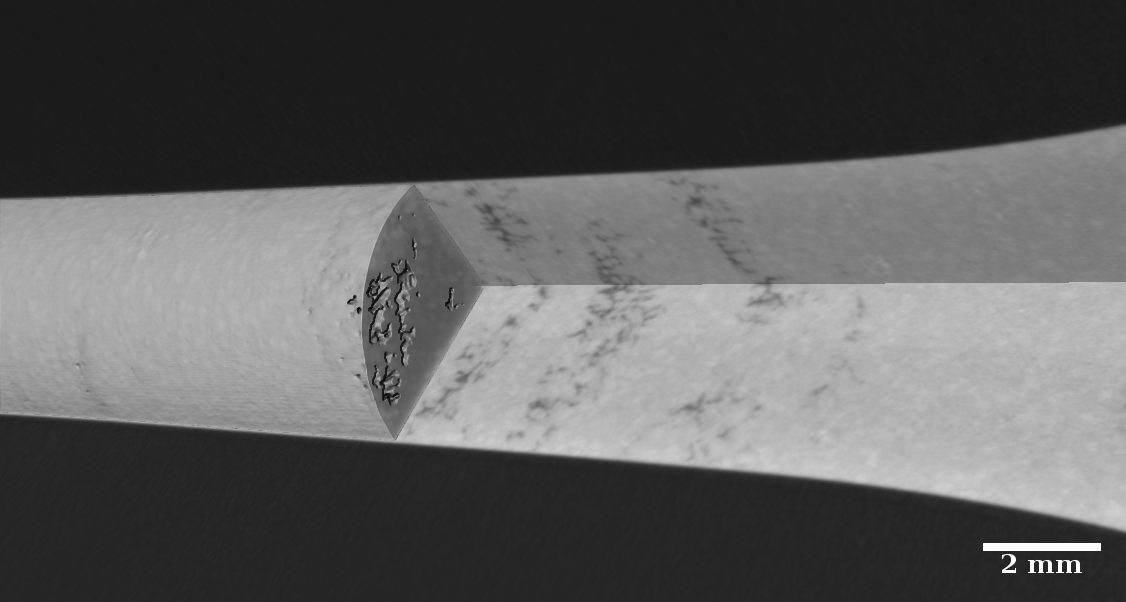}
    \caption{}
    \label{XCT}
    \end{subfigure}
  	\hspace{0.1cm}
    \begin{subfigure}{0.42\textwidth}
    	\includegraphics[width=\textwidth]{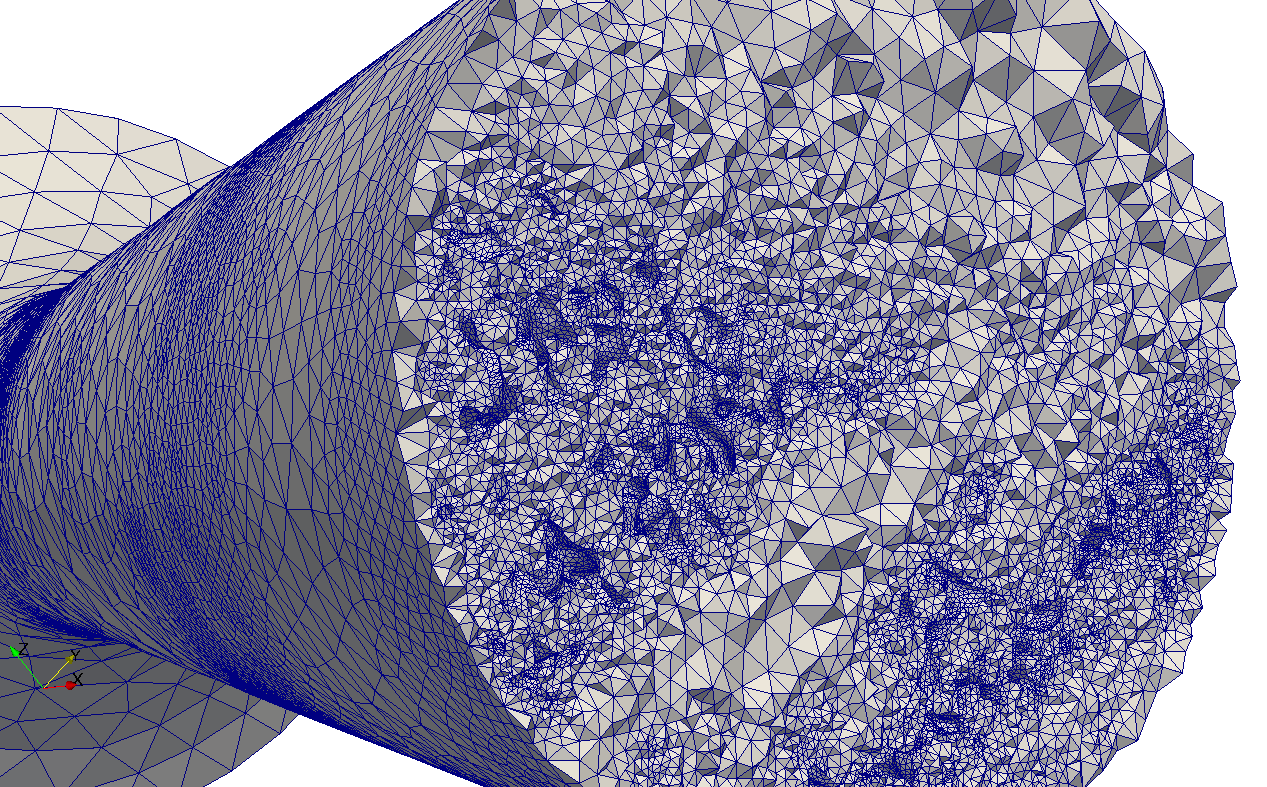}\label{IB}
    	\caption{}
    	 \end{subfigure}
    \caption{a) Volume rendered XCT image with ortho slices along the axis of the sample and b) Image-based FE model constructed from segmented defects of sample R1.}
    \label{fig:XCT_IB}
\end{figure}
\begin{figure}[H]
    \centering
    
    \begin{subfigure}{\textwidth}
    	\includegraphics[width=1.0\textwidth]{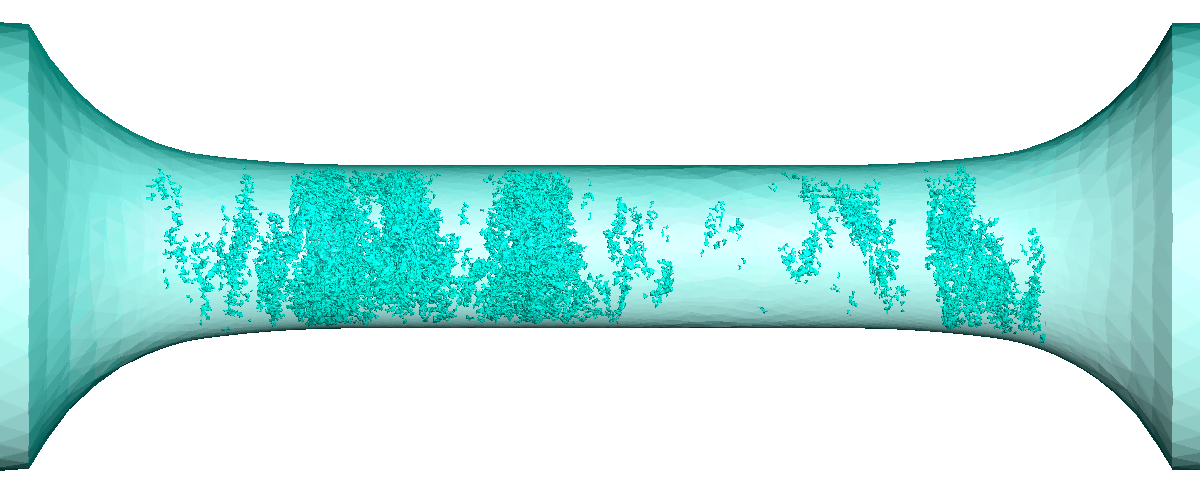}\caption{}\label{fig:IB_B4} \end{subfigure} \hfill
    \begin{subfigure}{\textwidth}
    	\includegraphics[width=1.0\textwidth]{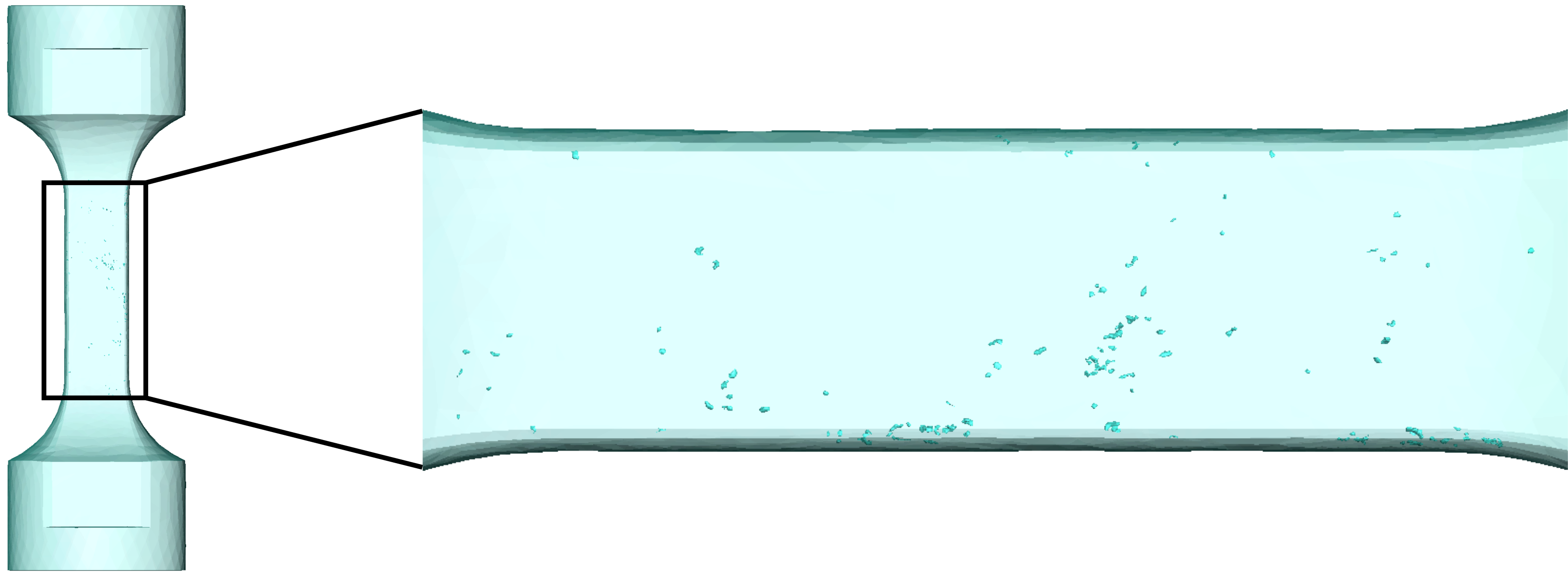}\caption{}\label{fig:IB_Epr12}
    	 \end{subfigure}
    \caption{Image based FE model of a) sample R3 (ASTM grade 8) with a very tortuous and large spongeous shrinkage and b) sample Rf1 with small pores and micro-shrinkages.}
    \label{fig:IBFE}
\end{figure}

Samples were imaged by X-ray computed tomography (XCT) to reveal the inner defect structure using a Nikon XT H 450 with a source voltage of 150 kV. A resolution of \SI{25}{\cubic \micro \meter} (size of voxel) was achieved for samples of IN100 and a resolution of around \SI{20}{\cubic \micro \meter} for the R125 samples as illustrated in Fig. \ref{XCT}. Flat field correction method was applied on raw XCT volumes to remove grey-scale irregularities and a gamma filter was applied to enhance the contrast. The local contrast was further enhanced by applying Contrast Limited Adaptive Histogram Equalization (CLAHE) process \cite{zuiderveld1994}. Finally, the processed image was segmented by Otsu thresholding. The entire segmentation process was validated manually by trial and error method and performed on an integrated platform between python and ImageJ \cite{fiji}. Segmented images were then processed to label each individual defect (based on strong connectivity). Each defect was characterized and its features analyzed (volume $\sqrt[3]{V_{defect}}$, surface area $A_p$, projection of defect in loading direction, sphericity $\Psi$, position in the sample). Surface meshes of the segmented defects were extracted via marching cubes algorithm and were slightly smoothened with Humphrey smoothing filter from the python trimesh module \cite{trimesh}. Surface smoothing is necessary to eliminate the sharp voxel morphology of the elements. The surface meshes of defects were inserted and registered in their actual position in the CAD model of the test sample. For the defects penetrating the surface of the test sample, a boolean operation was applied to unify the penetrating defect and test sample using GNU Triangulated surface library (GTS) module. Finally, volumic mesh was generated using gmsh library as shown in Fig. \ref{fig:XCT_IB}(b) and Fig. \ref{fig:IBFE} \cite{gmsh}.

\begin{figure}[H]
    \centering
    
    \includegraphics[width=0.8\textwidth]{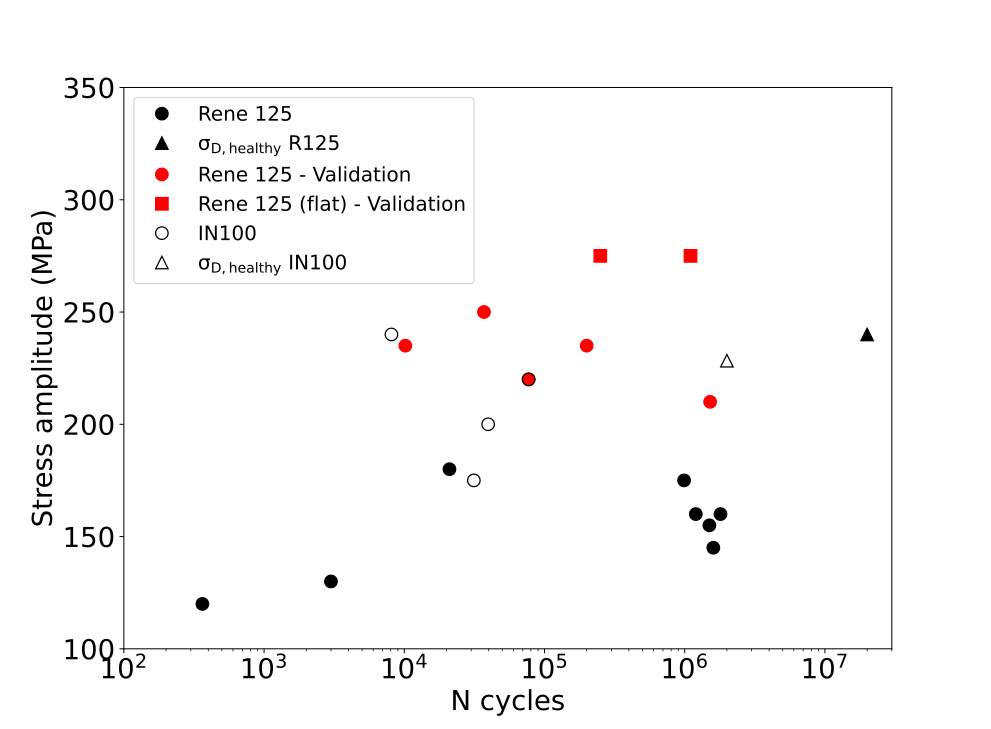}
    \caption{S-N plot for both IN100 and R125 samples. The fatigue lives of tested samples ranges between 300 and $2 \times 10^6$ cycles. The samples used for the validation of the proposed non-local model are marked in red.}
    \label{fig:Wohler}
\end{figure}

\subsection{Experimental Results}\label{x_results}
\begin{table}
    \centering
    \caption{Summary of results from experimental tests and XCT analysis (SS - Spongeous shrinkage, MS - Micro shrinkage, Inc - Inclusion). Reference volume (excluding mounting screw heads) of cylindrical IN100 samples was around \SI{230}{\cubic\mm}, cylindrical R125 samples around \SI{570}{\cubic\mm} and that of flat samples around \SI{130}{\cubic\mm}}.
    \label{table:Exp_results}
    \begin{adjustbox}{width=\textwidth,keepaspectratio}  
    \begin{tabular}{c ccc ccc c} 
    %\begin{tabular}{c c c c c c c c} 
        \hline
        \textbf{Sample} & \textbf{ASTM Grade} & \textbf{Type of defect} & \textbf{Largest defect size (mm)} & \textbf{Porosity (\%)} & \textbf{stress amplitude (MPa)} & \textbf{Fatigue life (cycles)} \\ [0.5ex] 
        %\hline
        %\textbf{Material: IN100} \\
        \hline
        \\
        \textbf{I1} & 5 & SS & 1.057 & 0.531 & 240 & $\mathrm{8\times10^3}$ \\
        \textbf{I2} & 5 & SS & 0.889 & 0.364 & 220 & $\mathrm{7.7\times10^4}$ \\
        \textbf{I3} & 6 & SS & 1.08 & 0.594 & 200 & $\mathrm{3.9\times10^4}$ \\
        \textbf{I4} & 6 & SS & 1.498 & 0.688 & 175 & $\mathrm{3.1\times10^4}$ \\
        \\
        %\hline
        %\textbf{Material: René 125} \\
        \hline
        \\
        \textbf{R1} & 7 & SS & 3.099 & 1.05 & 130 & $\mathrm{3.7\times10^3}$\\
        \textbf{R2} & 6 & SS & 1.272 & 0.304 & 160 & $\mathrm{1.2\times10^6}$ \\
        \textbf{R3} & 8 & SS & 3.128 & 1.714 & 120 & 338 \\
        \textbf{R4} & 7 & SS & 1.086 & 0.352 & 145 & $\mathrm{1.6\times10^6}$ \\
        \textbf{R5} & 7 & SS & 1.139 & 0.12 & 160 & $\mathrm{1.9\times10^6}$ \\
        \textbf{R6} & 5 & SS & 1.573 & 0.453 & 180 & $\mathrm{2.1\times10^4}$ \\
        \textbf{R7} & 6 & SS & 0.688 & 0.081 & 175 & $\mathrm{1\times10^6}$ \\
        \textbf{R8} & 6 & SS & 1.197 & 0.349 & 155 & $\mathrm{1.5\times10^6}$ \\
        \textbf{R9} & - & SS & 0.311 & 0.022 & 250 & $\mathrm{3.7\times10^4}$ \\
        \textbf{R10} & - & SS & 0.343 & 0.023 & 235 & $\mathrm{2\times10^5}$ \\
        \textbf{R11} & - & SS & 0.245 & 0.025 & 220 & $\mathrm{7.7\times10^4}$ \\
        \textbf{R12} & - & SS & 0.324 & 0.032 & 210 & $\mathrm{1.5\times10^6}$ \\
        \textbf{R13} & - & SS & 0.721 & 0.091 & 235 & $\mathrm{1\times10^4}$ \\
        \\
        \hline
        \\
        \textbf{Rf1} & - & MS & 0.162 & 0.038 & 250 & $\mathrm{2.5\times10^5}$ \\
        \textbf{Rf2} & - & Inc & 0.123 & 0.007 & 250 & $\mathrm{1\times10^6}$ \\
%        \textbf{Rf3} & - & - & - & 0.175 & 0.1068 & 250 & 28,850 \\
        \\
        \hline
    \end{tabular}
    \end{adjustbox}
\end{table}

Fatigue testing results of IN100 and R125 samples are presented in table \ref{table:Exp_results} and plotted in a S-N diagram in Fig. \ref{fig:Wohler}. Table \ref{table:Exp_results} also reports the porosity (void volume fraction) and largest defect size for each sample, from which it can be seen that the tested samples have different levels of porosity with varying defect population distributions. For simplicity, the tested samples can be classified based on porosity into two batches: a) highly or extremely porous (I1-I4 and R1-R8 among which samples R1 and R3 are very extremely porous) and b) less porous samples (R9-R13 and Rf1-Rf2). Some of these tested samples are graded as per ASTM E2660 standard which specifically addresses the cavity populations in the material estimated via radiography measurements of the test samples \cite{ASTME2660}. Rotella et al., demonstrated that fatigue lives are found to be dependent on ASTM grades \cite{rotellaFatigueLimitCast2018,rotellaInfluenceNaturalDefects2017}. It is difficult to make such a conclusion in our case due to the very large scatter in fatigue lives of tested samples: for instance, at a given ASTM grade 7, sample R4 and R5 have longer fatigue lives despite experiencing higher applied stress amplitudes compared to R1. ASTM grade 8 sample R3 was seen to fail at only 338 cycles with a stress amplitude less than half of the known endurance limit due to presence of a very large cluster of spongeous shrinkage, see Fig. \ref{fig:IB_B4}. It is therefore evident that criticality of samples are very different to one another. Magnitude of this criticality depends on various characteristics or features of defect population like porosity level, size of critical defect, etc. Nevertheless, from fig \ref{fig:Wohler}, it is clear that one S-N curve does not fit all samples of each material, at least in this macroscopic scale. All cylindrical samples contain important defect populations and have a very large dispersion in fatigue lives between as less as 300 cycles and up to 2 $\times 10^7$ cycles. On the other hand, the flat samples were tested at much higher stress amplitudes close to endurance limit of R125 material and were found to have fatigue lives of around $10^6$ cycles. While the fatigue strengths of less porous samples appear to be greater than those of the highly porous samples, it is currently not possible to establish a correlation between porosity and fatigue life or between the largest defect size and fatigue life. It is however evident from table \ref{table:Exp_results} that larger the porosity, larger is the size of the maximum defect. Therefore, whilst ASTM grades, porosity levels etc are efficient estimates to have an idea of sample criticality, they are insufficient and requires consideration of more local parameters such as defect morphology, their characteristics etc along with their inter-correlations \cite{yang2022,tijani2013,tijani2013detection,szmytka2020}.\\
\subsubsection{Damage mechanisms}
\begin{figure}[H]
    \centering
    \begin{subfigure}{0.3\textwidth}
    	\includegraphics[width=\textwidth]{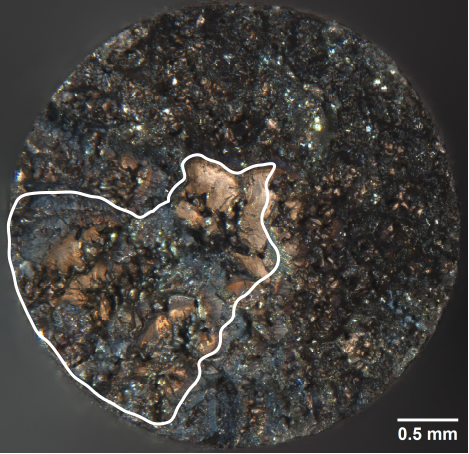}\caption{}
    	\end{subfigure}
    \begin{subfigure}{0.29\textwidth}\includegraphics[width=\textwidth]{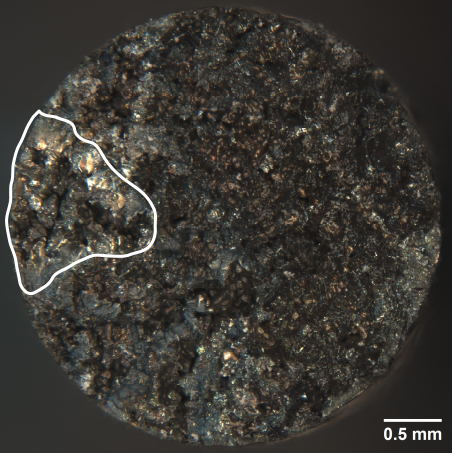}\caption{}\end{subfigure}
    \begin{subfigure}{0.289\textwidth}\includegraphics[width=\textwidth]{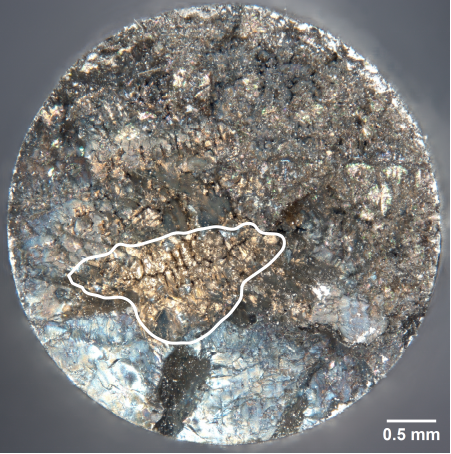}\caption{}\end{subfigure}
    \begin{subfigure}{0.291\textwidth}\includegraphics[width=\textwidth]{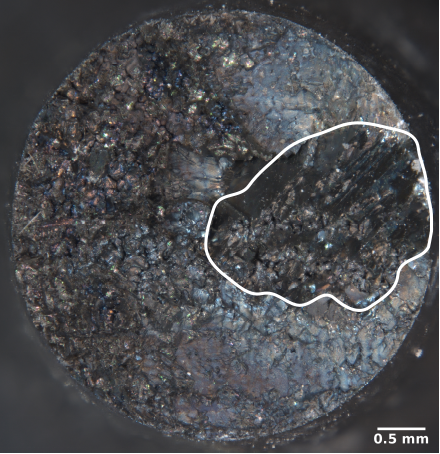}\caption{}\end{subfigure}
    \begin{subfigure}{0.31\textwidth}\includegraphics[width=\textwidth]{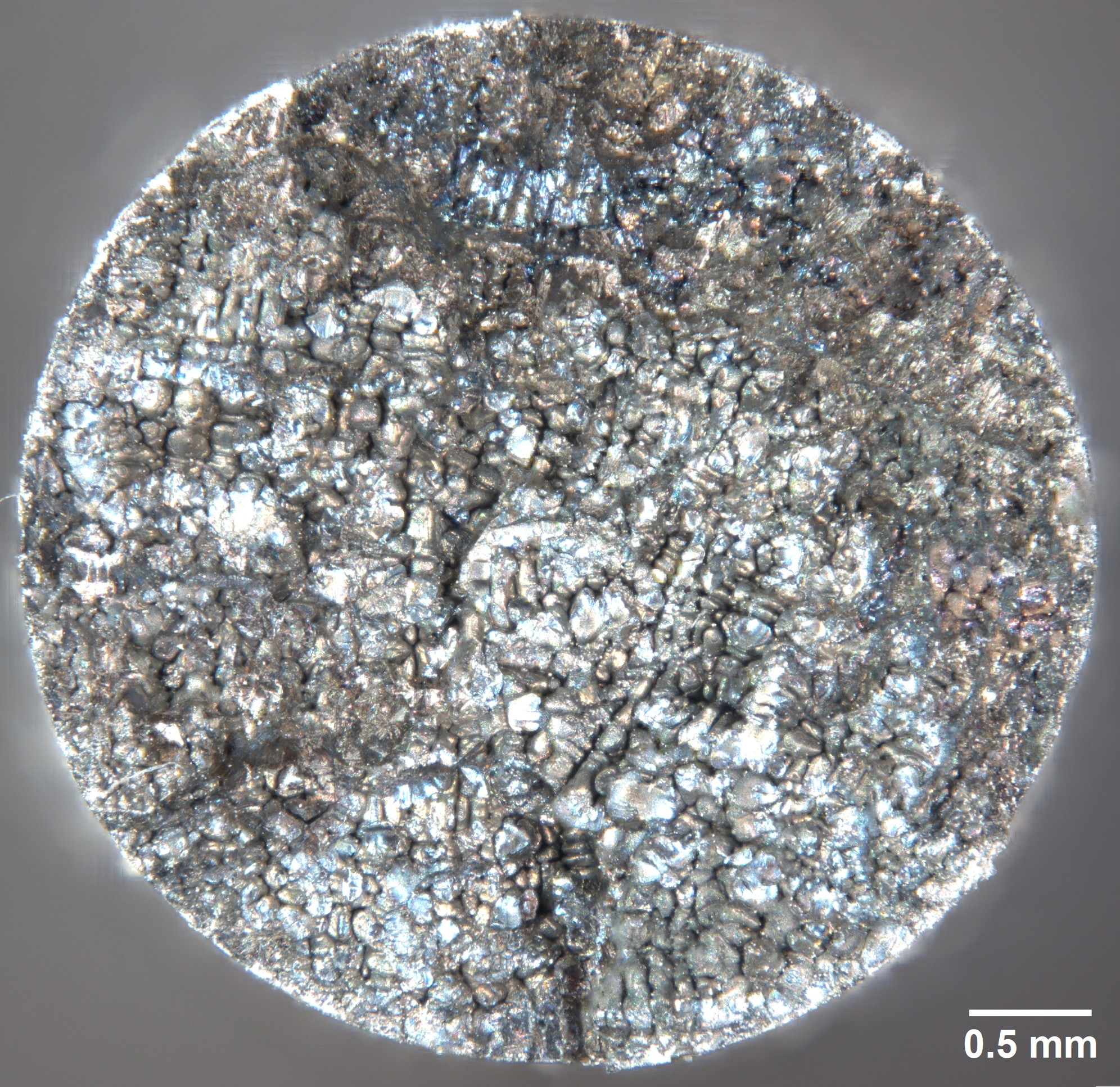}\caption{}\end{subfigure}
    \begin{subfigure}{0.1665\textwidth}\includegraphics[width=\textwidth]{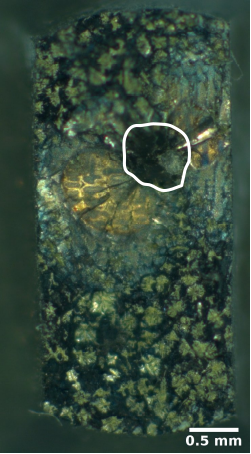}\caption{}\end{subfigure}
    \caption{Optical images of fracture surface of samples: a) I2 b) I4 c) R8 d) R4 e) R6 and f) Rf2. Critical defect that was responsible for failure is highlighted for samples I2, I4, R8, R4 and Rf2.}
    \label{fig:Fracto_optic}
\end{figure}

Figure \ref{fig:Fracto_optic} shows optical microscopy (OM) images of the fracture surface of some IN100 and R125 samples. Critical defect that initiated primary crack can be easily spotted in fracture surface of less porous cylindrical samples and even more easily in flat samples, see Fig. \ref{fig:Fracto_optic}(a)-(d) and (e). On the contrary, it is difficult to highlight the critical zone in very porous samples due to presence of large defects and thick clusters of defects, see Fig. \ref{fig:Fracto_optic}(e). \\ % and \ref{fig:Fracto_R1}. 
%\begin{figure}[H]
%    \centering
%    
%    \includesvg[width=1.0\textwidth]{2R1_fracto.svg}
%    \caption{a) Optical image of sample I3 along with b) scanning electron microscopic image of critical defect that was responsible for failure. Marked zones (red) indicate the possible crack-initiation sites describing a case of multiple crack initiations}
%    \label{fig:Fracto_I3}
%\end{figure}
\begin{figure}[H]
    \centering
    
    \begin{subfigure}{0.46\textwidth}\includegraphics[width=\textwidth]{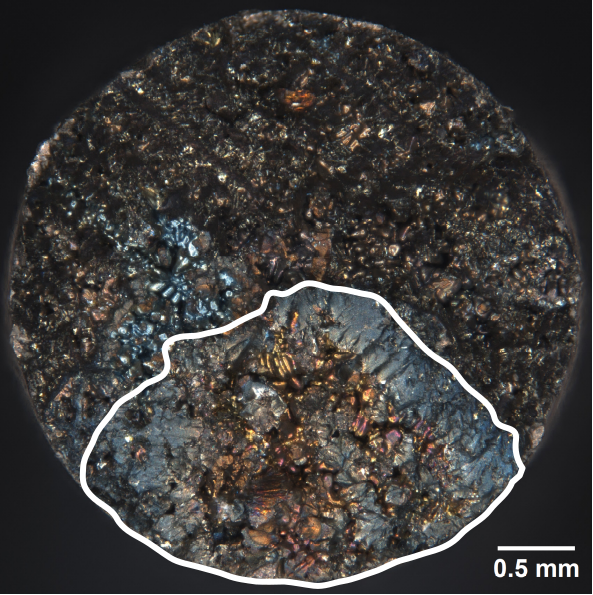}\caption{}\end{subfigure} \hspace{0.1cm}
    \begin{subfigure}{0.505\textwidth}\includegraphics[width=\textwidth]{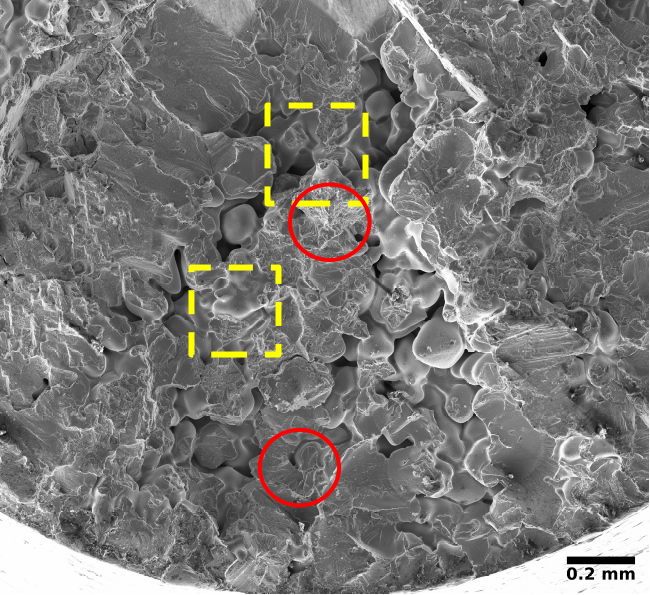}\caption{}\end{subfigure} \hfill
    \begin{subfigure}{0.419\textwidth}\includegraphics[width=\textwidth]{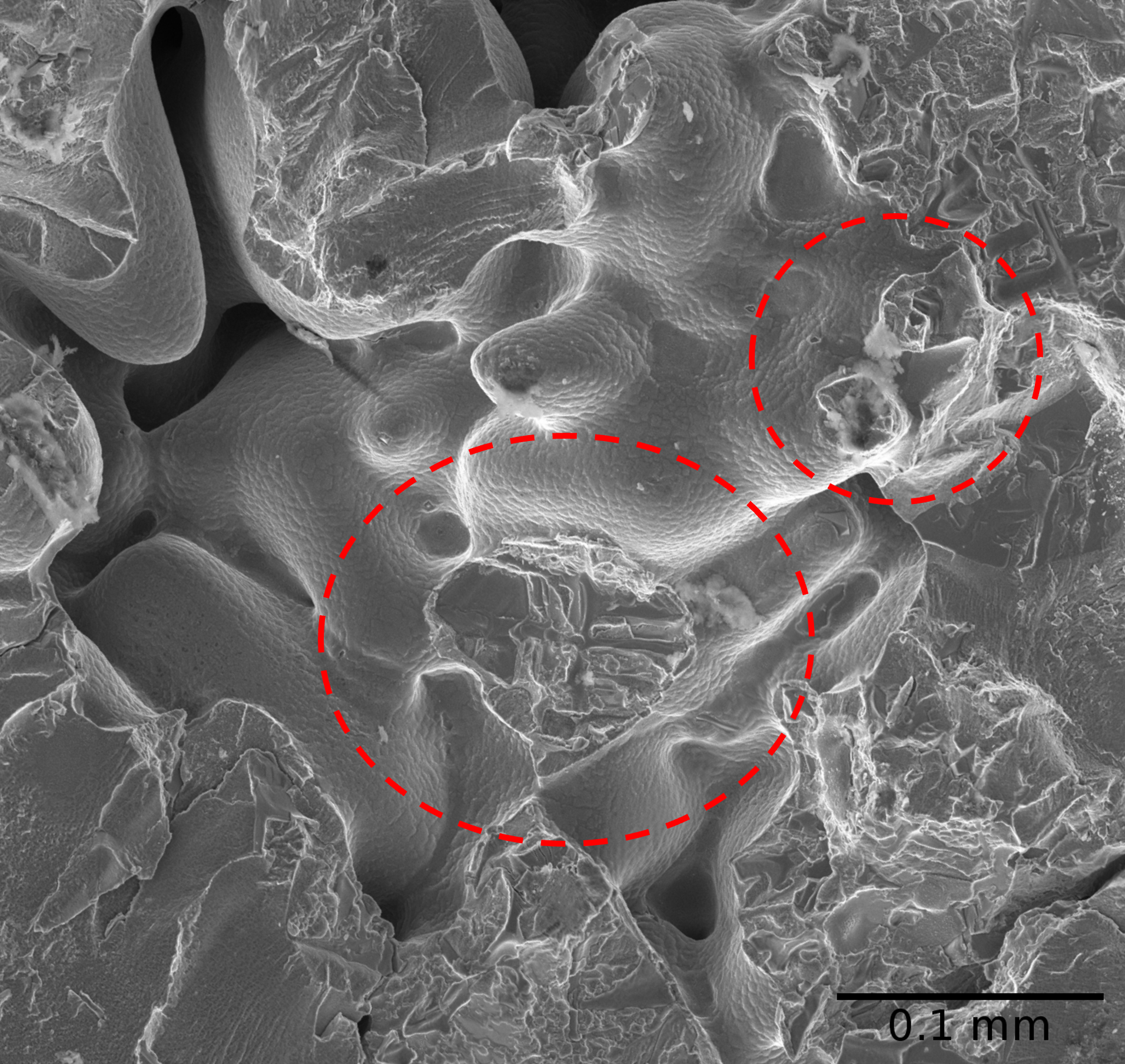}\caption{}\end{subfigure} \hspace{0.1cm}
    \begin{subfigure}{0.39\textwidth}\includegraphics[width=\textwidth]{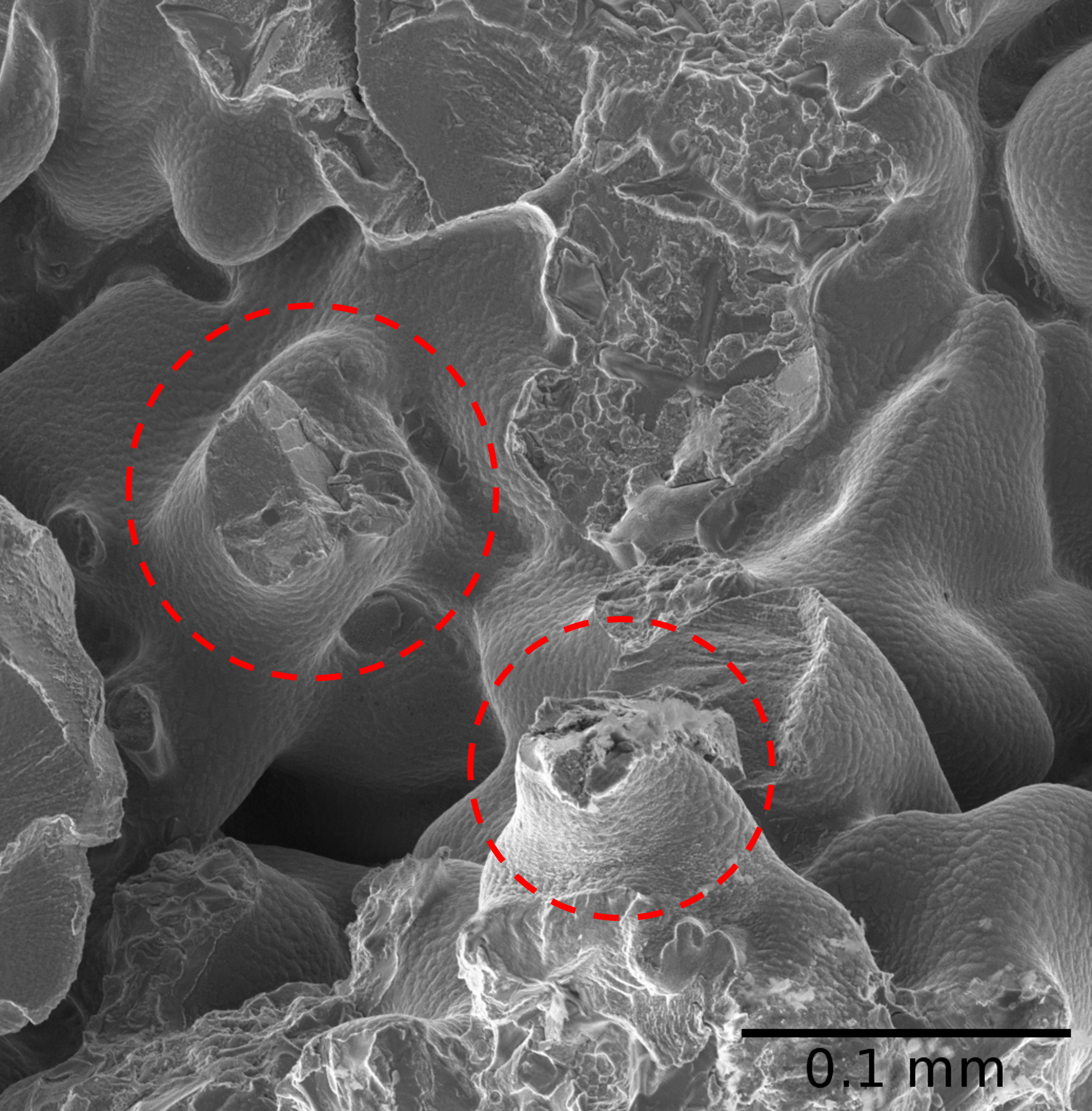}\caption{}\end{subfigure}
    \caption{SEM image of sample I3 a) displaying the fracture zone via optical microscopy b) SEM image of the fracture zone indicating the multiple crack-initiation zones marked in red. Some crack-initiation zones from small ligaments created by this tortuous defect are magnified in c) and d) and marked in yellow.}
    \label{fig:Fracto_ligaments}
\end{figure}

\begin{figure}[H]
    \centering
    
    \begin{subfigure}{0.545\textwidth}\includegraphics[width=\textwidth]{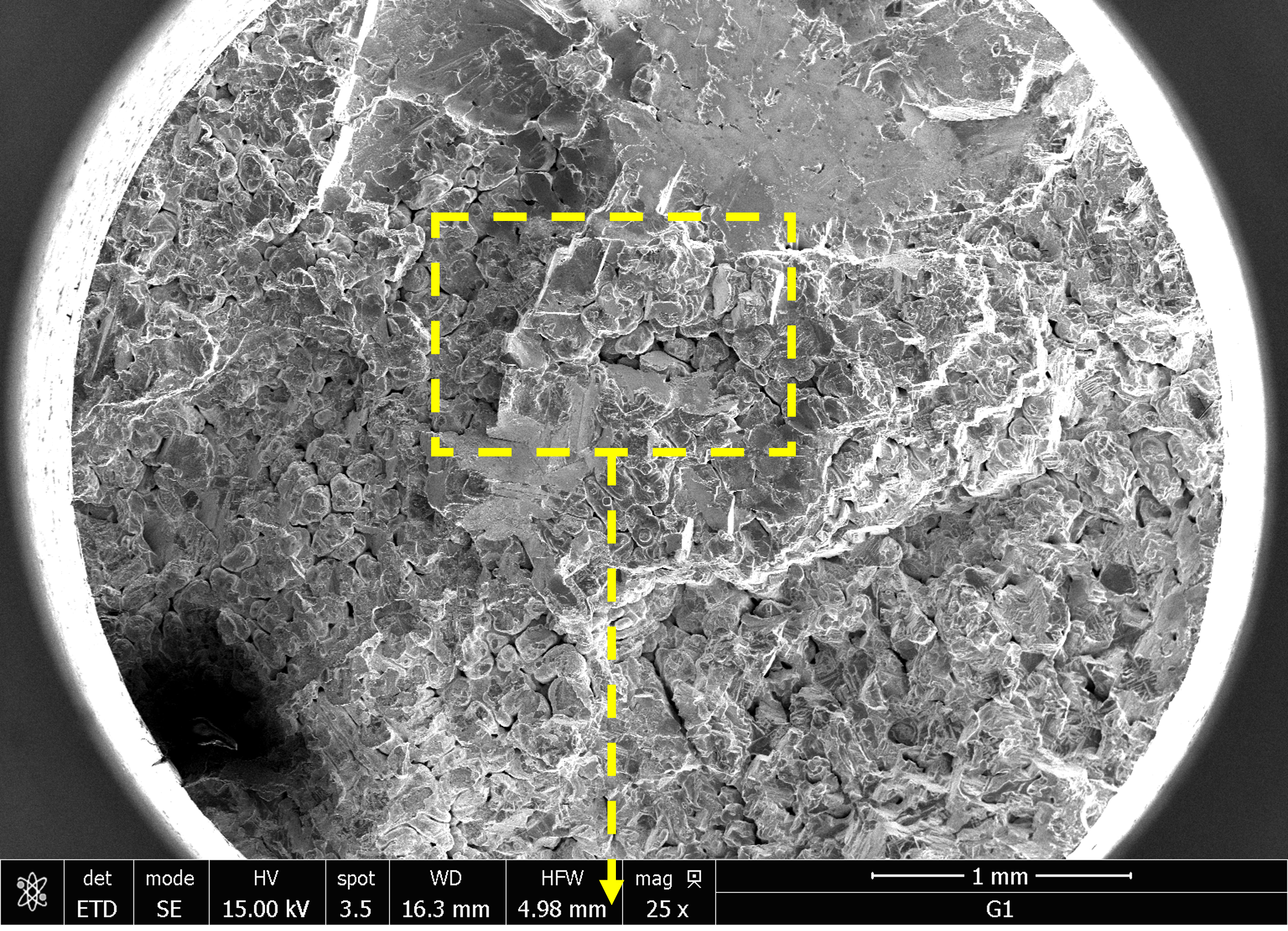}\caption{}\end{subfigure}\hspace{0.07cm}
    \begin{subfigure}{0.4\textwidth}\includegraphics[width=\textwidth]{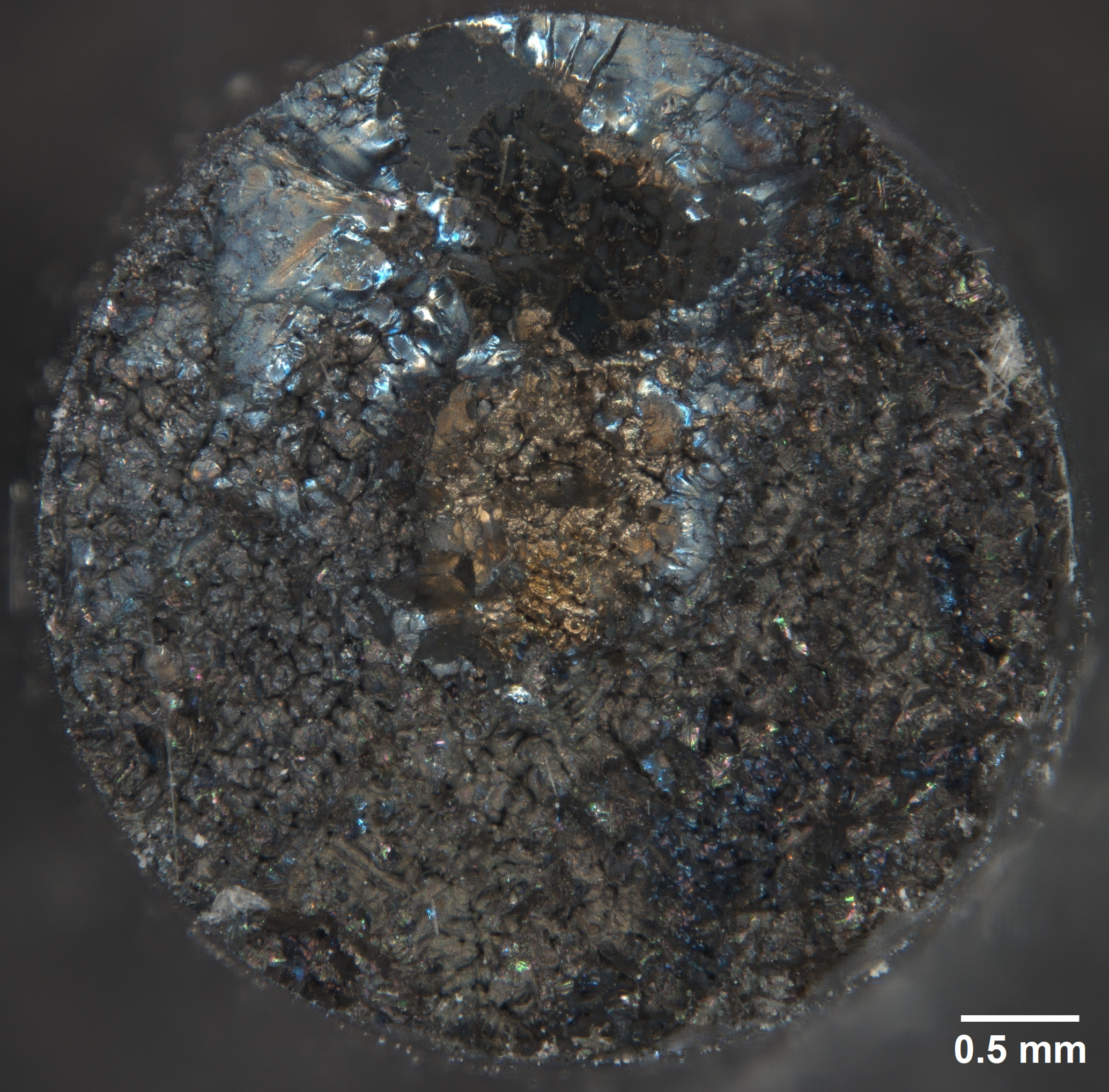}\caption{}\end{subfigure} 
    \begin{subfigure}{0.49\textwidth}\includegraphics[width=\textwidth]{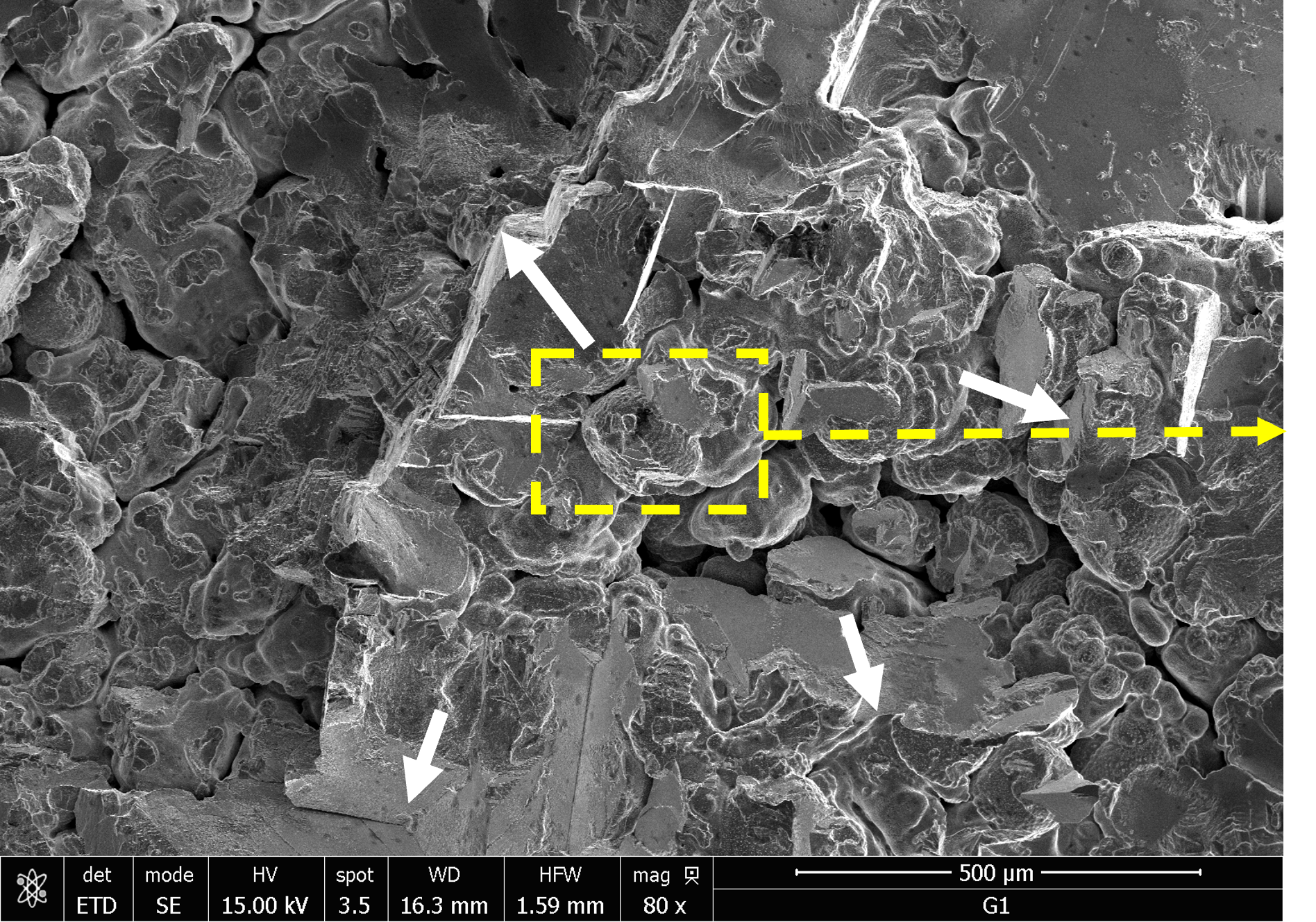}\caption{}\end{subfigure}
    \begin{subfigure}{0.485\textwidth}\includegraphics[width=\textwidth]{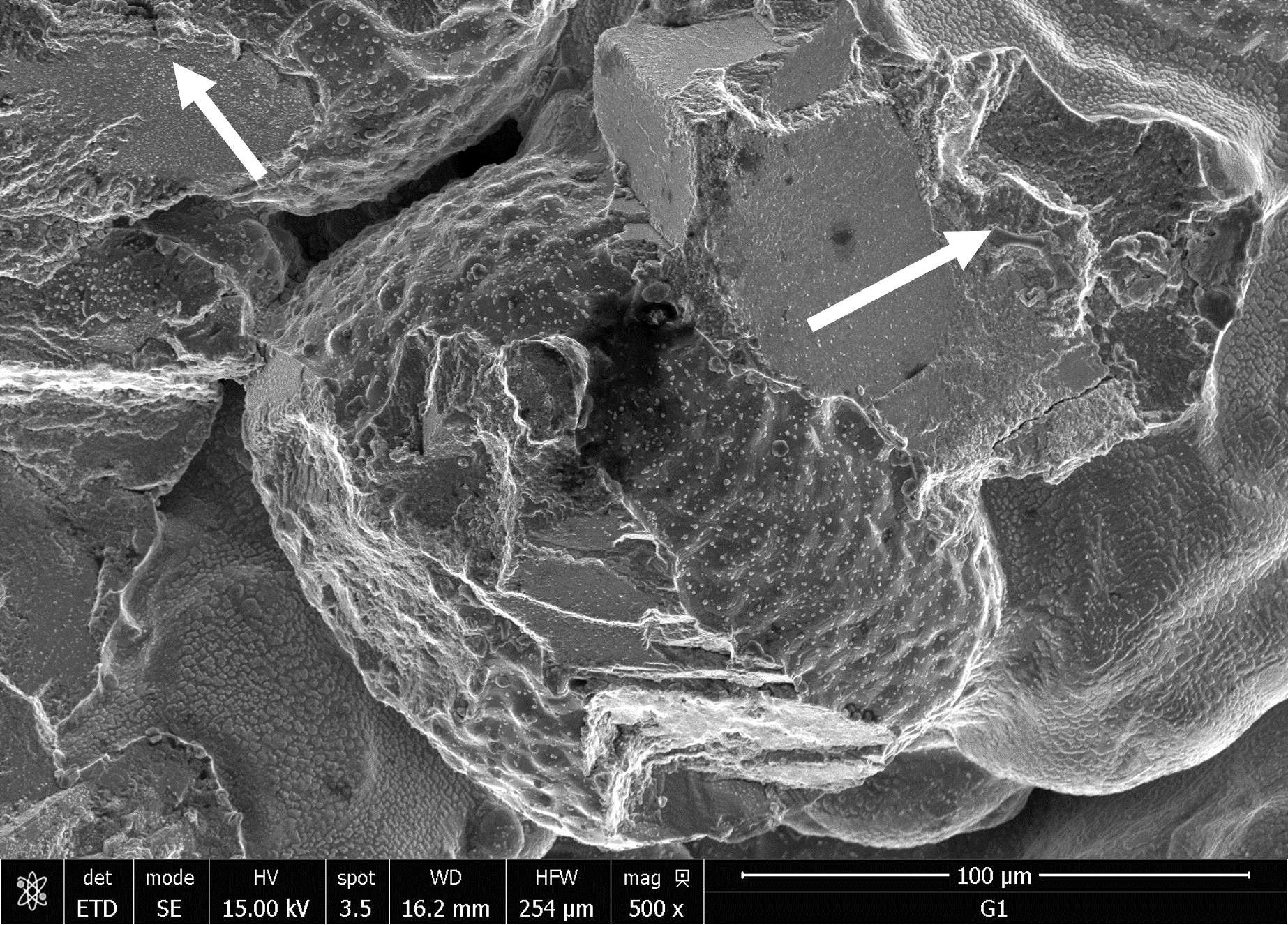}\caption{}\end{subfigure}
    \caption{Fractographic analysis of samples R7. a) SEM image of critical zone along with b) optical image c) Maginified SEM image of one of the critical zones or possible crack-initiation site. Crack is seen to propagate along crystallographic planes (white arrows) before coalescing with neighbouring defect in the defect cluster. d) Further magnified image of this crack-initiations site shows that propagation along crystallographic planes occurs even at much smaller scales (white arrows).}
    \label{fig:Fracto_R7}
\end{figure}

Due to the presence of multiple large defects in cylindrical samples, cracks are initiated from multiple defect volumes \cite{li2004crystallographic}. Conversely, in relatively less porous cylindrical samples, while the critical defect is readily identifiable through optical imaging, pinpointing the specific site of crack initiation becomes nearly unattainable due to the concurrent existence of multiple initiation sites surrounding the critical defect as shown in figure \ref{fig:Fracto_ligaments}. 
Furthermore, the tortuous morphology of defects engenders the formation of small ligaments within the defect cavity, resembling slender filaments suspended amidst the void: a characteristic of inter-dendritic and spongeous shrinkage. SEM images of sample I3 in Fig. \ref{fig:Fracto_ligaments} show the examples of these small ligaments on fracture surfaces. The crack fronts in these small ligaments describes the initiation of a subordinate crack which stopped propagating upon reaching the void surface or in other words, coalesced with the principal defect. Due to their very small cross-section area, and consequently due to their high local stresses, small ligaments most certainly broke within the very first cycles of loading or at least much before the primary crack was initiated. 
%In other sense, small ligaments initiate a small crack that breaks the ligament and then another crack is re-initiated at another small ligament. 
This either occurs one ligament after another or simultaneously, and this process continues until a crack is initiated at a stable hotspot which further develops into a primary crack.\\ 
Brueckner-Foit et al. \cite{brueckner2018} studied the crack initiations via multiple interrupted XCT scans in Al-Si-Cu alloys containing inter-dendritic shrinkage and reported similar results. They demonstrated that in cases of tortuous shrinkage defects, the crack-growth occurs in two phases: the first phase consists of the fracture of small ligaments which occurs even with moderate loads since the local stress concentration around these ligaments are very high and in the second phase, a stable crack emanates from one of the branches of inter-dendritic shrinkage. The small ligaments pose a serious problem even in numerical simulations where the real defect morphologies are incorporated via image-based FE models which will be discussed in later sections.\\
On the other hand, in almost all samples, the stable crack propagates along crystallographic plane before coalescing with neighbouring void surface of the defect cluster as seen in Fig. \ref{fig:Fracto_R7}. This coalescence phenomenon is also observable at smaller scales, as illustrated in Fig. \ref{fig:Fracto_R7}(c) and (d).
\subsubsection{Defect characteristics}
\begin{figure}[H]
    \centering
    
    \begin{subfigure}{0.45\textwidth}\includegraphics[width=\textwidth]{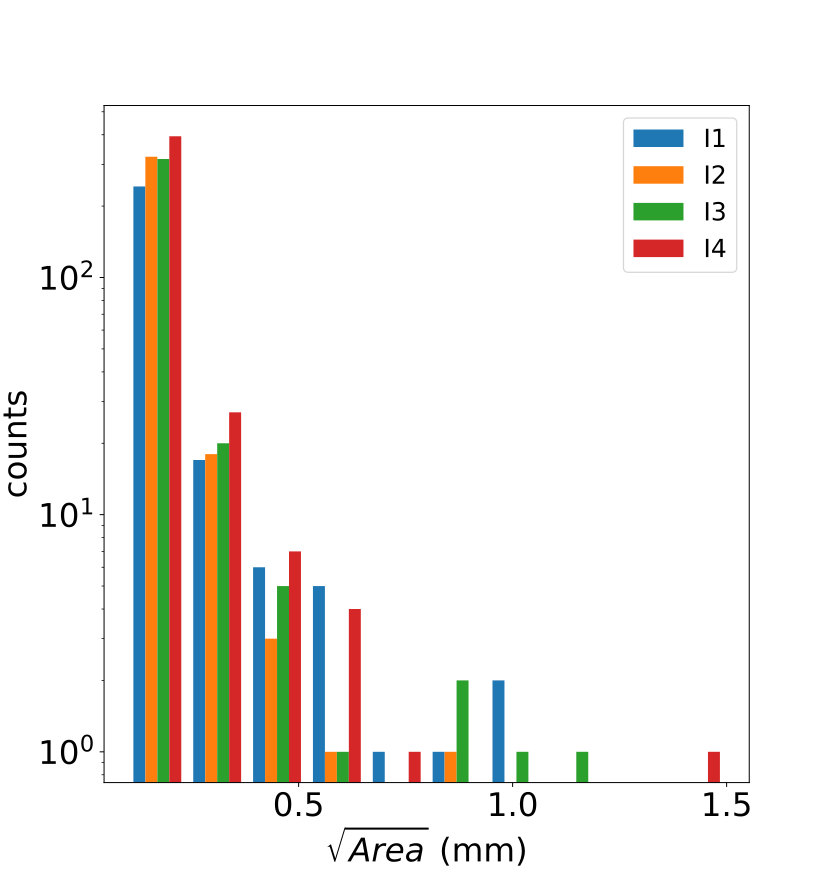}\label{fig:Size_IN100}\caption{}\end{subfigure}
    \begin{subfigure}{0.45\textwidth}\includegraphics[width=\textwidth]{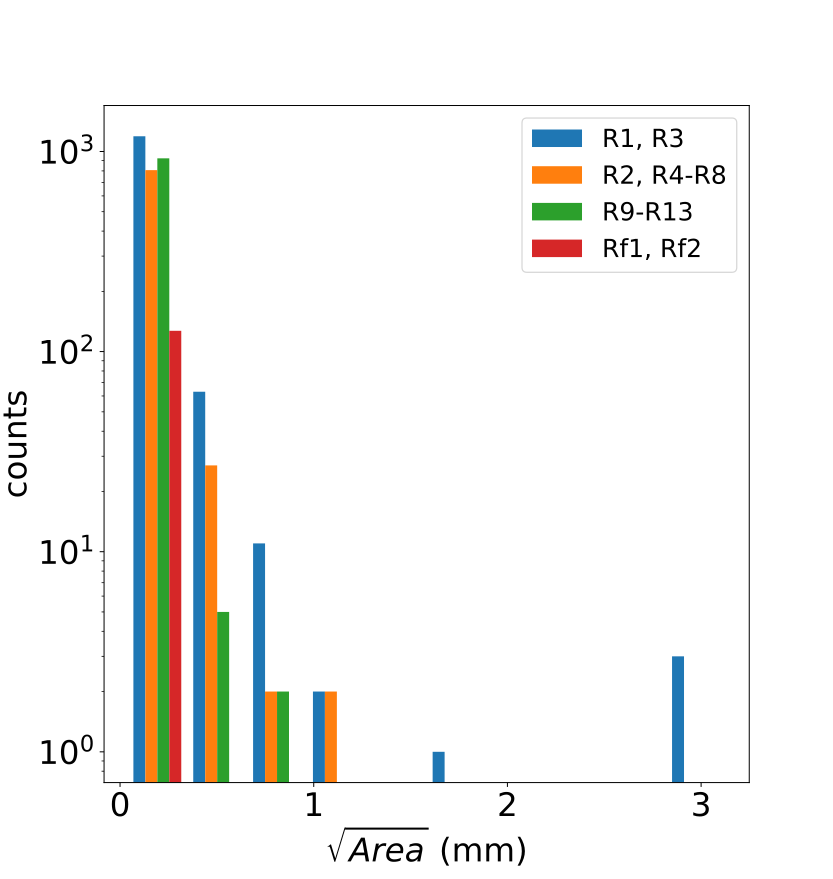}\label{fig:Size_R125}\caption{}\end{subfigure}
    \caption{Distribution of defect size in material material a) IN100 b) René 125; multiple samples of R125 have been grouped together to improve the graph readability.}
    \label{fig:Size}
\end{figure}

From the XCT images of these tested samples, each defect can be labelled and its characteristics can be estimated. The most important defect features playing a role in fatigue are the defect size, sphericity, aspect ratio and distance of the defect from the free surface \cite{maskeryQuantificationCharacterisationPorosity2016,Bao2021}. The defect size can be defined in three different ways: a) square root of the area of defect projected on a plane perpendicular to loading $\sqrt{Area}$. This measure of defect size was proposed by Murakami and takes into account the mode I cracking and finds its use in many LEFM models. b) cube root of volume $\sqrt[3]{V_{defect}}$ which considers the three dimensional aspects of defect and c) Finally, another way of defining a defect is the equivalent radius (radius of a sphere $r_{defect}$ whose volume is equated to that of the defect). From the findings of earlier research, it was seen that size defined as $\sqrt{Area}$ is always the largest amongst the three whilst $r_{defect}$ is the smallest. However, the three measures are found to be linearly related to each other \cite{raghavendra2022role}. Figure \ref{fig:Size} shows the defect size distributions of both IN100 and R125 samples. For the simplicity, R125 samples are represented by grouping together multiple samples classified based on their porosity and geometry. It can be seen that both materials contain important defects with some of the R125 samples containing much larger defects, i.e. samples of ASTM grade 7 and 8. Flat samples of R125 on the other hand had almost no prominent defects with largest defect in sample Rf1 being \SI{130}{\micro \meter} as also shown in Fig. \ref{fig:IB_Epr12}. Regarding the number of defects, IN100 samples contain almost same number of defects whilst the samples R1-R8 of R125 contain large number of defects followed by samples R9-R13 and then Rf1-Rf2. \\
\begin{figure}[H]
    \centering
    
    \begin{subfigure}{0.45\textwidth}\includegraphics[width=\textwidth]{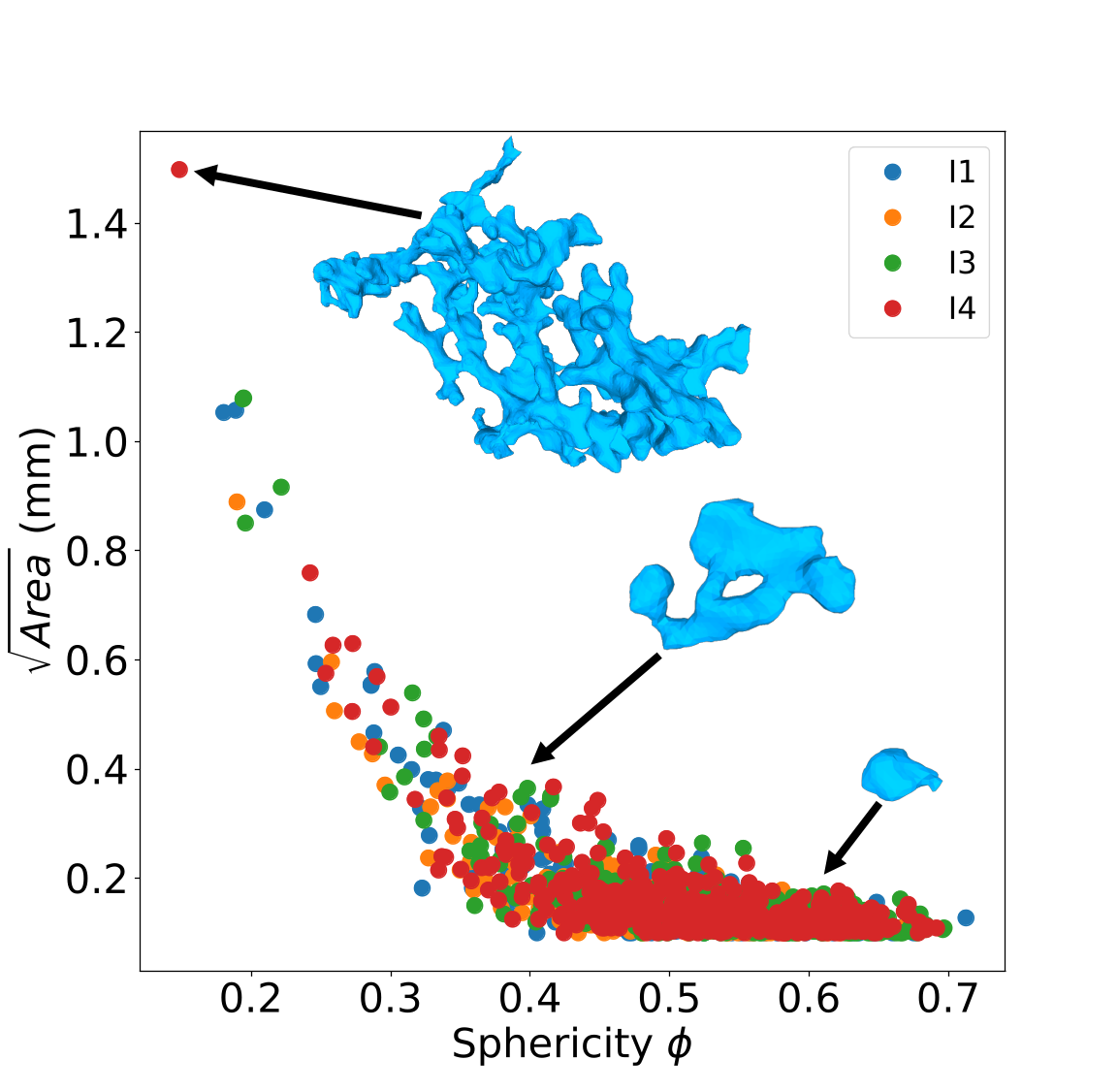}\label{fig:Spher_IN100}\caption{}\end{subfigure}
    \begin{subfigure}{0.45\textwidth}\includegraphics[width=\textwidth]{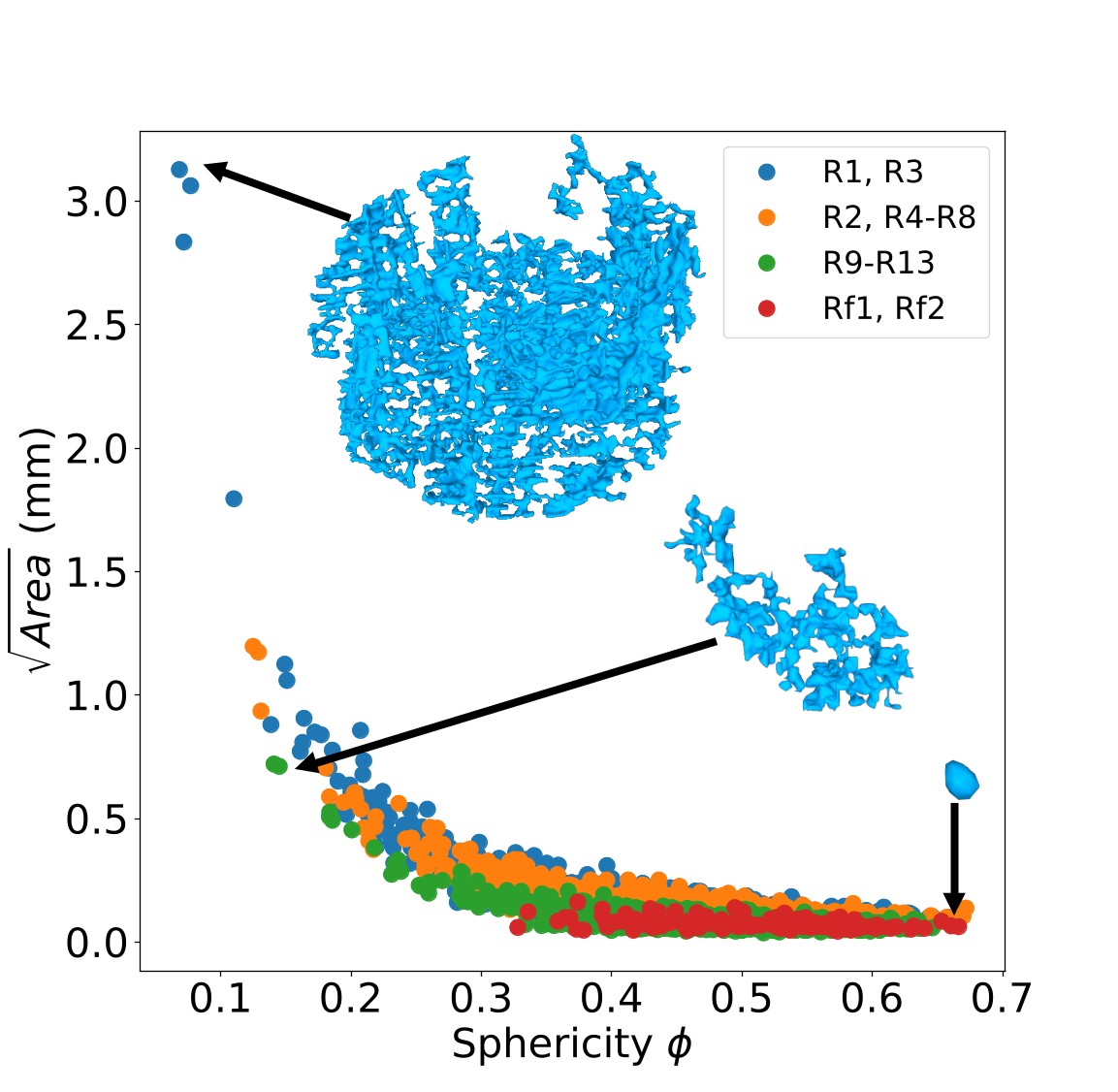}\label{fig:Spher_R125}\caption{}\end{subfigure}
    \caption{Variation of sphericity with respect to defect size for material a) IN100 b) René 125. Example surface meshes of some defects with varying characterisitcs are illustrated in the graph. Sphericity tends to 0 for a very tortuous shape and equals to 1 for a perfect sphere.}
    \label{fig:Sphericity}
\end{figure}

Sphericity on the other hand is a measure of how spherical a defect is: a value of 1 for perfect sphere and a value close to 0 for a tortuous shape \cite{matpadi2023generation}. For a defect with volume $V_{defect}$ and surface area $A_p$, sphericity $\phi$ can be defined as,
\begin{equation}
    \phi = \frac{(\pi)^{\frac{1}{3}}(6V_{defect})^\frac{2}{3}}{A_p}
\end{equation}
From Fig. \ref{fig:Sphericity}, it can be seen that sphericity of a defect decreases rapidly as their size increases. The larger defects which are essentially shrinkages are tortuous in shape and the smaller defects or pores are more spherical. Shrinkages are usually inter-dendritic defects with very complex shape as shown in Fig. \ref{fig:Sphericity}(b). %as also seen in figure \ref{fig:Fracto_R1}. 
Both shrinkages and pores interact in material medium to form spongeous shrinkages as detailed in our previous article \cite{matpadi2023generation}. It was shown via spatial point pattern analysis that a threshold in defect size can be defined at 0.4 mm to distinguish the shrinkages and pores. While this distinction is valid for the cases of spongeous shrinkages of the investigated materials, it is unclear if it holds true for other materials with different fabrication techniques resulting in different grain sizes. The tortuous morphology of these spongeous shrinkages with some defects illustrated in Fig. \ref{fig:Sphericity} is responsible for the multiple crack-initiation within the material and possess the small ligaments. Differences in the characteristics can also be noted from the different sample batches: while some of the cylindrical samples contained extremely tortuous defects, flat specimens contained more spherical pores due to their low porosity levels and size effect. \\

From the experimental campaign, it is seen that all cylindrical samples failed due to presence of an important defect associated to a spongeous shrinkage (or defect cluster). The flat R125 samples on the other hand contained no shrinkage but however had few micro-shrinkages and pores; Sample Rf1 failed due to an internal crack initiation whilst Rf2 failed due to an inclusion which initiated a crack as shown in Fig. \ref{fig:Fracto_optic}f that later propagated towards the nearest pore forming a long crack. Moreover, it was observed that the cavities contain critical sites responsible for the initiation of cracks. Among these critical sites are the small ligaments that initiate an early crack and breaks almost immediately. Due to these reasons, the fractographic analysis is cumbersome where even if the critical defect is identified, most of the initiation sites on this defect either break instantly or coalesce within the cluster of spongeous shrinkage without developing into a primary crack. Therefore, at this point, a simple criterion to rationalize the fatigue life of all tested samples (with both complex spongeous shrinkages and lesser porosity levels) does not emerge.

%simple criterion cannot be implied to explain both the complicated damage mechanism of samples containing extreme conditions of spongeous shrinkage and also samples containing lesser porosity levels.}

\section{Basis of the energy based non-local model}
\subsection{Material behaviour}
A digital representative of each sample was built from the XCT reconstructed volumes such that the image-based FE model includes the real defects at their respective positions as shown in Fig. \ref{fig:IBFE}. These image-based FE models were simulated using the Z-set software Suite \cite{Zebulon}. To perform a full field analysis considering the relaxation of stresses and evolution of plasticity in critical regions, an elasto-viscoplastic behaviour law with kinematic hardening and Norton flow of plasticity was adopted. The total strain $\underline{\underline{\varepsilon}}^{tot}$ consists of an elastic part $\underline{\underline{\varepsilon}}^{el}$, visco-plastic part $\underline{\underline{\varepsilon}}^{p}$ and thermal strain $\underline{\underline{\varepsilon}}^{th}$ which is elongation due to temperature, see table \ref{table:EVP_eqs}.
%\begin{equation} \label{eq1}
%    \underline{\underline{\varepsilon}} _{tot} = \underline{\underline{\varepsilon}} _{el} + \underline{\underline{\varepsilon}} _{p} + \underline{\underline{\varepsilon}} _{th}
%\end{equation}
%The elasto-viscoplastic strain rate is computed as per following equations, 
%\begin{equation} \label{eq2}
%    \dot {\underline{\underline{\varepsilon}}} _{p} = \left\langle \frac{J_2(\underline{\underline{\sigma}} - \underline{\underline{X}}) - R_0}{K}\right\rangle ^n  \text{sign}(\underline{\underline{\sigma}} - \underline{\underline{X}})
%\end{equation}
%\begin{equation} \label{eq3}
%    \underline{\underline{\dot{X}}} = C \dot {\underline{\underline{\epsilon}}} _{p} - DX|\dot {\epsilon} _{p}|
%\end{equation}
\begin{table}[h!]
\renewcommand{\arraystretch}{1.2}
    \centering
    \caption{Governing equations of elasto-viscoplastic behaviour used in FE simulations.}
    \label{table:EVP_eqs}
    \begin{tabular}{ll} 
    %\begin{tabular}{c c c c c c c c} 
        \hline
        Strain partitioning & $\underline{\underline{\varepsilon}} ^{tot} = \underline{\underline{\varepsilon}}^{el} + \underline{\underline{\varepsilon}}^{p} + \underline{\underline{\varepsilon}}^{th}$\\ [0.5ex] 
        \hline
        Yield function & $f=J_2(\underline{\underline{\sigma}} -\underline{\underline{X}})-R-\sigma _y$ \\
        Kinematic hardening & $\underline{\underline{\dot{X}}}=\dfrac{2}{3}C \underline{\underline{\dot{\alpha}}}\quad$ with $\quad\underline{\underline{\dot{\alpha}}}=\underline{\underline{\dot{\varepsilon}}}^p-\gamma \underline{\underline{\alpha}} \dot{p}$ \\
        Isotropic hardening (constant) & $R=R_0$ \\
        Plastic flow rate & $\dot{p}=\left\langle \dfrac{f}{K}\right\rangle ^n$ \\
        \hline
    \end{tabular}
\end{table}

The governing equations of the model are described in table \ref{table:EVP_eqs} where $R_0$ is the elastic limit which is constant for a given temperature, $X$ is the back stress for the kinematic hardening (with the state variable $\underline{\underline{\alpha}}$) and $K$, $n$, are the material constants related to Norton's flow while $C$ and $D$ are the material constants related to kinematic hardening for a given temperature. \\
\subsection{Strain energy criterion}
\begin{figure}[H]
    \centering
    \begin{subfigure}{0.35\textwidth}\includegraphics[width=\textwidth]{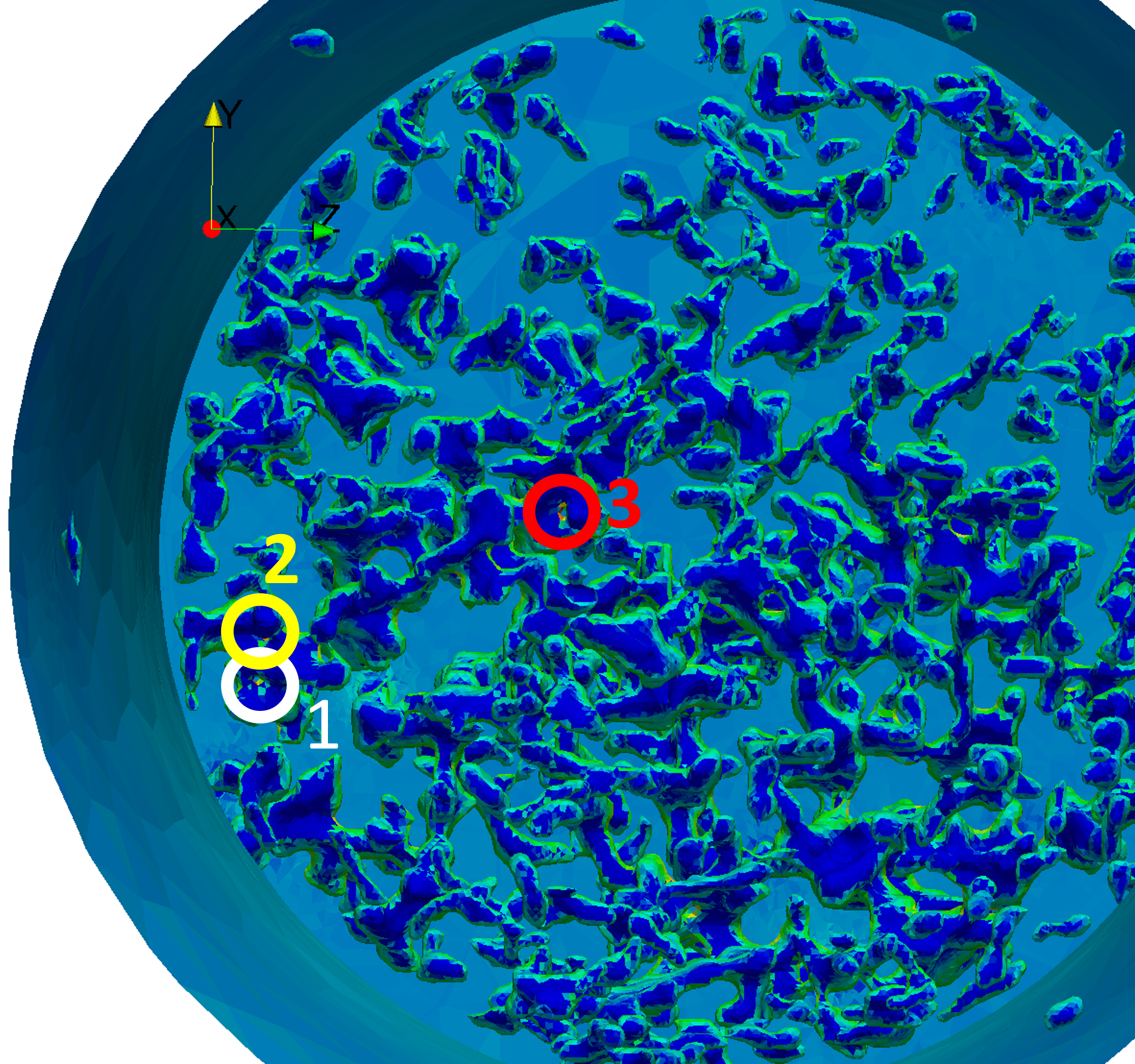}\caption{}\label{fig:illu_liga}
    \end{subfigure}
    \begin{subfigure}{0.24\textwidth}\includegraphics[width=\textwidth]{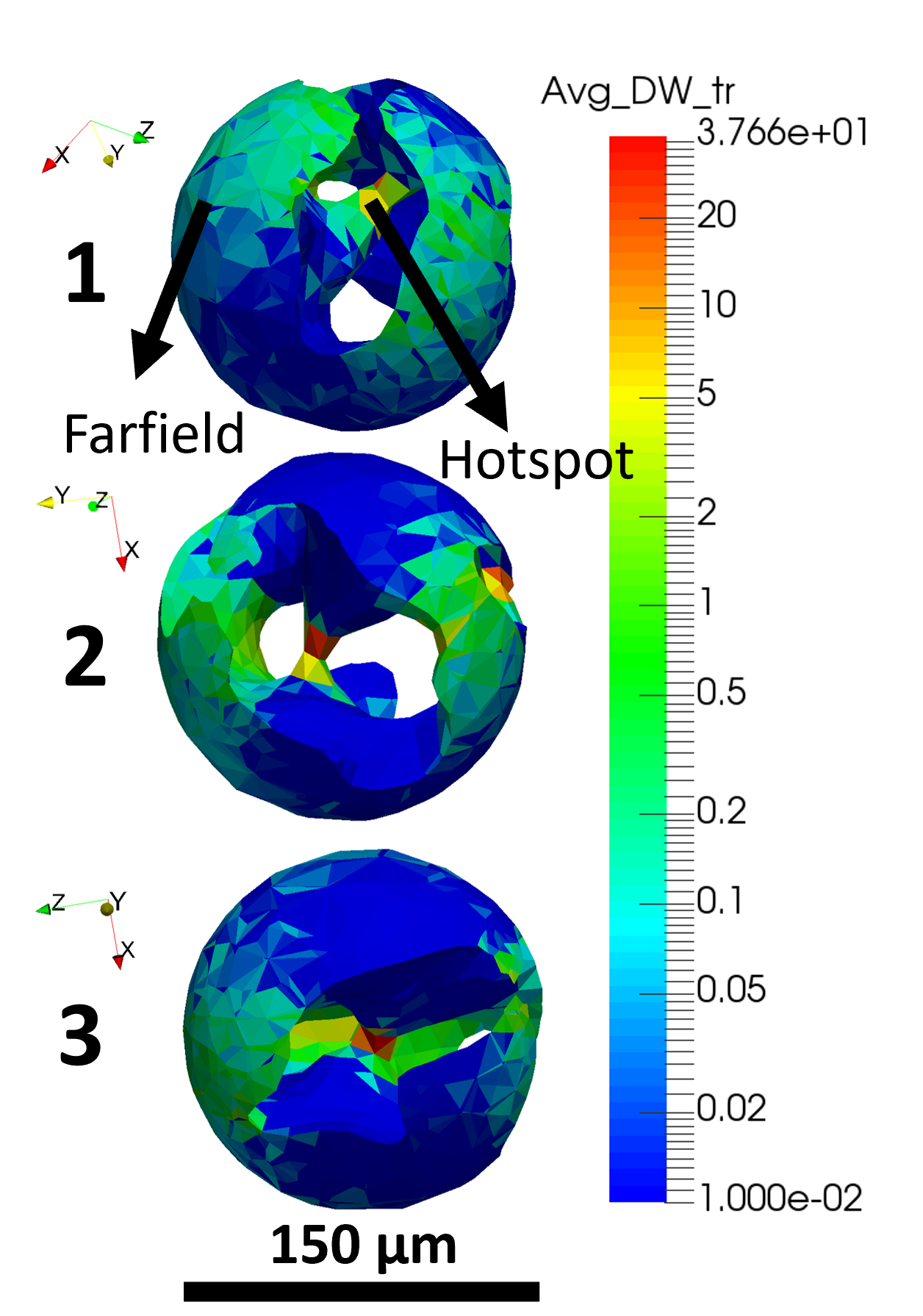}\label{fig:illu_liga}\caption{}
    \end{subfigure}
    \begin{subfigure}{0.39\textwidth}\includegraphics[width=\textwidth]{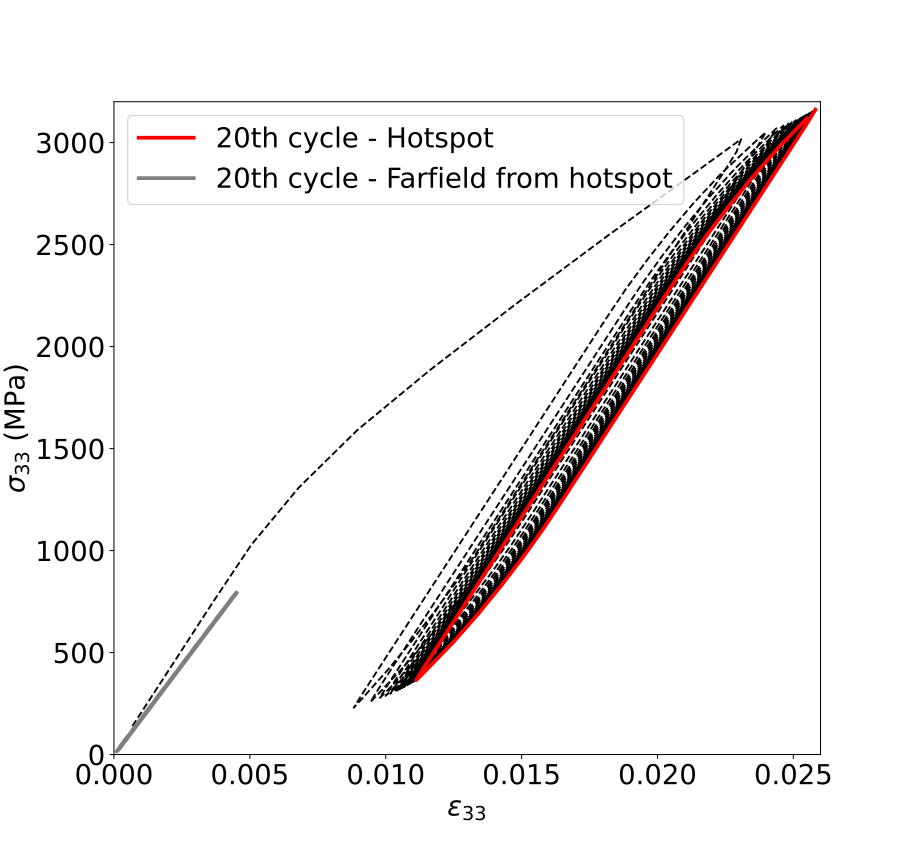}\label{fig:illu_hys}\caption{}
    \end{subfigure}
    \caption{a) Representation of the complex energy field of sample I4 in a plane perpendicular to loading (X-direction) along b) representation of averaging sphere centered on 3 hotspots that are small ligaments (marked as 1,2 and 3) in the defect cluster. c) Illustration of the evolution of stress and strain in the direction of loading or the hysterisis loop at hotspot (broken black line and marked in the averaging sphere) and at an integration point far away from the hotspot (grey line as marked in the averaging sphere) along with the comparison of the 20th stabilised cycle for hotspot (red) and an integration point away from the hotspot (grey).}
    \label{fig:FE_ligaments}
\end{figure}

In fatigue loading, energy is dissipated within the material due to plasticity, a part of which is absorbed and irrecoverable \cite{ellyin1988,charkaluk_fatigue_2002}. This dissipated plastic energy per cycle is the area inside the hysterisis loop which remains almost constant after a saturation point (plastic shakedown) until failure. Charkaluk et al. \cite{charkaluk_energetic_2000} performed LCF tests on cast iron samples and demonstrated that dissipated plastic energy which remains constant after a saturation point is a good damage parameter for fatigue estimations. Later, Ellyin and Golos \cite{ellyin1988} developed the idea to include elastic strain energy density in the energy based fatigue model. Similarly, Maurel et al. \cite{maurel_engineering_2009,maurel_fatigue_2017} proposed a fatigue crack propagation model using the same principle that used dissipated plastic energy and a hydrostatic part of elastic energy as damage indicators. This model was tested in LCF regime for Ni-based and Co-based superalloys and displayed excellent results \cite{maurel_engineering_2009,leost2023}. The use of strain energy comprising dissipated plastic energy as well as elastic energy can explain both phases of fatigue failure i.e. crack-initiation and propagation. Therefore energy-based (EB) models are used in this work to explain the mesoscopic damage. \\
Image-based Finite Element (FE) models are simulated under corresponding loading conditions as per experiments, involving identical stress amplitudes at the same temperature, frequency, and load ratio. Cyclic loads are applied for 20 cycles, considering that the hysteresis loop stabilizes even at the global hotspot of the FE models (i.e., the integration point with the maximum stress in the FE model). The stabilization of the hysteresis loop is crucial in EB models to obtain a closed curve \cite{maurel_engineering_2009}. Therefore, on the stabilised cycle i.e., the 20\textsuperscript{th} cycle, energy is computed at each integration point, see Fig. \ref{fig:FE_ligaments}(c). The energy is composed of hydrostatic part which plays a role in the opening of crack and a deviatoric part which is the plastic energy dissipated or observed by the material which can be expressed as,
\begin{equation} \label{eq4}
    W = W_{open} + W_{plas}
\end{equation}
where $W_{open}$ is the hydrostatic part of the energy given as \cite{maurel_engineering_2009,maurel_fatigue_2017},
\begin{equation} \label{eq5}
    W_{open} = \frac{1}{3}\int_{cycle} <tr(\underline{\underline{\sigma}})><tr(d \underline{\underline{\varepsilon}})>
\end{equation}
Due to the incompressibility of the plastic strain and since the total strain can be decomposed into elastic and plastic part, equation \ref{eq5} becomes,
\begin{equation} \label{eq6}
    W_{open} = \frac{1}{3}\int_{cycle} <tr(\underline{\underline{\sigma}})><tr(d \underline{\underline{\varepsilon}}^{el})>
\end{equation}
and the dissipated plastic energy part is given as \cite{maurel_engineering_2009,maurel_fatigue_2017,charkaluk_energetic_2000},
\begin{equation} \label{eq7}
    W_{plas} = \int_{cycle} \underline{\underline{s}}:d\underline{\underline{\varepsilon}}^p
\end{equation}
where $\underline{\underline{s}}$ is the deviatoric part of the stress tensor.\\

\subsection{Volumetric homogenization}
As demonstrated in section \ref{x_results}, a single relationship is insufficient to characterize the fatigue life of a material at a macroscopic scale, owing to the inherent variability among different samples. The criticality of these samples therefore needs to be estimated at a much more local scale, notably at the mesoscopic scale. Indeed, damage accumulation in a material occurs at a very mesoscopic scale, particularly in presence of stress concentrating heterogeneities like cavities. Therefore, non-local methodologies can be employed to predict fatigue lives at the mesoscopic scale \cite{le_investigation_2018,pedranz_new_2023}. One prevalent non-local approach involves volumetric homogenization, wherein a sphere is centered on the hotspot, and the field is volumetrically averaged within this sphere as given by,
\begin{equation}\label{eq7}
    W_{eq}=\frac{1}{V}\int _{V}W dV
\end{equation}
where $V$ is the volume within the averaging sphere of radius $r$ and $W_{eq}$ is the averaged equivalent energy. The resultant averaged field is subsequently utilized for fatigue calculations.
The characteristic length of the sphere (its radius $r$) in volumetric homogenization is identified as a material parameter representing the transition from a short crack to a long crack \cite{taylor_theory_2008,taylorGeometricalEffectsFatigue1999,lannay_stability_2023}. Below this length, the crack is considered to be a small crack that is dependent on microstructural aspects. In instances involving notches and defects, the lifespan for a short crack to initiate and propagate to the length of a long crack also relies on the local morphology of defects, specifically the local stress fields surrounding the initiation site of a small crack within highly stressed regions. Caton and Jha \cite{caton_small_2010} conducted a study on the initiation and propagation of small cracks in IN100 samples at \SI{650}{\celsius}, revealing a strong dependence of small crack initiation and propagation on the morphology of pores. Notably, the pores in their study were considerably smaller ($<$ \SI{100}{\micro \meter}) than the present work. Current work incorporates the actual morphologies of intricate defects into image-based finite element models, providing realistic full-field stress distributions. At hotspots or integration points with very high stress concentrations, the maximum local stress is sometimes even 7-8 times higher when compared to applied stress (see Fig. \ref{fig:FE_ligaments}(c)). These hotspots exhibit local ratcheting phenomenon and cyclic hardening, whereas an integration point at a far-field from the hotspot experiences a nearly linear stress-strain evolution with small plastic energy dissipation. Therefore in such a multiaxial stress-strain distribution conditions, an equivalent energy field computed from such local field is expected to describe crack-initiation life accurately. Generally, crack initiation can mean initiation of a small crack or initiation of long crack where the latter is predominantly used in the fracture mechanical approaches. Previously reported researches shows that around 90\% of the total fatigue life is dedicated to initiation of a small crack and its propagation up to long crack or to the initiation of a long crack \cite{musinski2012microstructure,christ2009propagation,wang1999technical}. Therefore, averaged equivalent energy $W_{eq}$ measured at a critical zone is equated to the fatigue life measured by experiments assuming that propagation life is negligible with a simple Basquin relationship and is defined as follows,
%as shown in Fig. \ref{fig:V03} where the equivalent energy of cylindrical R125 samples was measured with various parameters and is defined as follows, 
\begin{equation}\label{eq8}
    W_{eq}(p_1,p_2,\cdots,p_n) = A(N_f)^b
\end{equation}
where $N_f$ is the experimental fatigue life of samples and $A$, $b$ are constants of exponential relationship to be identified, $p_1,p_2,\cdots,p_n$ are the various parameters like averaging radius, stressed volume, hotspot etc in measuring equivalent energy which will be discussed in following subsections. By this relationship, it is assumed that irrespective of the macroscopic load applied, damage accumulation is a mesoscopic process and the energy per cycle measured at mesoscopic scale within a critical zone governs the crack initiation life. Therefore, the fatigue lives after neglecting the part dedicated to propagation of cracks in porous samples is estimated via cumulative mesoscopic energy per cycle which is assumed to remain constant until the initiation of the critical crack \cite{charkaluk_energetic_2000, maurel_engineering_2009,maurel_fatigue_2017,ellyin1988,charkaluk_fatigue_2002}.
\section{Novel non-local model for complex real defects}
While non-local methodologies have demonstrated efficacy in idealized scenarios such as notches or spherical pores, their application yields substantial measurement uncertainties when addressing real defects. This discrepancy becomes particularly pronounced when dealing with tortuous defects, such as spongeous shrinkage, which gives rise to clusters of defects. This limitation can be attributed to inherent deficiencies in the non-local model, specifically in defining parameters such as the stressed volume in the presence of defects and the identification of hotspots when confronted with the complex morphologies of actual defects. For example, in Fig. \ref{fig:FE_ligaments}(b), the application of volumetric homogenization is centered on hotspots which are small ligaments surrounded by unstressed volumes. The following sections address these issues to propose a generalised non-local model which can treat an extreme case of clusters of tortuous defects or even a simple case.

\subsection{Stressed volume}
\begin{figure}[H]
    \centering
    
    \includegraphics[width=1.0\textwidth]{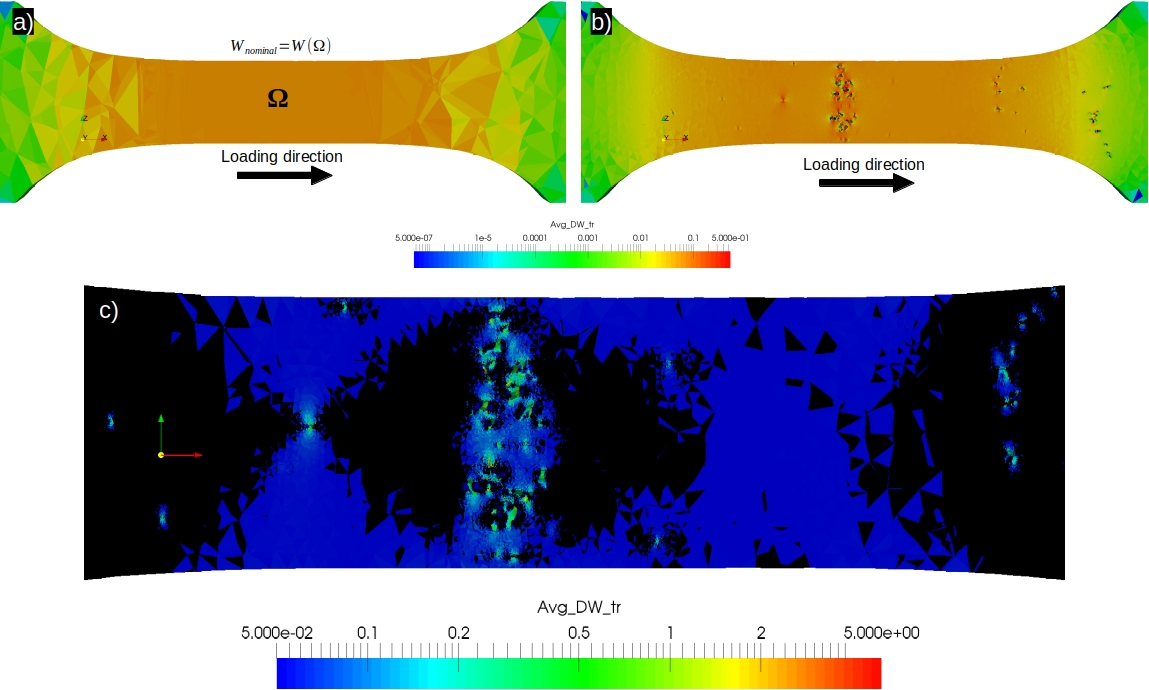}
    \caption{Illustration of the stressed volume concept for sample I4 a) the nominal energy field $W_{nominal}$ which is the energy in structure $\Omega$ in absence of defect (healthy sample) for the respective loading condition of the sample. b) energy field $W$ in presence of defects c) energy field of only the stressed volumes. The unstressed volume in the gauge section is coloured in black.}
    \label{fig:unstressed}
\end{figure}

The fundamental concept behind volumetric averaging lies in incorporating highly stressed regions around a hotspot when estimating fatigue. In the case of notches, the stressed volume surrounds the notch tip, designated as the hotspot, and the non-local approach utilizes an averaging sphere submerged within this stressed volume. When it comes to complex defects like shrinkages and pores, notion of stressed volume is not so direct. Complex defect volumes may contain numerous free surfaces where local stress (or energy) approaches zero due to the surface normal being parallel to the loading direction. Furthermore, shielding effect between defects in defect cluster can also create large zones where the local energy is less than nominal value, see Fig. \ref{fig:unstressed}a and b. The nominal energy is defined as the energy at a particular location of a structure $\Omega$ at the same macroscopic loading conditions in the absence of stress concentrating irregularities, cavity defects in this case. Figure \ref{fig:unstressed}a shows the energy field in the cross-section of the sample in absence of defect compared to the case with defects highlighting all the unstressed volumes in Fig. \ref{fig:unstressed}c (or the elements whose value is less than nominal - in black). It is crucial to emphasize that unlike far-field values, the term "nominal" is structure-dependent and not a constant. Unstressed volumes, characterized by energy levels below the nominal value, do not contribute to crack development and do not align with the principles of volumetric averaging that focus on highly stressed regions. Therefore a modification is proposed in equation \ref{eq7} of volumetric averaging to consider only the stressed volume as follows,
%\begin{equation}\label{eq9}
    %W_{eq}=\frac{1}{V}\int _{cycle}W_{total}dV \quad \forall \quad W > W_{nominal}
%\end{equation}
\begin{equation}\label{eq9}
    W_{eq}=\frac{1}{V}\int _{V}W I(\Omega)dV \qquad I(\Omega) \rightarrow \left\{
    \begin{array}{ll}
        0, & \text{if} \ W(\Omega)< W_{nominal}(\Omega)\\
        1, & \text{otherwise}
    \end{array}
    \right.
\end{equation}
\begin{figure}[H]
    \centering
    
    \begin{subfigure}{0.45\textwidth}\includegraphics[width=\textwidth]{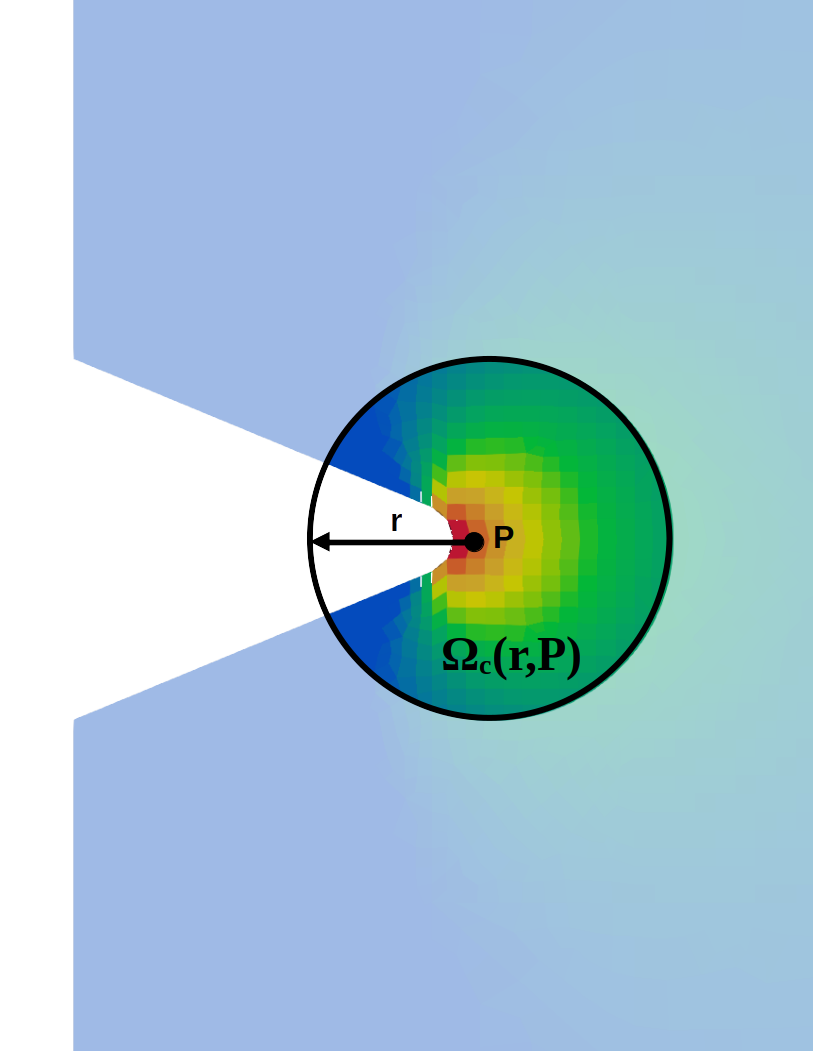}\caption{}\end{subfigure}
    \begin{subfigure}{0.45\textwidth}\includegraphics[width=\textwidth]{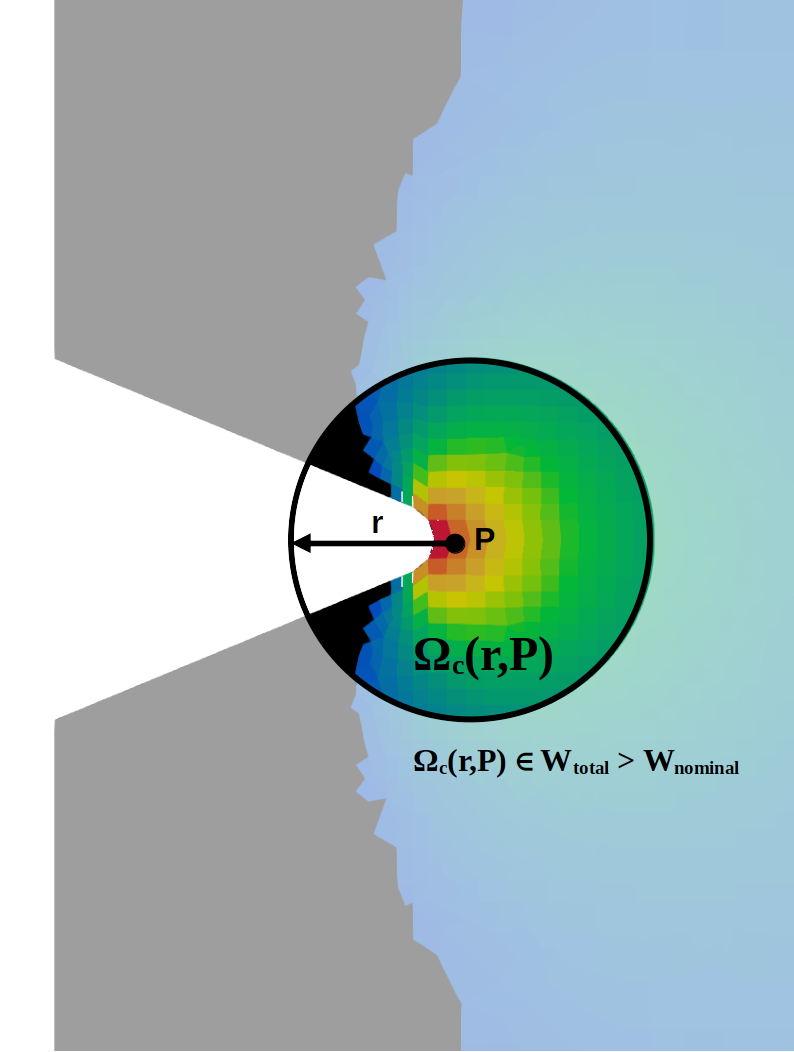}\caption{}\end{subfigure}
    \caption{a) Volumetric homogenization b) Modified volumetric homogenization defining the stressed volume.}
    \label{fig:illu_unstressed}
\end{figure}

where the nominal is a structure dependent value in the absence of defect enabling the application of the method to any geometry irrespective of its complexity. Figure \ref{fig:illu_unstressed} illustrates this modification for the simple case of a plate containing a notch with an applied stress of 350 MPa. Since the chosen plate in absence of notch is of uniform geometry, nominal energy is constant at all Gauss points and equal to 0.05 J as per the applied stress. %Verify 
In the volumetric homogenization approach, all integration points within the critical zone $\Omega _c(r,P)$ are considered for averaging as shown in Fig. \ref{fig:illu_unstressed}a. The proposed modification, illustrated in Fig. \ref{fig:illu_unstressed}, advocates for the inclusion of only those elements whose local values are higher than nominal value in the averaging process. A similar proposal was made by Morel and Palin-Luc \cite{morel_non-local_2002} where they considered only those elements above a certain critical stress of the material to perform volumetric averaging. It has to be noted that ideally the critical zone is much smaller than illustrated in Fig. \ref{fig:illu_unstressed} and is usually submerged in the stressed zone. But the same argument is not true for the case of defects due to complex morphology and spatial arrangement as shown in Fig. \ref{fig:unstressed}. Presence of defects concentrates stress in certain regions, particularly at surfaces whose normal is perpendicular to loading in uni-axial loading case whilst reduces stress than the nominal value in certain regions due to free surface and shielding effects as shown in Fig. \ref{fig:unstressed}. This modification therefore filters the unstressed volume and considers only the stressed regions for fatigue estimations.

\subsection{Identification of hotspots}
The small crack emanates from the critical hotspot and propagates until it develops into a long crack. In the case of notches, there is only one hotspot which is the notch tip whilst the same is not true when treating real cavity defects. There are previous research works where the critical hotspot was defined by performing averaging at all integration points and selecting the integration point yielding the maximum averaged value to be the hotspot \cite{le_investigation_2018}. However, the reliability of this maximum average method is compromised for two primary reasons: a) it incorporates local values from unstressed regions, and b) the averaging sphere is not accurately centered on the critical hotspot but rather in between two maxima. This method relies on identifying an integration point that provides the maximum averaged field value. Due to tortuous shape of the observed defects, the critical zone determined using this approach tends to be centered on an integration point with a low local value, while the actual hotspots are situated at the surface of the averaging sphere. \\
\begin{figure}[H]
    \centering
    
    \includegraphics[width=0.6\textwidth]{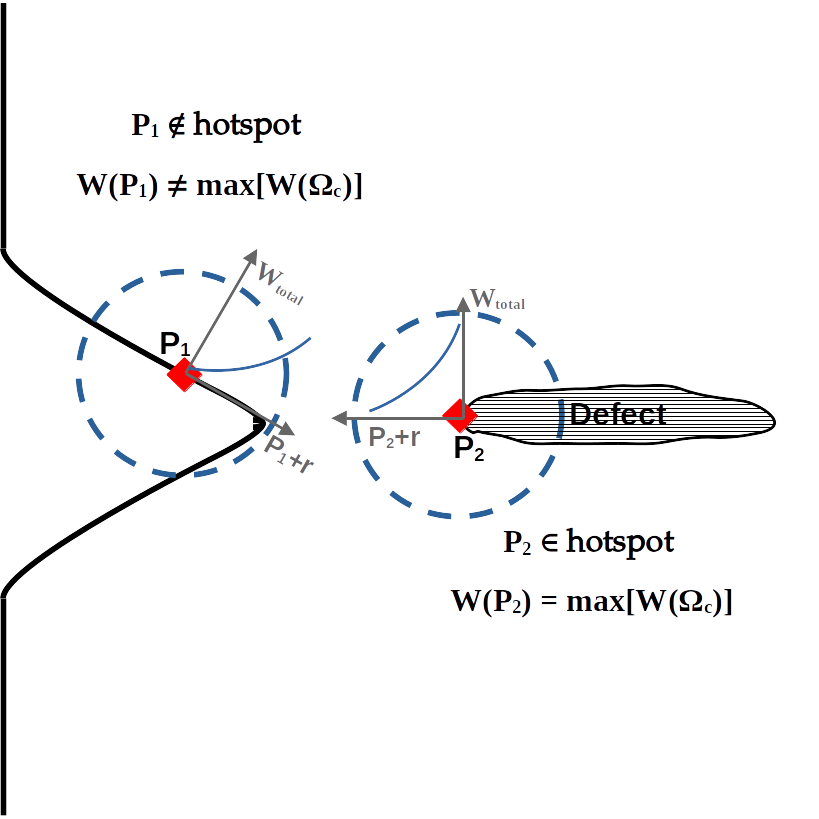}
    \caption{Illustration of method to extract all local maxima or hotspots. Illustrating a case of a defect situated very close to a notch.}
    \label{fig:illu_hotspots}
\end{figure}

\begin{figure}[H]
    \centering
    
    \includegraphics[width=0.6\textwidth]{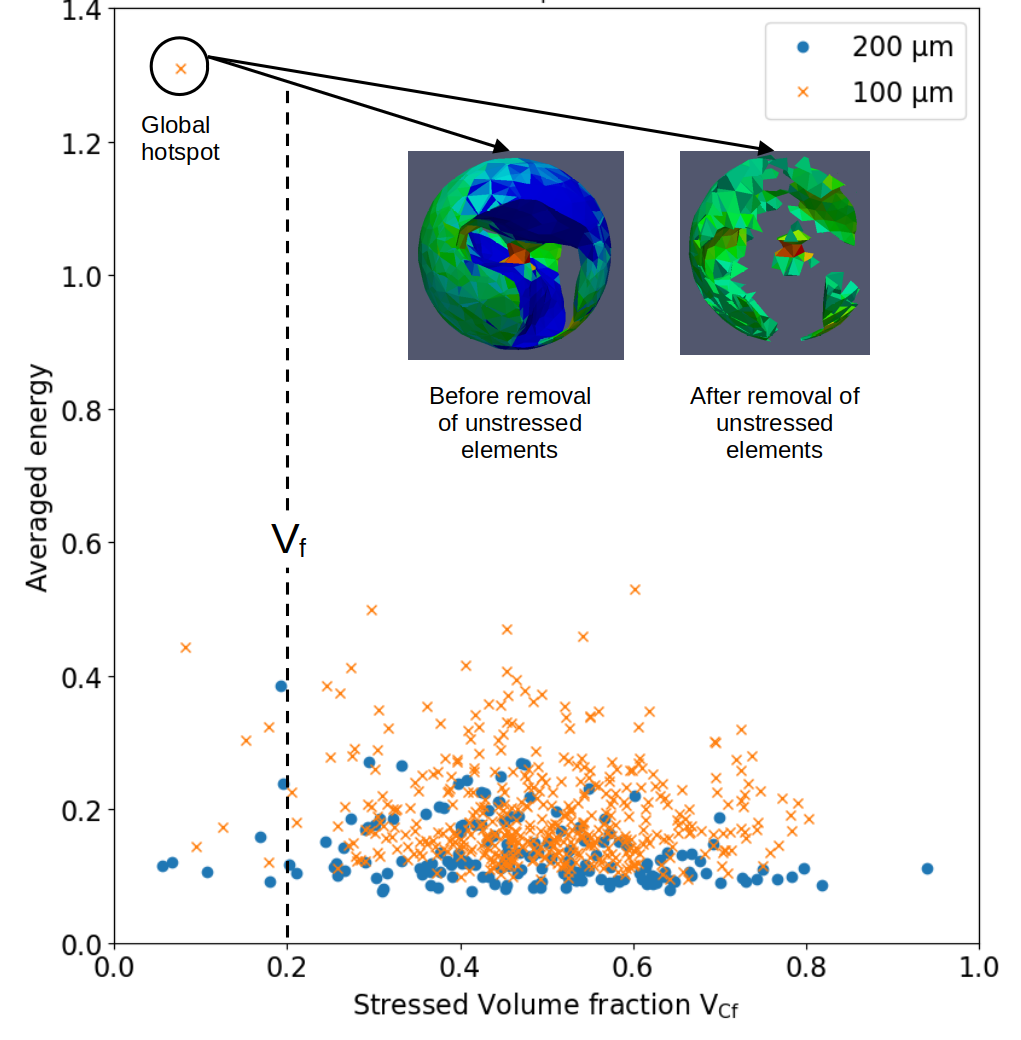}
    \caption{All hotspots of sample I3 extracted for two different radius (\SI{200}{\micro \meter} and \SI{100}{\micro \meter}). Control sphere centered on the global hotspot which is located on a small ligament is represented for the cases before and after deleting unstressed elements displaying that stressed volume is confined to a very small zone. The line also represents chosen threshold volume fraction parameter for this material.}
    \label{fig:hotspots}
\end{figure}

In the general context, the energy field constitutes a continuous medium, and hotspots can be defined as local maxima within this continuous field. Given a specified averaging radius, all hotspots within the continuous energy medium can be identified. As illustrated in Fig. \ref{fig:illu_hotspots}, for an averaging radius $r$, one can verify if an integration point $P$ (center of the averaging sphere $\Omega _c$) is the maximum field value within the averaging sphere i.e., $W(P)=\max[W(\Omega _c)]$. Figure \ref{fig:illu_hotspots} illustrates two obvious cases which was used by Gillham et al \cite{gillham_tailoring_2023}, integration point $P_1$ which isn't the maximum value within the averaging sphere $\Omega _c$ or among the elements within a distance of $P_1+r$. On the other hand, $P_2$ is a hotspot since it is a local maxima within the averaging sphere $\Omega _c$. Therefore, the energy gradient will be oriented outwards from the center of averaging sphere as illustrated in Fig. \ref{fig:illu_hotspots} for point $P_2$ whereas the gradient will be oriented inward for point $P_1$. This process is similar to gradient ascent which is much like climbing a hill to reach the peak. \\
From Fig. \ref{fig:illu_unstressed}, one can imagine that there would be one hotspot for a notch which is the notch tip since the averaging sphere centered on other integration points around notch will never be centered on a local maxima. % unless local field values are constant within $\Omega _c$ or in other words, in absence of gradient. 
Importance of gradient in non-local approaches has been stressed by many researchers in the past \cite{QYLAFKU1999,peerlings2000gradient}. %Add example of researches here
%In the case of real defects, it is a cumbersome process to consider gradient of field in averaging but however, the hotspots can be defined as explained above considering if the gradient is negative (flowing outward from the center) or positive within the critical zones. 
In the context of real defects, incorporating the gradient of the field into the averaging process can be complicated. However, the determination of hotspots can still be achieved through the methodology explained earlier, considering the direction of the gradient (negative, indicating outward flow from the center, or positive) within the critical zones.\\
Figure \ref{fig:hotspots} shows all the hotspots determined for sample I3 along with the evolution of averaged energy $W_{avg}$ at each hotspot with respect to stressed volume fraction within the control sphere (averaging sphere). Due to deletion of unstressed volumes and also presence of cavities, the volume of remaining stressed elements within the averaging sphere is not a constant and varies largely as shown in Fig. \ref{fig:hotspots}. Stressed volume fraction of the control sphere $V_{Cf}$ is given as,
\begin{equation}\label{eq10}
    V_{Cf} = \frac{V_i(\Omega _c)}{V_C(r)}
\end{equation}

where $V_i(\Omega _c)$ is the total volume of stressed elements $(\Omega _c)$ within the control sphere of radius $r$ at integration point $i$ and $V_C(r)$ is the volume of a sphere of radius $r$. Interestingly, one can see that there are a large number of hotspots with a stressed volume less than 50 \% of the volume of control sphere. Figure \ref{fig:hotspots} also shows that when the radius of averaging $r$ is increased, the number of hotspots are also reduced as some of the hotspots are consumed by a neighbouring larger value. 
Finally, for each radius of averaging, hotspots can be classified as global hotspot and secondary hotspots. Global hotspot is the one which has the highest local value in the structure whilst the rest are secondary hotspots or other local maxima.

\subsection{Small ligaments}
As demonstrated in section \ref{x_results}, shrinkage defects lead to the formation of small ligaments, which appear as thin filaments suspended within the void. In instances of spongeous shrinkage, the manifestation of small ligaments extends beyond tortuous defect volumes to clusters formed by closely arranged defects. The existence of these small ligaments presents a serious challenge in numerical Finite Element (FE) analysis, as the maximum stress concentration consistently localizes at one of these ligaments, potentially causing oversight of the hotspot initiating primary crack.
Most of the hotspots (obtained from numerical analysis) where the stressed region is confined to a very small volume fraction are these small ligaments as shown in Fig. \ref{fig:FE_ligaments}. This phenomenon is further elucidated in Fig. \ref{fig:hotspots}, wherein the elimination of unstressed volumes results in the isolation of the stressed region from the sphere, confining it to an extremely diminutive volume within the ligament. Owing to its high averaged energy, the crack initiates much earlier and breaks the ligament almost quickly. This process continues in parallel with all the small ligaments. In almost all the cases related to tortuous defects, the hotspots with very small stressed volume fraction are associated to these small ligaments. From Fig. \ref{fig:hotspots}, the hotspot with maximum averaged energy will most certainly be centered on a global hotspot (due to the fact that now unstressed volumes are not considered in averaging) which typically localizes on a small ligament in cases characterized by tortuous defects. Therefore, the global hotspot of the sample can also not be considered as a critical hotspot that might initiate a long crack.\\
To summarize, the determination of the stressed volume within materials containing casting defects has been discussed. Historically, the concept of the highly stressed volume (HSV) has been prevalent in fatigue estimations, representing the volume subjected to stress levels exceeding 90\% of the maximum stress \cite{lin1998}. The non-local estimations applied on notched samples is always submerged within this highly stressed volume \cite{wang2017combined,elkhoukhi2021} and as demonstrated earlier, this is not the case in presence of casting defects due to their tortuous morphology. The same HSV concept cannot be applied readily on porous samples for two reasons. First, it requires redefining the threshold with respect to maximum stress since with the 90\% classically used, all the elements above this threshold might be small ligaments. Second, this threshold for the HSV concept, even if identified for porous samples, is not expected to be constant for different porosity levels and other factors depending on pore distributions. Defining the stressed volume via the nominal stress allows to avoid these problems as well as to accommodate any structure geometry as demonstrated earlier. Furthermore, the definitions of hotspots is absolutely necessary due to the presence of multiple local maxima in the energy field of porous samples. The two processes, definition of stressed volumes and the identification of potential critical points, facilitates and makes it easy to locate all the small ligaments, characterised by their small stressed volume which, as discussed, should not be potential long crack initiators.

\subsection{Fatigue criterion: Parameters calibration}
In complex arrangement of defects as in the case of spongeous shrinkage, failure occurs via multiple crack initiations and coalescence in all scales where some initiate and coalesce much earlier than the rest. Therefore the challenge is to identify the critical zone where an initiated small crack develops at least upto the length of transition of a short crack to long crack. Average energy associated to the initiation of a long crack is then equated to experimental fatigue life as defined by equation \ref{eq8}.
In a broader sense, all hotspots initiating a crack which would coalesce before becoming a long crack is termed to be a small ligament. \\
\begin{figure}[h]
    \centering
    
    \begin{subfigure}{0.475\textwidth}\includegraphics[width=\textwidth]{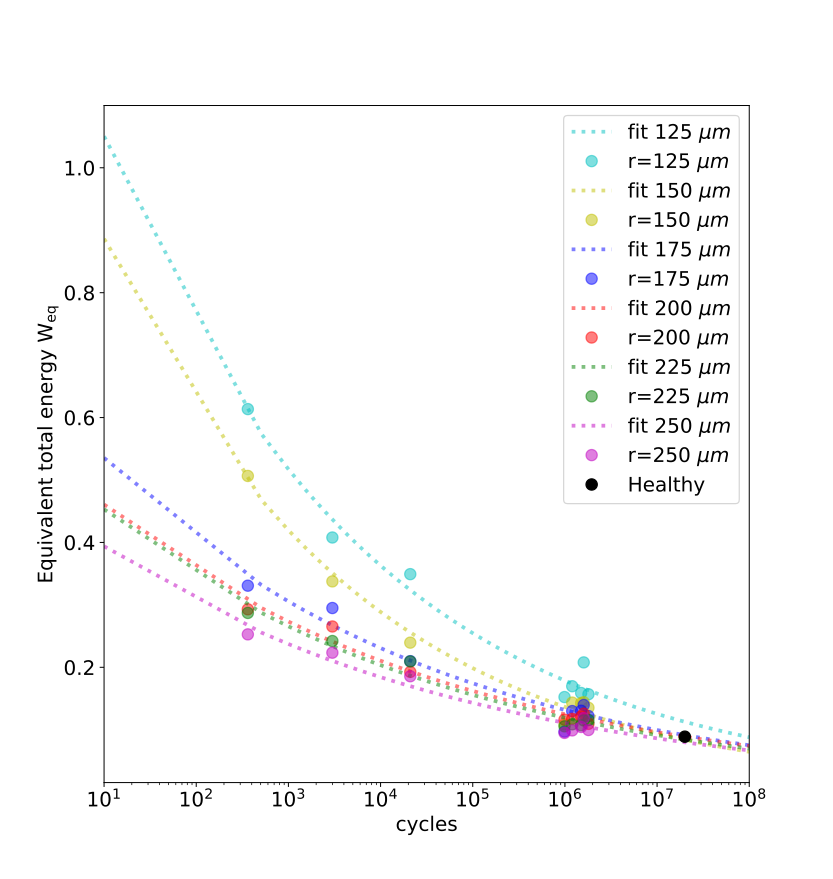}\label{fig:Wohler_V03}\caption{}\end{subfigure}
    \begin{subfigure}{0.475\textwidth}\includegraphics[width=\textwidth]{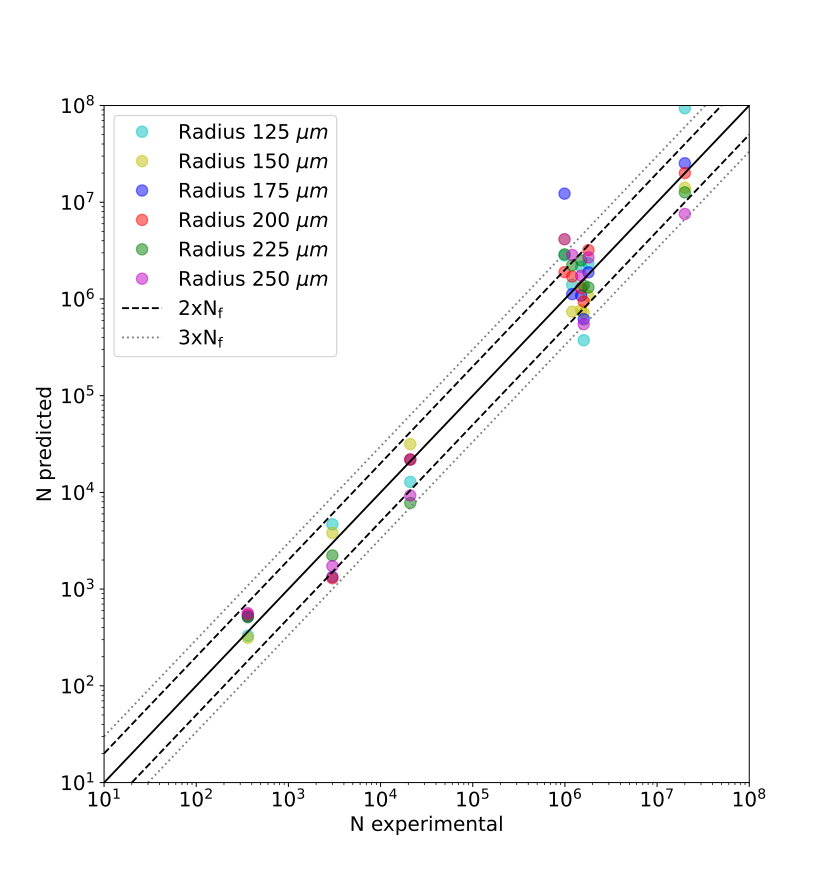}\label{fig:Pred_V03}\caption{}\end{subfigure}
    \caption{a) Fitted curve of equivalent energy $W_{eq}$ vs experimental life $N_f$ at all radius of averaging $r$ and threshold volume fraction $V_f=0.3$ for cylindrical samples of R125 and b) graph of predicted life vs experimental life from the fitted curves.}
    \label{fig:V03}
\end{figure}

Therefore to eliminate all the hotspots which belongs to small ligaments, a parameter of threshold volume fraction $V_f$ is introduced. Threshold volume fraction $V_f$ is the stressed volume fraction of hotspots below which a hotspot is termed as a small ligament. Eliminating small ligaments with the use of stressed volume fraction is proposed since a small confined stressed volume is the main characteristic of small ligaments. %This is a parameter which is case specific, i.e., to a case of very tortuous defects. It is however to be noted .....
Parameters radius of averaging $r$ and threshold volume fraction $V_f$ are identified via a grid search method on all samples except those used for validation (samples R9-R13 and Rf1-Rf2). The process can be broken down to the following steps: \\
\begin{enumerate}
    \item Create a list of parameters i.e. start with an initial value and proceed in an increasing order such as, $r = [100, 125, 150, 175, 200, \cdots]$\SI{}{\micro \meter} and $V_f = [0.0,0.1,0.2,\cdots,0.8]$.
    \item For each $r$ value, all potential hotspots are extracted from each numerical sample.
    \item Starting with initial $V_f$, the hotspot with maximum averaged energy and the one which has a volume fraction $V_{Cf} >$ threshold volume fraction $V_f$ is selected as the critical hotspot (and thereby, the associated energy as the equivalent energy $W_{eq}$ of that sample).
    \item Repeat the process for all samples and using equation \ref{eq11} find the coefficients $A$ and $b$ of the exponential relationship to fit a Basquin-type curve by minimizing the least square error.
    \begin{equation}\label{eq11}
        W_{eq}(r,V_f) = A(N_f)^b
    \end{equation}
    \item From the fitted curve for a given $r$ and $V_f$, measure the factor of difference (FOD) between experimental fatigue life to that predicted by fitted curve for all samples, see equation \ref{eq:FOD}. Since the endurance limit of both materials are known, the predicted fatigue life from the fitted curve at energy corresponding to endurance limit and consequently its FOD is also measured. Generally, a factor of 2 or 3 is accepted in fatigue modelling where FOD is one for a perfect fit. FOD is given as,
    \begin{equation}\label{eq:FOD}
        \textrm{FOD} = \max\left(\frac{N_{pred}}{N_f},\frac{N_f}{N_{pred}}\right)
    \end{equation}
    where $N_{pred}$ is the fatigue life as predicted by the fitted curve for a given sample. Figure \ref{fig:V03}(a) shows the fitted curve for R125 samples at $V_f = 0.3$ along with comparison of predicted and experimental fatigue life as shown in Fig. \ref{fig:V03}(b). It can be seen that FOD for some samples are larger than 3, particularly for smaller averaging radius $r$.
    \item Compute the mean absolute error (MAE) between FODs of samples (along with the FOD at endurance limit) to that of a perfect fit i.e., equal to 1. MAE is given as, \\
    \begin{equation}\label{eq:MAE}
        \text{MAE}(r,V_f) = \frac{\Sigma _{i=1}^{n}|\textrm{FOD}_i(r,V_f)-1|}{n}
    \end{equation}
    where $\textrm{FOD}_i(r,V_f)$ is the $\textrm{FOD}$ of sample $i$ for parameters $r$ and $V_f$ and $n$ is the number of samples. The MAE equals to zero for a perfect fit.
    \item Increase the threshold volume fraction $V_f$ and repeat the above steps until final value of $V_f$.
    \item Repeat the above steps for all radius of averaging $r$.
\end{enumerate}
\begin{figure}[h]
    \centering
    
    \begin{subfigure}{0.45\textwidth}\includegraphics[width=\textwidth]{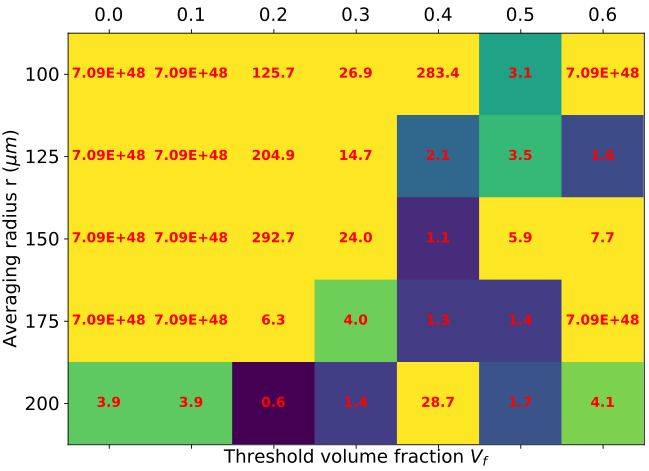}\caption{}\end{subfigure}
    \begin{subfigure}{0.49\textwidth}\includegraphics[width=\textwidth]{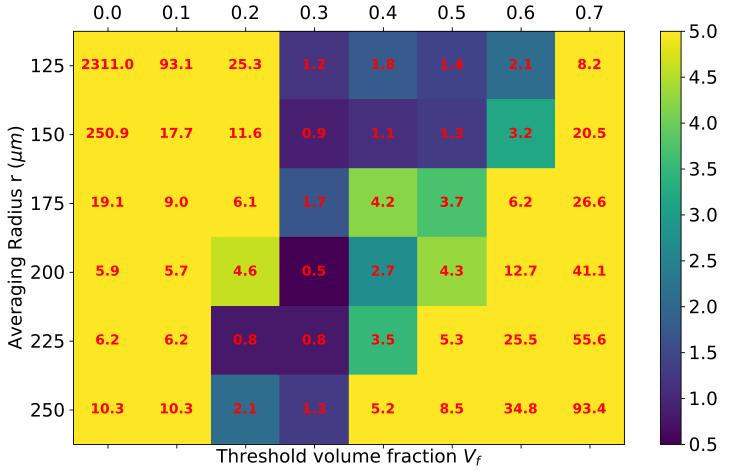}\caption{}\end{subfigure}
    \caption{The error map (MAE) measured using equation \ref{eq:MAE} for each radius of averaging $r$ with respect to the threshold volume fraction $V_f$ for cylindrical samples of material a) IN100 and b) R125. The colour of each parametric cell represents the measured error which is thresholded to a maximum value of 5 for better contrast. Consequently, a value equal to zero represents a perfect fit (or a prediction with an average factor of one with respect to experimental fatigue lives). The exact values of the MAE are represented in each cell.}
    \label{fig:Emap}
\end{figure}

This process generates an error map grid, representing the Mean Absolute Error corresponding to each $r$ and $V_f$ as illustrated in Fig. \ref{fig:Emap} for both materials. Equation \ref{eq11} is utilized to determine the best-fitting curve concerning individual samples, without explicitly considering the endurance limit of healthy sample. Nevertheless, the error assessment of the fitted curve also incorporates the error related to endurance limit $\sigma _{D,healthy}$. Observing Fig. \ref{fig:V03}(a), the averaged energy at $V_f = 0.3$ for R125 samples reduces rapidly as the averaging radius increases signifying the stretch of the energy gradient from the hotspot. In the absence of such a gradient, the averaged value is evidently the local values. Consequently, the estimation of mesoscopic damage through equivalent energy $W_{eq}$ must align with the known endurance limit of the material. This alignment ensures that even in the absence of a gradient, the predicted endurance limit by this model remains consistent with experiments. For IN100 as well as R125, there exists a diagonal band of parameters which fits well with tested samples as seen in Fig. \ref{fig:Emap}. On either sides of this diagonal are the parameters that fits poorly. These three regions can be explained as follows:
\begin{itemize}
    \item The first region with poor fit on the left side of the diagonal with smaller volume fractions consists of hotspots which are associated with small ligaments. Due to their small volume fractions and consequently high local energy density around these hotspots, an early crack is initiated which then breaks the ligament much before the crack could develop into a long crack, see Fig. \ref{fig:Fracto_ligaments}.
    \item A nearly diagonal band with low errors (values less than 3-4) consists of parameters by which stable hotspots could be selected. Stable hotspots, in this context, are defined as those where an initiated small crack develops into a long crack.
    \item and finally there exists hotspots with higher volume fractions (those which are found on the right of the diagonal) which most certainly do not initiate a crack due to their low averaged energy.
\end{itemize}
The presence of the diagonal band suggests a stability in the critical volume of the hotspots associated with reduced errors. As the radius of the averaging sphere increases, the volume fraction producing good fits for that specific radius decreases, as observed in the error maps (Fig. \ref{fig:Emap}). This prompts the question of whether the widely employed critical length theory is indeed a critical volume, implying that, a constant control volume could be a better descriptor.
However, implementing such a control volume method in finite element models would be cumbersome, given that the shape of this control volume is unknown. One can start with control volume method assuming its shape is a sphere. Nevertheless, parameters that produce the minimum error are chosen to measure the averaged equivalent energy $W_{eq}$ and are fitted with experimental fatigue lives using equation \ref{eq11} i.e., $r=200$\SI{}{\micro \meter}, $V_f=0.2$ for IN100, and $r=200$\SI{}{\micro \meter} and $V_f=0.3$ for R125 material.
Although the hotspot with maximum equivalent energy having a volume fraction greater than $V_f$ is chosen as the critical hotspot, other hotspots with sufficiently high energy might initiate a crack too. However, the one with the maximum $W_{eq}$ might initiate the long crack faster than other hotspots. 

\section{Numerical results and Validation}
The identification of the critical zone initiating the primary long crack was accomplished through the methodologies presented in the previous sections. The significance of the chosen parameters is evident through the following observations:
\begin{enumerate}
    \item The selected hotspots with their respective threshold volume fraction $V_f$ do not correspond to small ligaments. This implies that these critical zones effectively capture the initiation of a long crack.
    \item The chosen parameters yield the best fit for the fatigue lives of samples at the mesoscopic scale, i.e., the averaged energy estimated within a critical zone defined by a sphere.
    \item The fitted curve with the chosen parameters traverses through the endurance limit of the respective healthy which is evident due to the incorporation of endurance limit in estimating errors and consequently the best fitting parameters. This signifies that in the absence of defects or heterogeneities, the proposed non-local model predicts the same endurance limit for healthy samples as experimentally known.
\end{enumerate}
\begin{figure}[H]
    \centering
    
    \begin{subfigure}{0.475\textwidth}\includegraphics[width=\textwidth]{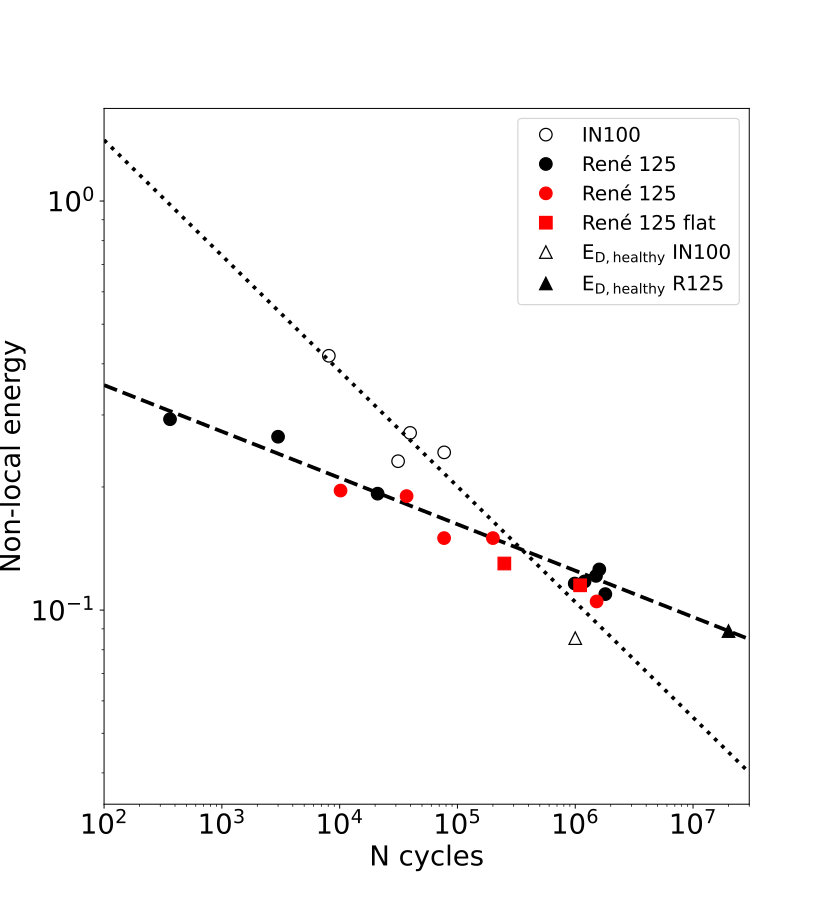}\label{fig:Wohler_pred}\caption{}\end{subfigure}
    \begin{subfigure}{0.475\textwidth}\includegraphics[width=\textwidth]{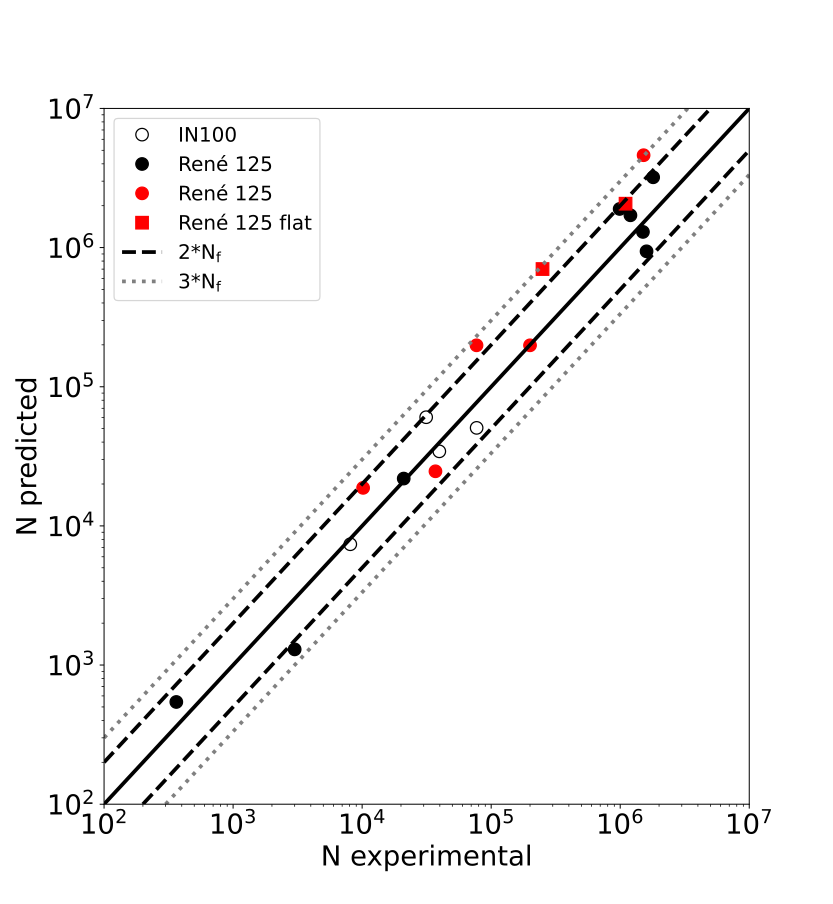}\label{fig:Pred_error}\caption{}\end{subfigure}
    \caption{The mesoscopic model with determined parameters $r$ and $V_f$ a) Wohler fit of basquin type as given by equation \ref{eq11} b) Prediction vs experimental fatigue life from the fit of mesoscopic energy $W_{eq}$.}
    \label{fig:Pred}
\end{figure}

The endurance limit of a healthy sample is a crucial microstructure-specific characteristic of a material. Therefore, consideration of this in the non-local model is crucial for the transition from macroscale to mesoscopic scale. \\
The identified non-local parameters were also applied on the validation samples of R125 (samples R9-R13 and Rf1-Rf2, see Table \ref{table:Exp_results}) which had considerably lesser levels of porosity as compared to other samples and consequently smaller defects and higher fatigue lives at elevated stress amplitudes. From Fig. \ref{fig:Pred}, it can be seen that the proposed EB non local model predicts the fatigue lives of the validation samples within a factor of 3. Nevertheless, from Fig. \ref{fig:Pred}, it can be seen that one single relationship can be established per material between the damage parameter and fatigue lives at mesoscopic scale despite the large variability in the porosity, defect distributions and fatigue lives among the tested samples which wasn't true in the macroscopic scale. Therefore, fatigue life of a model can be better described by energy measured at mesoscopic scale within a critical zone that governs the crack initiation.\\ 
The mesoscopic curve fits well at LCF domain however, the error increases as we approach the HCF domain. At low stress amplitudes, the damage is very microsctructure dependent since the plasticity accumulates in a very minute scale which is also principal reason for the increase of scatter magnitude as stress amplitude decreases. The effect of microstructural heterogeneities is also exaggerated in samples of low porosity when compared to porous samples which explains the error of around factor 3 for the validation samples, for example, sample Rf2 was found to initiate a crack from an oxide inclusion which later propagated towards the nearest pore forming a long crack. However, the proposed model is shown to work well not only for an extreme case of clustered defects (within an average factor less than 2) but also for less porous or nearly healthy samples of completely different size and geometry. 
%However, even for samples with less porosity, the failure zone by experimental and numerical approaches are found to be same. 

\section{Discussion}
% Explain that model works for large fatigue life variance as well for sample with low porosity and different geometry. 
The presence of casting defects can drastically reduce fatigue life of a sample, often by several orders of magnitude, for example, as in the case of sample R3 (see Fig. \ref{fig:Wohler} and table \ref{table:Exp_results}) that failed due to a very large defect. As the maximum defect size in the sample increases, it becomes more and more tortuous creating small ligaments of material. These small ligaments accommodate local hotspots which are essentially not the critical zones to be incorporated in non-local models due to the large discontinuity of stress where the stressed region is confined to a very small volume. Non-local models work with an idea of capturing the initiation of long crack whilst the ones initiated at small ligaments never develop into one. Furthermore, in contrast to straightforward cases, the tortuous defect is enveloped by a sizable unstressed zone, a consequence of its distinctive morphology. While these considerations are pre-established and may induce fewer errors in measurements when dealing with ideal cases such as notched sample or with an idealised spherical or ellipsoidal defect, their significance becomes more pronounced when addressing scenarios involving spongeous shrinkages or when incorporating the real morphology of defects. These drawbacks have hindered the construction of a fatigue model considering the shape of defects, particularly due to small ligaments. This work therefore emphasizes the crucial importance of these aspects and proposes a novel method to define stressed volume, hotspots and eliminate small ligaments from the identified critical zones.

\begin{figure}[h]
    \centering
    
    \begin{subfigure}{0.48\textwidth}\includegraphics[width=\textwidth]{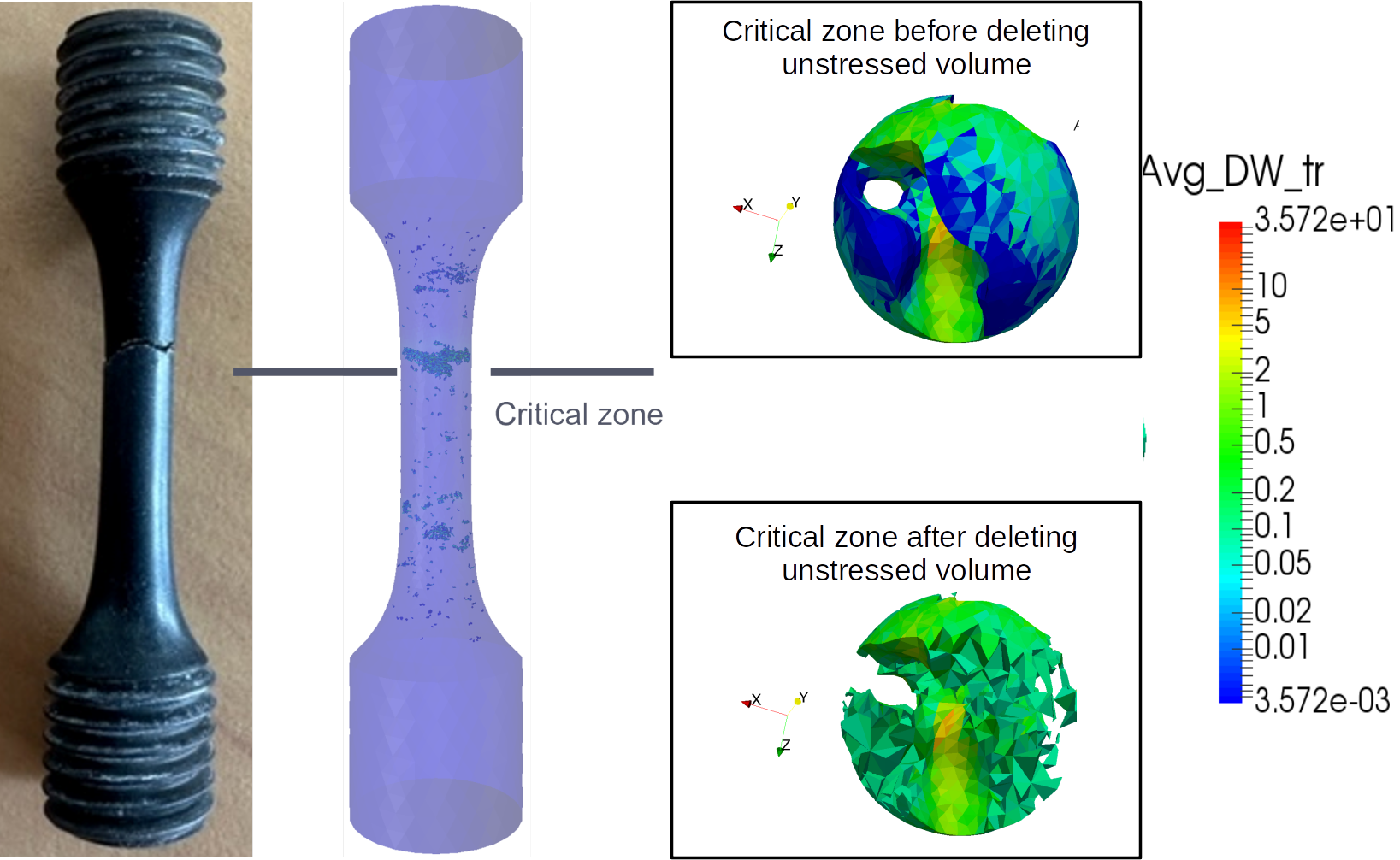}\label{fig:2R1_comp}\caption{}\end{subfigure}
    \begin{subfigure}{0.48\textwidth}\includegraphics[width=\textwidth]{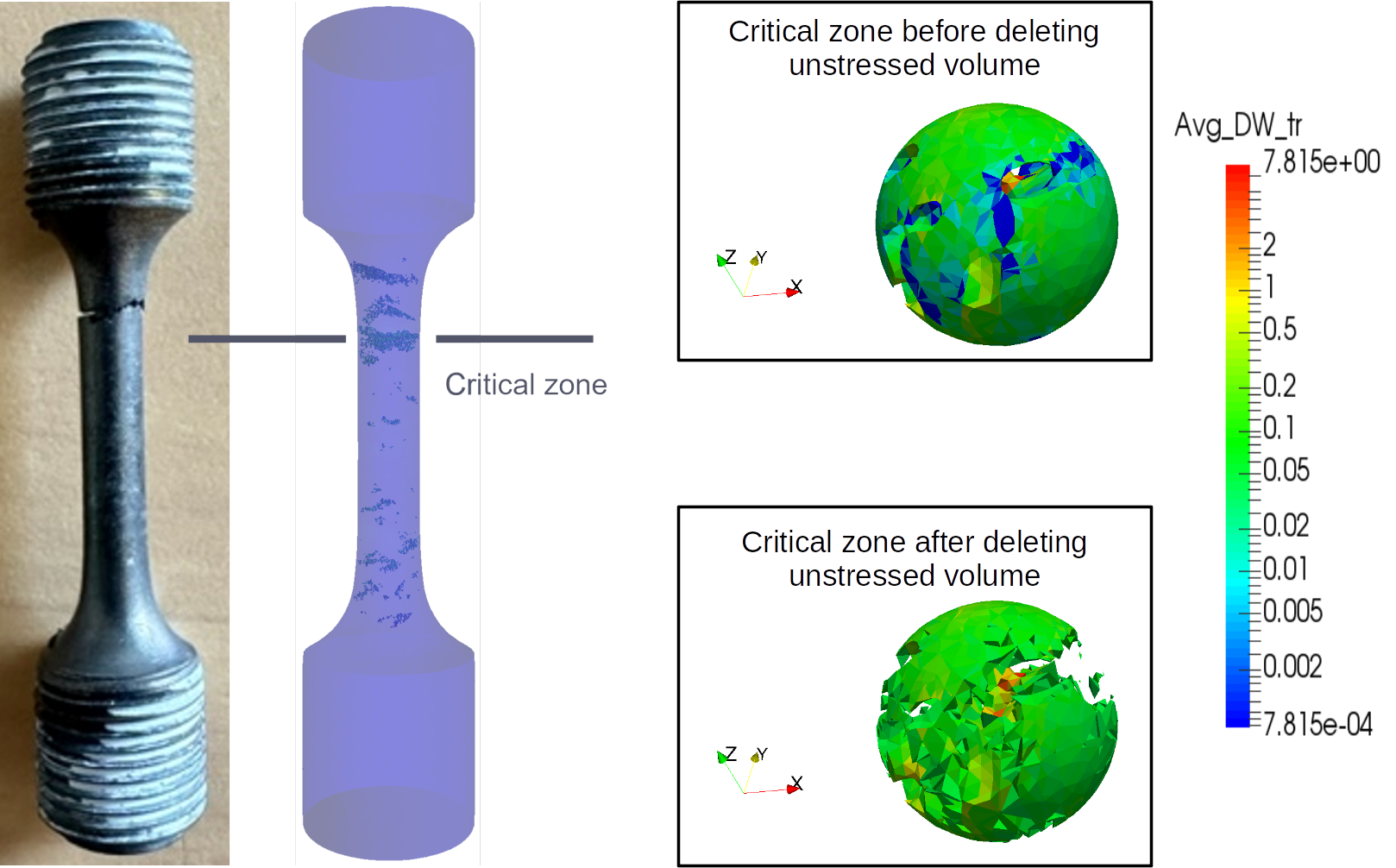}\label{fig:A4_comp}\caption{}\end{subfigure} \hfill
    \begin{subfigure}{0.483\textwidth}\includegraphics[width=\textwidth]{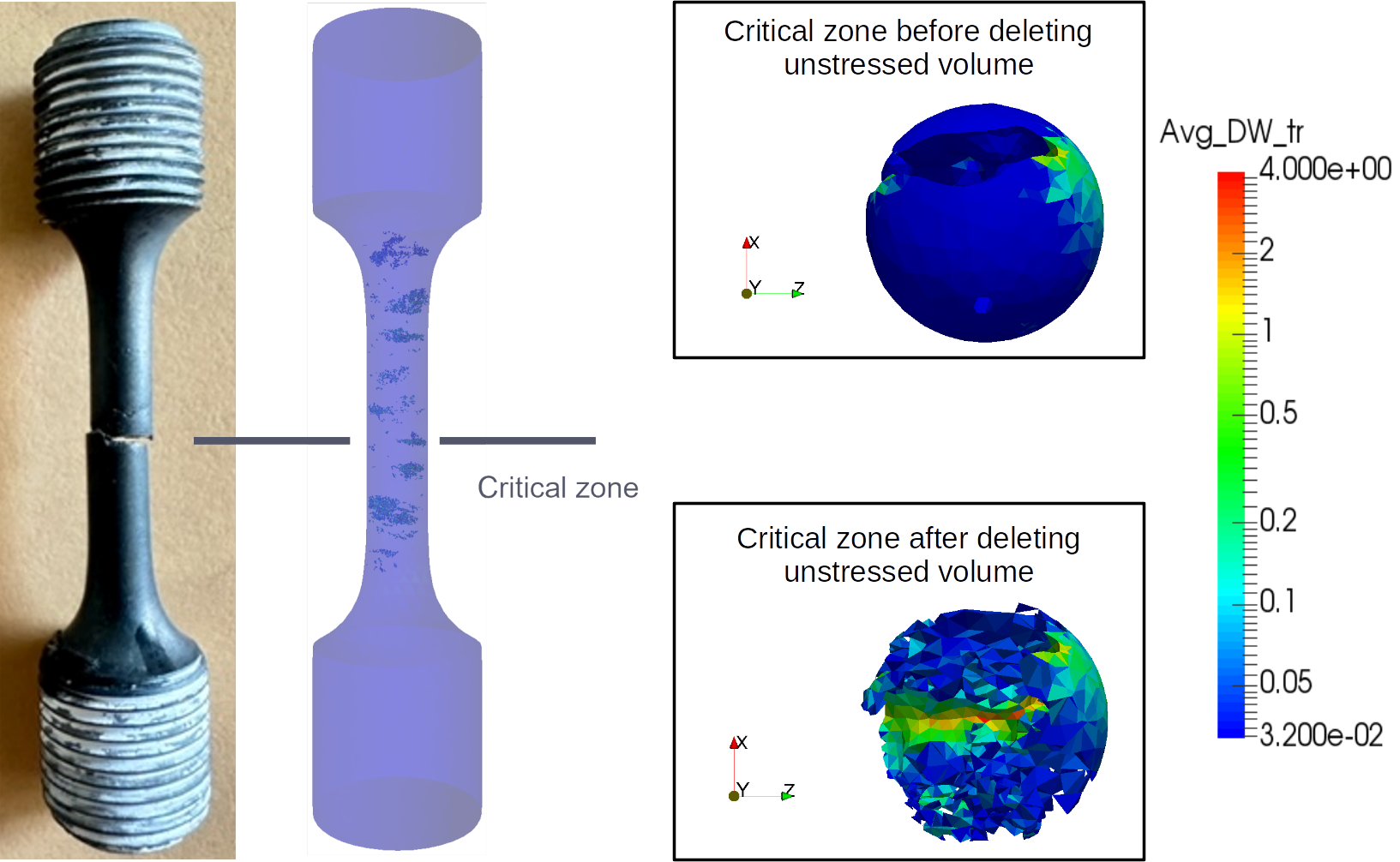}\label{fig:B5_comp}\caption{}\end{subfigure}
    \begin{subfigure}{0.483\textwidth}\includegraphics[width=\textwidth]{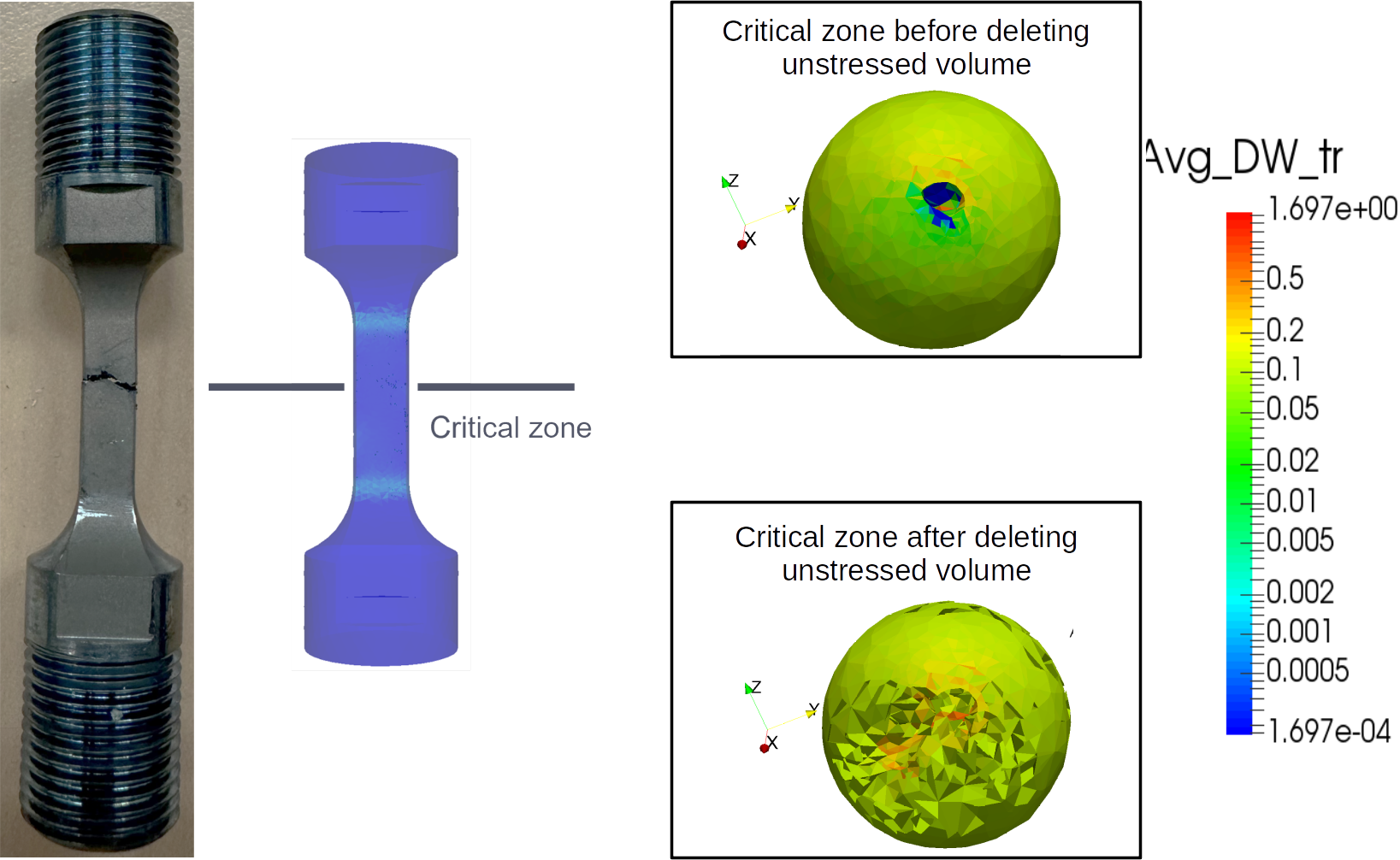}\label{fig:Epr12_comp}\caption{}\end{subfigure}
    \caption{Comparison of experimental failure zone with position of critical zone along with the respective critical zone before and after removing unstressed volumes for samples a) I3 b) R2 c) R4 and d) Rf1. The horizontal line represents the location of critical zone found via numerical predictions.}
    \label{fig:CZ_comp}
\end{figure}

The proposed EB non-local model uses two parameters, namely, radius of averaging sphere $r$ which is associated to the transition length of short crack to long crack and the threshold stressed volume fraction $V_f$ which is a parameter below which the hotspots are deemed to be small ligaments. Parameter $V_f$ is mostly useful only in the presence of tortuous defects and has no significance on smaller defects since they generally do not contain hotspots with stressed volume fraction less than $V_f$. Approaches based on the transition length from short to long cracks estimate the fatigue life required to initiate a long crack while ignoring the cycles dedicated to the propagation of this crack to failure. In this work, the fatigue life dedicated to crack propagation is assumed to be negligible. However, a comprehensive fatigue analysis that separates crack-initiation and propagation lives and applies the proposed model is necessary and will be addressed in an upcoming article.
In methods that incorporate true microstructural aspects, such as casting defects, into the FE model, the spatial resolution of the XCT scan significantly impacts the quality of the results. While higher resolution yields finer results, the computational cost increases exponentially with finer spatial resolution. Therefore, a balanced approach is needed to trade-off between computational cost and spatial resolution, ensuring that the XCT scan is detailed enough to capture all morphological aspects of the microstructure, such as small ligaments and interconnectivity between different defect volumes. The influence of spatial resolution on the results of image-based FE models requires further study. Concerning the computation cost, it is also possible to perform fast elasto-viscoplastic computations with approaches based on Neuber \cite{palchoudhary2024} that can significantly reduce the computation cost and is interesting for applications based on image-based FE models. Additionally, advances in modern science now allow for the generation of synthetic microstructures through machine learning techniques, which can explicitly include realistic casting defects \cite{matpadi2023generation}. A fatigue model that accounts for the realistic shapes of casting defects can be directly applied to such synthetic microstructures, eliminating the need for tomography on multiple samples for certain statistical analyses.\\

Figure \ref{fig:CZ_comp} shows that positions of critical zone found by numerical approaches correlates well with the fracture zone of samples. This is evident as firstly, the stress concentration is large at defect cluster and secondly, the large tortuous defect which most certainly is the critical defect is found within the defect cluster as also explained in a previous paper \cite{matpadi2023generation}. On the other hand, the numerical failure zone also corresponds well with experimental findings for less porous samples as shown in Fig. \ref{fig:CZ_comp}(d). The proposed method uses the nominal energy which is dependent on the location in a structure to define the stressed volumes in presence of microstructural heterogeneities. Although cylindrical and rectangular cross section samples were used in this work, the critical defects were also found outside the homogeneous sections of the samples as shown in Fig. \ref{fig:CZ_comp}(c) for example. The nominal energy in such cases or in any complex structures can be found by simply performing numerical simulations on defect-free samples and later correlating them with defective structures to define the stressed volumes.\\
While calibrating the parameters of the model, note that utmost importance is paid to verify if the fitted curve passes through the endurance limit of a defect free material such that predicted endurance limit by the mesoscopic model remains unchanged in absence of defects. However, an intriguing question arises regarding the potential existence of a distinct critical energy value for a defect-free material, above which damage initiation occurs at the mesoscopic scale, as alluded to by Morel and Palin-Luc \cite{morel_non-local_2002}. This topic requires further attention and is not treated in this work.
%Impact of the XCT spatial resolution on the statistical analysis of casting defects and numerical analysis of the respective image-based FE model is also another important aspect that needs to be investigated. 
Moreover, the endurance limit of a healthy sample, utilized in the calibration of the mesoscopic model, is inherently a probabilistic value (defined at 50 \% survivability of the material). The magnitude of scatter in this domain is typically substantial, influenced by microstructural factors such as variations in short-long crack transition length, grain orientation, grain size, and related effects. It would be a great idea to incorporate the microstructural scatter into the mesoscopic damage model to enable a probabilistic prediction at mesoscopic scale rather than being deterministic. Furthermore, probabilistic models can be formulated by calibrating a Weibull-based model, ensuring the survival of each local volume within the mesoscopic critical zones identified with the proposed method \cite{zhu2022probabilistic}. However, it is important to note that constructing such models would necessitate a more extensive number of tested samples. Nevertheless, it is possible to augment the fatigue response database using synthetic microstructures to construct probabilistic fatigue life estimation methods with techniques such as Monte-Carlo \cite{elkhoukhi2020}, or other statistical approaches.\\
Nevertheless, this work focuses on addressing the issues encountered when applying non-local theory to non-ideal cases and proposes an important generalisation of the non-local approach such that it can be even applied to real defects by defining the stressed volumes and associated hotspots. It has been shown that the model calibrated on the extreme case of clustered defects also works on less porous samples and of different size and geometry. 
%The entire method was applied via a python code and the FE model was decomposed into multiple domains for a rapid parallel computation. 

\section{Conclusion}
This study introduces a generalized treatment of the non-local approach to fatigue failure tailored for addressing real microstructural heterogeneities. Two sample sets, each containing casting defects of distinct materials, Inconel 100 and René 125, were employed to demonstrate the versatility of this approach. Utilizing X-ray tomographic image volumes enabled the visualization and segmentation of casting defects within the microstructures, facilitating the creation of digital representations through image-based finite element models. Experimental determination of fatigue life for these samples was conducted and subsequently compared with numerical simulations performed on the image-based finite element models under identical conditions to the experiments. The following conclusions were drawn from an extensive analysis:\\
\begin{enumerate}
    \item In the presence of large clusters of spongeous shrinkages, cracks initiate on the defect surface associated to one of the clusters and coalesce with neighboring defects after propagating along favorable crystal slip planes. This coalescence phenomenon is observed across all scales due to the high porosity of the samples. Large tortuous defects give rise to small ligaments that initiate the first few cracks due to their high local stress concentration, breaking the ligament well before it can evolve into a long crack.
    \item To address these findings, an innovative method is developed and a novel energy-based non-local model is proposed that incorporates the real morphology of casting defects, validated through comparisons of numerical simulations from image-based finite element (FE) models with experimental results. This model predicts fatigue life at the mesoscopic scale using equivalent energy as the damage parameter. It enhances existing non-local methods by considering only stressed volumes and identifying hotspots likely to evolve into long cracks during the averaging process.
    \item The model parameters were calibrated using highly porous samples, yielding accurate fatigue life predictions at mesoscopic scale within a factor of 3 of the experimental results. The model was subsequently validated on samples with significantly lower porosity demonstrating its effectiveness not only in the case of clustered defects but also for samples of different geometries, sizes, and less porous conditions.
    \item This development enables more in-depth fatigue analyses in the fields of crack-propagation, fretting-fatigue etc promoting the use of true morphology of microstructural heterogeneities. Additionally, it facilitates statistical fatigue analyses on synthetic microstructures containing realistic casting defects.
\end{enumerate}
\section*{Acknowledgements}
The authors extend their sincere gratitude to Besnik Sadriji and Sandrine Charles of Safran Aircraft Engines for stimulating discussions and assistance with the necessary resources for this project. Additionally, authors extend their appreciation to Matthieu Rambaudon of Centre des Matériaux -- Mines Paris for his invaluable support in conducting fatigue testing on the samples. Special gratitude is also expressed to René Cluzet and Frédéric Coutard of Centre des Matériaux -- Mines Paris for their assistance with sample machining.
%\section{Author Contributions} 
%\textbf{Arjun Kalkur MATPADI RAGHAVENDRA: }Conceptualization, Methodology, Software, Investigation, Writing - Original Draft, Writing - Review \& Editing. \textbf{Vincent MAUREL: }Methodology, Investigation, Writing - Review \& Editing, Supervision. \textbf{Lionel MARCIN: }Supervision, Investigation, Resources, Writing - Review \& Editing. \textbf{Henry PROUDHON: }Supervision, Investigation, Project administration, Writing - Review \& Editing.

\section*{References}
%\bibliographystyle{elsarticle-num}
%\bibliography{My_references.bib}

\begin{thebibliography}{10}
	\expandafter\ifx\csname url\endcsname\relax
	\def\url#1{\texttt{#1}}\fi
	\expandafter\ifx\csname urlprefix\endcsname\relax\def\urlprefix{URL }\fi
	\expandafter\ifx\csname href\endcsname\relax
	\def\href#1#2{#2} \def\path#1{#1}\fi
	
	\bibitem{buffiereExperimentalStudyPorosity2001}
	J.-Y. Buffi{\`e}re, S.~Savelli, P.~Jouneau, E.~Maire, R.~Foug{\`e}res,
	Experimental study of porosity and its relation to fatigue mechanisms of
	model {{Al}}\textendash{{Si7}}\textendash{{Mg0}}.3 cast {{Al}} alloys,
	Materials Science and Engineering: A 316~(1-2) (2001) 115--126.
	\newblock \href {http://dx.doi.org/10.1016/S0921-5093(01)01225-4}
	{\path{doi:10.1016/S0921-5093(01)01225-4}}.
	
	\bibitem{NADOT2022106531}
	Y.~Nadot,
	\href{https://www.sciencedirect.com/science/article/pii/S0142112321003856}{Fatigue
		from defect: Influence of size, type, position, morphology and loading},
	International Journal of Fatigue 154 (2022) 106531.
	\newblock \href
	{http://dx.doi.org/https://doi.org/10.1016/j.ijfatigue.2021.106531}
	{\path{doi:https://doi.org/10.1016/j.ijfatigue.2021.106531}}.
	\newline\urlprefix\url{https://www.sciencedirect.com/science/article/pii/S0142112321003856}
	
	\bibitem{rotellaInfluenceNaturalDefects2017}
	A.~Rotella, Y.~Nadot, M.~Piellard, R.~Augustin, Influence of natural defects on
	the fatigue limit of a cast {{Al}}-{{Si}} alloy, Procedia Structural
	Integrity 7 (2017) 513--520.
	\newblock \href {http://dx.doi.org/10.1016/j.prostr.2017.11.120}
	{\path{doi:10.1016/j.prostr.2017.11.120}}.
	
	\bibitem{KUNZ201247}
	L.~Kunz, P.~Lukáš, R.~Konečná, S.~Fintová,
	\href{https://www.sciencedirect.com/science/article/pii/S0142112311003173}{Casting
		defects and high temperature fatigue life of in 713lc superalloy},
	International Journal of Fatigue 41 (2012) 47--51.
	\newblock \href
	{http://dx.doi.org/https://doi.org/10.1016/j.ijfatigue.2011.12.002}
	{\path{doi:https://doi.org/10.1016/j.ijfatigue.2011.12.002}}.
	\newline\urlprefix\url{https://www.sciencedirect.com/science/article/pii/S0142112311003173}
	
	\bibitem{wangFatigueBehaviorA356T62001}
	Q.~G. Wang, D.~Apelian, D.~A. Lados, Fatigue behavior of a356-t6 aluminum cast
	alloys.part i. effect of casting defects, Journal of Light Metals (2001) 12.
	
	\bibitem{elkhoukhi2019}
	D.~El~Khoukhi, F.~Morel, N.~Saintier, D.~Bellett, P.~Osmond, V.-D. Le,
	J.~Adrien, Experimental investigation of the size effect in high cycle
	fatigue: {{Role}} of the defect population in cast aluminium alloys,
	International Journal of Fatigue 129 (2019) 105222.
	\newblock \href {http://dx.doi.org/10.1016/j.ijfatigue.2019.105222}
	{\path{doi:10.1016/j.ijfatigue.2019.105222}}.
	
	\bibitem{bercelliProbabilisticApproachHigh2021}
	L.~Bercelli, S.~Moyne, M.~Dhondt, C.~Doudard, S.~Calloch, J.~Beaudet, A
	probabilistic approach for high cycle fatigue of {{Wire}} and {{Arc Additive
			Manufactured}} parts taking into account process-induced pores, Additive
	Manufacturing 42 (2021) 101989.
	\newblock \href {http://dx.doi.org/10.1016/j.addma.2021.101989}
	{\path{doi:10.1016/j.addma.2021.101989}}.
	
	\bibitem{tammas-williamsXCTAnalysisInfluence2015}
	S.~{Tammas-Williams}, H.~Zhao, F.~L{\'e}onard, F.~Derguti, I.~Todd,
	P.~Prangnell, {{XCT}} analysis of the influence of melt strategies on defect
	population in {{Ti}}\textendash{{6Al}}\textendash{{4V}} components
	manufactured by {{Selective Electron Beam Melting}}, Materials
	Characterization 102 (2015) 47--61.
	\newblock \href {http://dx.doi.org/10.1016/j.matchar.2015.02.008}
	{\path{doi:10.1016/j.matchar.2015.02.008}}.
	
	\bibitem{murrEFFECTBUILDPARAMETERS}
	L.~Murr, S.~Gaytan, F.~Medina, E.~Martinez, D.~Hernandez, L.~Martinez,
	M.~Lopez, R.~Wicker, S.~Collins, Effect of build parameters and build
	geometries on residual microstructures and mechanical properties of ti-6al-4v
	components built by electron beam melting (ebm).
	
	\bibitem{murakamiEffectsDefectsInclusions1994}
	Y.~Murakami, M.~Endo, Effects of defects, inclusions and inhomogeneities on
	fatigue strength, International Journal of Fatigue 16~(3) (1994) 163--182.
	\newblock \href {http://dx.doi.org/10.1016/0142-1123(94)90001-9}
	{\path{doi:10.1016/0142-1123(94)90001-9}}.
	
	\bibitem{le_investigation_2018}
	V.-D. Le, N.~Saintier, F.~Morel, D.~Bellett, P.~Osmond,
	\href{https://linkinghub.elsevier.com/retrieve/pii/S0142112317303778}{Investigation
		of the effect of porosity on the high cycle fatigue behaviour of cast
		{Al}-{Si} alloy by {X}-ray micro-tomography}, International Journal of
	Fatigue 106 (2018) 24--37.
	\newblock \href {http://dx.doi.org/10.1016/j.ijfatigue.2017.09.012}
	{\path{doi:10.1016/j.ijfatigue.2017.09.012}}.
	\newline\urlprefix\url{https://linkinghub.elsevier.com/retrieve/pii/S0142112317303778}
	
	\bibitem{yamashitaDefectAnalysisFatigue2018}
	Y.~Yamashita, T.~Murakami, R.~Mihara, M.~Okada, Y.~Murakami, Defect analysis
	and fatigue design basis for {{Ni}}-based superalloy 718 manufactured by
	selective laser melting, International Journal of Fatigue 117 (2018)
	485--495.
	\newblock \href {http://dx.doi.org/10.1016/j.ijfatigue.2018.08.002}
	{\path{doi:10.1016/j.ijfatigue.2018.08.002}}.
	
	\bibitem{yang2022}
	S.~Yang, W.~Hu, Z.~Zhan, Q.~Meng, Novel quantification of porosity defects on
	fatigue behavior for cast aluminum-silicon alloys by x-ray micro-tomography,
	Materials Science and Engineering: A 856 (2022) 143992.
	
	\bibitem{tijani2013}
	Y.~Tijani, A.~Heinrietz, T.~Bruder, H.~Hanselka, Quantitative evaluation of
	fatigue life of cast aluminum alloys by non-destructive testing and parameter
	model, International journal of fatigue 57 (2013) 73--78.
	
	\bibitem{tijani2013detection}
	Y.~Tijani, A.~Heinrietz, W.~Stets, P.~Voigt, Detection and influence of
	shrinkage pores and nonmetallic inclusions on fatigue life of cast aluminum
	alloys, Metallurgical and Materials Transactions A 44 (2013) 5408--5415.
	
	\bibitem{szmytka2020}
	F.~Szmytka, E.~Charkaluk, A.~Constantinescu, P.~Osmond, Probabilistic low cycle
	fatigue criterion for nodular cast-irons, International Journal of Fatigue
	139 (2020) 105701.
	
	\bibitem{matpadi2023generation}
	A.~K. Matpadi~Raghavendra, L.~Lacourt, L.~Marcin, V.~Maurel, H.~Proudhon,
	Generation of synthetic microstructures containing casting defects: a machine
	learning approach, Scientific Reports 13~(1) (2023) 11852.
	
	\bibitem{raghavendra2022role}
	A.~K.~M. Raghavendra, T.~Armanni, S.~Charles, L.~Marcin, Role of defects in
	fatigue performance of in100, Engineering Fracture Mechanics 261 (2022)
	108224.
	
	\bibitem{hangaiClusteredShrinkagePores2010}
	Y.~Hangai, O.~Kuwazuru, T.~Yano, T.~Utsunomiya, Y.~Murata, S.~Kitahara,
	S.~Bidhar, N.~Yoshikawa, Clustered {{Shrinkage Pores}} in
	{{Ill}}-{{Conditioned Aluminum Alloy Die Castings}}, MATERIALS TRANSACTIONS
	51~(9) (2010) 1574--1580.
	\newblock \href {http://dx.doi.org/10.2320/matertrans.MAW201032}
	{\path{doi:10.2320/matertrans.MAW201032}}.
	
	\bibitem{Nadot_ellipsoid}
	M.~Vincent, C.~Nadot-Martin, Y.~Nadot, A.~Dragon,
	\href{https://www.sciencedirect.com/science/article/pii/S014211231300248X}{Fatigue
		from defect under multiaxial loading: Defect stress gradient (dsg) approach
		using ellipsoidal equivalent inclusion method}, International Journal of
	Fatigue 59 (2014) 176--187.
	\newblock \href
	{http://dx.doi.org/https://doi.org/10.1016/j.ijfatigue.2013.08.027}
	{\path{doi:https://doi.org/10.1016/j.ijfatigue.2013.08.027}}.
	\newline\urlprefix\url{https://www.sciencedirect.com/science/article/pii/S014211231300248X}
	
	\bibitem{afroz2023}
	L.~Afroz, S.~Inverarity, M.~Qian, M.~Easton, R.~Das, Analysing the effect of
	defects on stress concentration and fatigue life of l-pbf alsi10mg alloy
	using finite element modelling, Progress in Additive Manufacturing (2023)
	1--19.
	
	\bibitem{bleicher2017}
	C.~Bleicher, R.~Wagener, H.~Kaufmann, T.~Melz, Fatigue assessment of nodular
	cast iron with material imperfections, SAE International Journal of Engines
	10~(2) (2017) 340--349.
	
	\bibitem{hardin2009}
	R.~A. Hardin, C.~Beckermann, Prediction of the fatigue life of cast steel
	containing shrinkage porosity, Metallurgical and Materials Transactions A 40
	(2009) 581--597.
	
	\bibitem{serrano-munoz_casting_2018}
	I.~Serrano-Munoz, J.-Y. Buffiere, C.~Verdu,
	\href{https://linkinghub.elsevier.com/retrieve/pii/S0142112318303918}{Casting
		defects in structural components: {Are} they all dangerous? {A} {3D} study},
	International Journal of Fatigue 117 (2018) 471--484.
	\newblock \href {http://dx.doi.org/10.1016/j.ijfatigue.2018.08.019}
	{\path{doi:10.1016/j.ijfatigue.2018.08.019}}.
	\newline\urlprefix\url{https://linkinghub.elsevier.com/retrieve/pii/S0142112318303918}
	
	\bibitem{pedranz_new_2023}
	M.~Pedranz, V.~Fontanari, S.~Raghavendra, C.~Santus, F.~Zanini, S.~Carmignato,
	D.~Lusuardi, F.~Berto, M.~Benedetti,
	\href{https://linkinghub.elsevier.com/retrieve/pii/S0142112322007411}{A new
		energy based highly stressed volume concept to investigate the notch-pores
		interaction in thick-walled ductile cast iron subjected to uniaxial fatigue},
	International Journal of Fatigue 169 (2023) 107491.
	\newblock \href {http://dx.doi.org/10.1016/j.ijfatigue.2022.107491}
	{\path{doi:10.1016/j.ijfatigue.2022.107491}}.
	\newline\urlprefix\url{https://linkinghub.elsevier.com/retrieve/pii/S0142112322007411}
	
	\bibitem{maireQuantitativeXrayTomography2014}
	E.~Maire, P.~J. Withers, Quantitative {{X}}-ray tomography, International
	Materials Reviews 59~(1) (2014) 1--43.
	\newblock \href {http://dx.doi.org/10.1179/1743280413Y.0000000023}
	{\path{doi:10.1179/1743280413Y.0000000023}}.
	
	\bibitem{dezecot3DCharacterizationModeling2017}
	S.~Dezecot, V.~Maurel, J.-Y. Buffiere, F.~Szmytka, A.~Koster, {{3D}}
	characterization and modeling of low cycle fatigue damage mechanisms at high
	temperature in a cast aluminum alloy, Acta Materialia 123 (2017) 24--34.
	\newblock \href {http://dx.doi.org/10.1016/j.actamat.2016.10.028}
	{\path{doi:10.1016/j.actamat.2016.10.028}}.
	
	\bibitem{dezecotSitu3DCharacterization2016}
	S.~Dezecot, J.-Y. Buffiere, A.~Koster, V.~Maurel, F.~Szmytka, E.~Charkaluk,
	N.~Dahdah, A.~El~Bartali, N.~Limodin, J.-F. Witz, In situ {{3D}}
	characterization of high temperature fatigue damage mechanisms in a cast
	aluminum alloy using synchrotron {{X}}-ray tomography, Scripta Materialia 113
	(2016) 254--258.
	\newblock \href {http://dx.doi.org/10.1016/j.scriptamat.2015.11.017}
	{\path{doi:10.1016/j.scriptamat.2015.11.017}}.
	
	\bibitem{taylorGeometricalEffectsFatigue1999}
	D.~Taylor, Geometrical effects in fatigue: A unifying theoretical model,
	International Journal of Fatigue 21~(5) (1999) 413--420.
	\newblock \href {http://dx.doi.org/10.1016/S0142-1123(99)00007-9}
	{\path{doi:10.1016/S0142-1123(99)00007-9}}.
	
	\bibitem{taylor_theory_2008}
	D.~Taylor,
	\href{https://linkinghub.elsevier.com/retrieve/pii/S0013794407002172}{The
		theory of critical distances}, Engineering Fracture Mechanics 75~(7) (2008)
	1696--1705.
	\newblock \href {http://dx.doi.org/10.1016/j.engfracmech.2007.04.007}
	{\path{doi:10.1016/j.engfracmech.2007.04.007}}.
	\newline\urlprefix\url{https://linkinghub.elsevier.com/retrieve/pii/S0013794407002172}
	
	\bibitem{krzyzak_fatigue_2014}
	D.~Krzyżak, T.~Łagoda,
	\href{https://linkinghub.elsevier.com/retrieve/pii/S0142112313003526}{Fatigue
		life estimation of notched elements with use of non-local volumetric method},
	International Journal of Fatigue 61 (2014) 59--66.
	\newblock \href {http://dx.doi.org/10.1016/j.ijfatigue.2013.12.004}
	{\path{doi:10.1016/j.ijfatigue.2013.12.004}}.
	\newline\urlprefix\url{https://linkinghub.elsevier.com/retrieve/pii/S0142112313003526}
	
	\bibitem{berto2011multiaxial}
	F.~Berto, P.~Lazzarin, J.~Yates, Multiaxial fatigue of v-notched steel
	specimens: a non-conventional application of the local energy method, Fatigue
	\& Fracture of Engineering Materials \& Structures 34~(11) (2011) 921--943.
	
	\bibitem{lazzarin_neubers_2005}
	P.~Lazzarin, F.~Berto,
	\href{http://link.springer.com/10.1007/s10704-005-4393-x}{From {Neuber}’s
		{Elementary} {Volume} to {Kitagawa} and {Atzori}’s {Diagrams}: {An}
		{Interpretation} {Based} on {Local} {Energy}}, International Journal of
	Fracture 135~(1-4) (2005) L33--L38.
	\newblock \href {http://dx.doi.org/10.1007/s10704-005-4393-x}
	{\path{doi:10.1007/s10704-005-4393-x}}.
	\newline\urlprefix\url{http://link.springer.com/10.1007/s10704-005-4393-x}
	
	\bibitem{moghtaderi_review_2023}
	S.~H. Moghtaderi, A.~Jedi, A.~K. Ariffin,
	\href{https://www.mdpi.com/1996-1944/16/2/831}{A {Review} on {Nonlocal}
		{Theories} in {Fatigue} {Assessment} of {Solids}}, Materials 16~(2) (2023)
	831.
	\newblock \href {http://dx.doi.org/10.3390/ma16020831}
	{\path{doi:10.3390/ma16020831}}.
	\newline\urlprefix\url{https://www.mdpi.com/1996-1944/16/2/831}
	
	\bibitem{liao2020}
	D.~Liao, S.-P. Zhu, J.~A. Correia, A.~M. De~Jesus, F.~Berto, Recent advances on
	notch effects in metal fatigue: A review, Fatigue \& Fracture of Engineering
	Materials \& Structures 43~(4) (2020) 637--659.
	
	\bibitem{adair_identification_2013}
	B.~S. Adair, W.~S. Johnson, S.~D. Antolovich, A.~Staroselsky,
	\href{https://onlinelibrary.wiley.com/doi/10.1111/j.1460-2695.2012.01715.x}{Identification
		of fatigue crack growth mechanisms in {IN100} superalloy as a function of
		temperature and frequency: {IDENTIFICATION} {OF} {FATIGUE} {CRACK} {GROWTH}
		{MECHANISMS}}, Fatigue \& Fracture of Engineering Materials \& Structures
	36~(3) (2013) 217--227.
	\newblock \href {http://dx.doi.org/10.1111/j.1460-2695.2012.01715.x}
	{\path{doi:10.1111/j.1460-2695.2012.01715.x}}.
	\newline\urlprefix\url{https://onlinelibrary.wiley.com/doi/10.1111/j.1460-2695.2012.01715.x}
	
	\bibitem{wang2016dissolution}
	T.~Wang, X.~Wang, Z.~Zhao, Z.~Zhang, Dissolution behaviour of the $\gamma$'
	precipitates in two kinds of ni-based superalloys, Materials at High
	Temperatures 33~(1) (2016) 51--57.
	
	\bibitem{li2004crystallographic}
	K.~Li, N.~Ashbaugh, A.~Rosenberger, Crystallographic initiation of nickel-base
	superalloy in100 at rt and 538 c under low cycle fatigue conditions,
	Superalloys 2004 251.
	
	\bibitem{JOUIAD2016}
	M.~Jouiad, E.~Marin, R.~Devarapalli, J.~Cormier, F.~Ravaux, C.~{Le Gall}, J.-M.
	Franchet,
	\href{https://www.sciencedirect.com/science/article/pii/S0264127516305238}{Microstructure
		and mechanical properties evolutions of alloy 718 during isothermal and
		thermal cycling over-aging}, Materials \& Design 102 (2016) 284--296.
	\newblock \href
	{http://dx.doi.org/https://doi.org/10.1016/j.matdes.2016.04.048}
	{\path{doi:https://doi.org/10.1016/j.matdes.2016.04.048}}.
	\newline\urlprefix\url{https://www.sciencedirect.com/science/article/pii/S0264127516305238}
	
	\bibitem{kantzos2000high}
	J.~Gayda, R.~Miner, J.~Telesman, P.~Dickerson, High cycle fatigue crack
	initiation study of case blade alloy rene 125, Tech. rep. (2000).
	
	\bibitem{vincentIdentificationElastoviscoplasticityDamage2009}
	S.~Vincent, R.~Billardon, Identification of an elasto-viscoplasticity damage
	model for a copper alloy submitted to extreme thermo-mechanical loading
	(2009) 6.
	
	\bibitem{wojcikIdentificationChabocheLemaitre2021}
	M.~W{\'o}jcik, A.~Skrzat, Identification of {{Chaboche}}\textendash{{Lemaitre}}
	combined isotropic\textendash kinematic hardening model parameters assisted
	by the fuzzy logic analysis, Acta Mechanica 232~(2) (2021) 685--708.
	\newblock \href {http://dx.doi.org/10.1007/s00707-020-02851-z}
	{\path{doi:10.1007/s00707-020-02851-z}}.
	
	\bibitem{gongDeterminationMaterialProperties2010}
	Y.~P. Gong, C.~J. Hyde, W.~Sun, T.~H. Hyde, Determination of material
	properties in the {{Chaboche}} unified viscoplasticity model, Proceedings of
	the Institution of Mechanical Engineers, Part L: Journal of Materials: Design
	and Applications 224~(1) (2010) 19--29.
	\newblock \href {http://dx.doi.org/10.1243/14644207JMDA273}
	{\path{doi:10.1243/14644207JMDA273}}.
	
	\bibitem{zuiderveld1994}
	K.~Zuiderveld, Contrast limited adaptive histogram equalization, in: Graphics
	gems IV, 1994, pp. 474--485.
	
	\bibitem{fiji}
	J.~Schindelin, I.~Arganda-Carreras, E.~Frise, V.~Kaynig, M.~Longair,
	T.~Pietzsch, S.~Preibisch, C.~Rueden, S.~Saalfeld, B.~Schmid, et~al., Fiji:
	an open-source platform for biological-image analysis, Nature methods 9~(7)
	(2012) 676--682.
	
	\bibitem{trimesh}
	{Dawson-Haggerty}, \href{https://trimsh.org/}{trimesh}.
	\newline\urlprefix\url{https://trimsh.org/}
	
	\bibitem{gmsh}
	C.~Geuzaine, J.-F. Remacle, Gmsh: A 3-d finite element mesh generator with
	built-in pre-and post-processing facilities, International journal for
	numerical methods in engineering 79~(11) (2009) 1309--1331.
	
	\bibitem{ASTME2660}
	A.~standard E2660-17, Standard digital reference images for investment steel
	castings for aerospace applications (2017).
	
	\bibitem{rotellaFatigueLimitCast2018}
	A.~Rotella, Y.~Nadot, M.~Piellard, R.~Augustin, M.~Fleuriot, Fatigue limit of a
	cast {{Al}}-{{Si}}-{{Mg}} alloy ({{A357}}-{{T6}}) with natural casting
	shrinkages using {{ASTM}} standard {{X}}-ray inspection, International
	Journal of Fatigue 114 (2018) 177--188.
	\newblock \href {http://dx.doi.org/10.1016/j.ijfatigue.2018.05.026}
	{\path{doi:10.1016/j.ijfatigue.2018.05.026}}.
	
	\bibitem{brueckner2018}
	A.~Brueckner-Foit, M.~Luetje, M.~Wicke, I.~Bacaicoa, A.~Geisert, M.~Fehlbier,
	On the role of internal defects in the fatigue damage process of a cast
	al-si-cu alloy, International Journal of Fatigue 116 (2018) 562--571.
	
	\bibitem{maskeryQuantificationCharacterisationPorosity2016}
	I.~Maskery, N.~Aboulkhair, M.~Corfield, C.~Tuck, A.~Clare, R.~Leach,
	R.~Wildman, I.~Ashcroft, R.~Hague, Quantification and characterisation of
	porosity in selectively laser melted
	{{Al}}\textendash{{Si10}}\textendash{{Mg}} using {{X}}-ray computed
	tomography, Materials Characterization 111 (2016) 193--204.
	\newblock \href {http://dx.doi.org/10.1016/j.matchar.2015.12.001}
	{\path{doi:10.1016/j.matchar.2015.12.001}}.
	
	\bibitem{Bao2021}
	H.~Bao, S.~Wu, Z.~Wu, G.~Kang, X.~Peng, P.~J. Withers, A machine-learning
	fatigue life prediction approach of additively manufactured metals,
	Engineering Fracture Mechanics 242 (2021) 107508.
	\newblock \href {http://dx.doi.org/10.1016/j.engfracmech.2020.107508}
	{\path{doi:10.1016/j.engfracmech.2020.107508}}.
	
	\bibitem{Zebulon}
	\href{http://www.zset-software.com}{Zebulon z-set}.
	\newline\urlprefix\url{http://www.zset-software.com}
	
	\bibitem{ellyin1988}
	F.~Ellyin, K.~Golos, Multiaxial fatigue damage criterion, Journal of
	Engineering Materials and Technology 110~(1) (1988) 63--68.
	
	\bibitem{charkaluk_fatigue_2002}
	E.~Charkaluk, A.~Bignonnet, A.~Constantinescu, K.~Dang~Van,
	\href{http://doi.wiley.com/10.1046/j.1460-2695.2002.00612.x}{Fatigue design
		of structures under thermomechanical loadings: {THERMOMECHANICAL} {FATIGUE}
		{DESIGN}}, Fatigue \& Fracture of Engineering Materials \& Structures 25~(12)
	(2002) 1199--1206.
	\newblock \href {http://dx.doi.org/10.1046/j.1460-2695.2002.00612.x}
	{\path{doi:10.1046/j.1460-2695.2002.00612.x}}.
	\newline\urlprefix\url{http://doi.wiley.com/10.1046/j.1460-2695.2002.00612.x}
	
	\bibitem{charkaluk_energetic_2000}
	E.~Charkaluk, A.~Constantinescu,
	\href{http://www.tandfonline.com/doi/full/10.1179/mht.2000.17.3.001}{An
		energetic approach in thermomechanical fatigue for silicon molybdenum cast
		iron}, Materials at High Temperatures 17~(3) (2000) 373--380.
	\newblock \href {http://dx.doi.org/10.1179/mht.2000.17.3.001}
	{\path{doi:10.1179/mht.2000.17.3.001}}.
	\newline\urlprefix\url{http://www.tandfonline.com/doi/full/10.1179/mht.2000.17.3.001}
	
	\bibitem{maurel_engineering_2009}
	V.~Maurel, L.~Rémy, F.~Dahmen, N.~Haddar,
	\href{https://linkinghub.elsevier.com/retrieve/pii/S0142112308002193}{An
		engineering model for low cycle fatigue life based on a partition of energy
		and micro-crack growth}, International Journal of Fatigue 31~(5) (2009)
	952--961.
	\newblock \href {http://dx.doi.org/10.1016/j.ijfatigue.2008.09.004}
	{\path{doi:10.1016/j.ijfatigue.2008.09.004}}.
	\newline\urlprefix\url{https://linkinghub.elsevier.com/retrieve/pii/S0142112308002193}
	
	\bibitem{maurel_fatigue_2017}
	V.~Maurel, A.~Köster, L.~Rémy, M.~Rambaudon, D.~Missoum-Benziane,
	V.~Fontanet, F.~Salgado-Goncalves, A.~Heudt, H.~Wang, M.~Trabelsi,
	\href{https://linkinghub.elsevier.com/retrieve/pii/S0142112317301391}{Fatigue
		crack growth under large scale yielding condition: {The} need of a
		characteristic length scale}, International Journal of Fatigue 102 (2017)
	184--201.
	\newblock \href {http://dx.doi.org/10.1016/j.ijfatigue.2017.03.021}
	{\path{doi:10.1016/j.ijfatigue.2017.03.021}}.
	\newline\urlprefix\url{https://linkinghub.elsevier.com/retrieve/pii/S0142112317301391}
	
	\bibitem{leost2023}
	N.~Leost, A.~K{\"o}ster, D.~Missoum-Benziane, M.~Rambaudon, L.~Cameriano,
	F.~Comte, B.~Le~Pannerer, V.~Maurel, Full-field analysis of damage under
	complex thermomechanical loading, International Journal of Fatigue 170 (2023)
	107513.
	
	\bibitem{lannay_stability_2023}
	H.~Lannay, S.~Fouvry, B.~Berthel, C.~Gandiolle, J.~Neggers, G.~Paramucchio,
	\href{https://linkinghub.elsevier.com/retrieve/pii/S0301679X23003572}{Stability
		of the critical distance method for fretting cracking prediction: influence
		of microstructure and crack nucleation size}, Tribology International 186
	(2023) 108570.
	\newblock \href {http://dx.doi.org/10.1016/j.triboint.2023.108570}
	{\path{doi:10.1016/j.triboint.2023.108570}}.
	\newline\urlprefix\url{https://linkinghub.elsevier.com/retrieve/pii/S0301679X23003572}
	
	\bibitem{caton_small_2010}
	M.~Caton, S.~Jha,
	\href{https://linkinghub.elsevier.com/retrieve/pii/S0142112310000289}{Small
		fatigue crack growth and failure mode transitions in a {Ni}-base superalloy
		at elevated temperature}, International Journal of Fatigue 32~(9) (2010)
	1461--1472.
	\newblock \href {http://dx.doi.org/10.1016/j.ijfatigue.2010.01.015}
	{\path{doi:10.1016/j.ijfatigue.2010.01.015}}.
	\newline\urlprefix\url{https://linkinghub.elsevier.com/retrieve/pii/S0142112310000289}
	
	\bibitem{musinski2012microstructure}
	W.~D. Musinski, D.~L. McDowell, Microstructure-sensitive probabilistic modeling
	of hcf crack initiation and early crack growth in ni-base superalloy in100
	notched components, International Journal of Fatigue 37 (2012) 41--53.
	
	\bibitem{christ2009propagation}
	H.-J. Christ, O.~D{\"u}ber, C.-P. Fritzen, H.~Knobbe, P.~K{\"o}ster, U.~Krupp,
	B.~K{\"u}nkler, Propagation behaviour of microstructural short fatigue cracks
	in the high-cycle fatigue regime, Computational materials science 46~(3)
	(2009) 561--565.
	
	\bibitem{wang1999technical}
	Q.~Wang, J.~Berard, S.~Rathery, C.~Bathias, Technical note high-cycle fatigue
	crack initiation and propagation behaviour of high-strength sprin steel
	wires, Fatigue \& Fracture of Engineering Materials \& Structures 22~(8)
	(1999) 673--677.
	
	\bibitem{morel_non-local_2002}
	F.~Morel, T.~Palin-Luc,
	\href{http://doi.wiley.com/10.1046/j.1460-2695.2002.00527.x}{A non-local
		theory applied to high cycle multiaxial fatigue: {A} {NON}-{LOCAL} {THEORY}
		{IN} {MULTIAXIAL} {FATIGUE}}, Fatigue \& Fracture of Engineering Materials \&
	Structures 25~(7) (2002) 649--665.
	\newblock \href {http://dx.doi.org/10.1046/j.1460-2695.2002.00527.x}
	{\path{doi:10.1046/j.1460-2695.2002.00527.x}}.
	\newline\urlprefix\url{http://doi.wiley.com/10.1046/j.1460-2695.2002.00527.x}
	
	\bibitem{gillham_tailoring_2023}
	B.~Gillham, A.~Yankin, F.~McNamara, C.~Tomonto, C.~Huang, J.~Soete,
	G.~O'Donnell, D.~Trimble, S.~Yin, D.~Taylor, R.~Lupoi,
	\href{https://linkinghub.elsevier.com/retrieve/pii/S0142112323001032}{Tailoring
		the theory of critical distances to better assess the combined effect of
		complex geometries and process-inherent defects during the fatigue assessment
		of {SLM} {Ti}-{6Al}-{4V}}, International Journal of Fatigue 172 (2023)
	107602.
	\newblock \href {http://dx.doi.org/10.1016/j.ijfatigue.2023.107602}
	{\path{doi:10.1016/j.ijfatigue.2023.107602}}.
	\newline\urlprefix\url{https://linkinghub.elsevier.com/retrieve/pii/S0142112323001032}
	
	\bibitem{QYLAFKU1999}
	G.~Qylafku, Z.~Azari, N.~Kadi, M.~Gjonaj, G.~Pluvinage,
	\href{https://www.sciencedirect.com/science/article/pii/S0142112399000468}{Application
		of a new model proposal for fatigue life prediction on notches and
		key-seats}, International Journal of Fatigue 21~(8) (1999) 753--760.
	\newblock \href
	{http://dx.doi.org/https://doi.org/10.1016/S0142-1123(99)00046-8}
	{\path{doi:https://doi.org/10.1016/S0142-1123(99)00046-8}}.
	\newline\urlprefix\url{https://www.sciencedirect.com/science/article/pii/S0142112399000468}
	
	\bibitem{peerlings2000gradient}
	R.~H. Peerlings, W.~M. Brekelmans, R.~de~Borst, M.~G. Geers, Gradient-enhanced
	damage modelling of high-cycle fatigue, International Journal for Numerical
	Methods in Engineering 49~(12) (2000) 1547--1569.
	
	\bibitem{lin1998}
	C.-K. Lin, W.-J. Lee, Effects of highly stressed volume on fatigue strength of
	austempered ductile irons, International journal of fatigue 20~(4) (1998)
	301--307.
	
	\bibitem{wang2017combined}
	R.~Wang, D.~Li, D.~Hu, F.~Meng, H.~Liu, Q.~Ma, A combined critical distance and
	highly-stressed-volume model to evaluate the statistical size effect of the
	stress concentrator on low cycle fatigue of ta19 plate, International Journal
	of Fatigue 95 (2017) 8--17.
	
	\bibitem{elkhoukhi2021}
	D.~El~Khoukhi, F.~Morel, N.~Saintier, D.~Bellett, P.~Osmond, V.-D. Le,
	Probabilistic modeling of the size effect and scatter in {{High Cycle
			Fatigue}} using a {{Monte}}-{{Carlo}} approach: {{Role}} of the defect
	population in cast aluminum alloys, International Journal of Fatigue 147
	(2021) 106177.
	\newblock \href {http://dx.doi.org/10.1016/j.ijfatigue.2021.106177}
	{\path{doi:10.1016/j.ijfatigue.2021.106177}}.
	
	\bibitem{palchoudhary2024}
	A.~Palchoudhary, S.~Peter, V.~Maurel, C.~Ovalle, P.~Kerfriden, A plastic
	correction algorithm for full-field elasto-plastic finite element simulations
	: critical assessment of predictive capabilities and improvement by machine
	learning (2024).
	\newblock \href {http://arxiv.org/abs/2402.06313} {\path{arXiv:2402.06313}}.
	
	\bibitem{zhu2022probabilistic}
	S.-P. Zhu, Y.-L. Wu, X.~Yi, S.~Fu, J.~A. Correia, Probabilistic fatigue
	assessment of notched components under size effect using generalized
	weakest-link model, International Journal of Fatigue 162 (2022) 107005.
	
	\bibitem{elkhoukhi2020}
	E.~K. Driss, S.~Nicolas, M.~Franck, B.~Daniel, O.~Pierre, L.~Viet-Duc, Spatial
	point pattern methodology for the study of pores 3d patterning in two casting
	aluminium alloys, Materials Characterization 177.
	\newblock \href {http://dx.doi.org/10.1016/j.matchar.2021.111165}
	{\path{doi:10.1016/j.matchar.2021.111165}}.
	
\end{thebibliography}

\end{document}